\documentclass[manuscript,screen]{acmart}

\PassOptionsToPackage{dvipsnames,svgnames,table,xcdraw}{xcolor}

\usepackage{colortbl}

\usepackage{etoolbox}
\usepackage{placeins} 
\usepackage{soul}
\usepackage{graphicx}
\usepackage{float} 
\usepackage{tcolorbox}
\usepackage{multirow} 
\usepackage{array}
\usepackage[normalem]{ulem} 
\usepackage{cellspace}
\usepackage{tikz}
\usepackage{lipsum}
\usepackage{natbib}
\usepackage{hyperref}
\usepackage{pifont}
\newcommand{\cmark}{\ding{51}}  
\newcommand{\xmark}{\ding{55}}  
\AtBeginDocument{%
  }

\tcbset{
    on line, 
    boxsep=1pt, 
    left=0pt, 
    right=3pt, 
    top=0pt, 
    bottom=0pt, 
    colframe=red, 
    colback=red!10,
    fontupper=\rmfamily\bfseries
}

\definecolor{babyblue}{rgb}{0.82, 0.96, 0.936} 
\definecolor{lightred}{rgb}{0.965, 0.776, 0.733}
\definecolor{lightgreen}{rgb}{0.608, 0.780, 0.620}
\definecolor{lightblue}{rgb}{0.776, 0.800, 0.984}
\definecolor{lightyellow}{rgb}{0.976, 0.855, 0.616}

\newcommand{\driven}{\tcbox[colframe=lightred,colback=lightred,nobeforeafter]{\textsf{DRIVEN}}\ignorespaces}

\newcommand{\roads}{\tcbox[colframe=lightgreen, colback=lightgreen]{\textsf{TICKING ROADS}}\unskip}%

\newcommand{\questions}{\tcbox[colframe=lightyellow,colback=lightyellow,nobeforeafter]{\textsf{DRIVING QUESTIONS}}\ignorespaces}

\newcommand{\trivia}{\tcbox[colframe=lightblue,colback=lightblue,nobeforeafter]{\textsf{TRIVIA RIDE}}\ignorespaces}

\newcommand{\elim}[1]{\hyperref[#1]{\tcbox[colframe=lightgray, colback=lightgray]{\textsf{#1}}}}

\setcopyright{acmlicensed}
\copyrightyear{2018}
\acmYear{2018}
\acmDOI{XXXXXXX.XXXXXXX}

\acmConference[Conference acronym 'XX]{Make sure to enter the correct conference title from your rights confirmation email}{June 03--05, 2018}{Woodstock, NY}

\acmISBN{978-1-4503-XXXX-X/18/06}
\renewcommand\footnotetextcopyrightpermission[1]{}

\begin{document}
\title{Beyond Riding: Passenger Engagement with Driver Labor through Gamified Interactions}  

\author{Jane Hsieh} \orcid{0000-0002-4933-1156} 
\affiliation{%
\institution{Carnegie Mellon University}   \city{Pittsburgh}   \country{United States} } \email{jhsieh2@andrew.cmu.edu}  \author{Emmie Regan} %
\orcid{0000-0002-1504-7640} \affiliation{%
\institution{Wellesley College}   \city{Wellesley}   \country{United States} } \email{er111@wellesley.edu}  \author{Jose Elizalde} %
\orcid{0000-0002-1877-1845} \affiliation{%
\institution{University of California, Los Angeles}   \city{Los Angeles}   \country{United States} } \email{joseelizalde02@g.ucla.edu}  \author{Haiyi Zhu} \orcid{0000-0001-7271-9100} \affiliation{%
\institution{Carnegie Mellon University}   \city{Pittsburgh}   \country{United States} } \email{haiyiz@andrew.cmu.edu}

\renewcommand{\shortauthors}{Hsieh et al.}
\newcommand{\todo}[1]{\textcolor{red}{#1}}
\newcommand{\add}[1]{\textcolor{green}{\uwave{#1}}}
\newcommand{\del}[1]{{\textcolor{red}{\sout{#1}}}}
\newcommand{\janeAdd}[1]{\textcolor{blue}{{#1}}}
\renewcommand{\janeAdd}[1]{{#1}}
\renewcommand{\del}[1]{}

\begin{abstract}
Modern cities increasingly rely on ridesharing services for on-demand transportation, which offer consumers convenience and mobility across the globe. 
However, these marketed consumer affordances give rise to burdens and vulnerabilities that drivers shoulder alone, without adequate infrastructures \del{for regulatory protection }for \janeAdd{labor regulations or} consumer-led advocacy. \del{ and mutual support. To promote civic engagement and oversight around driver protections and conditions, }
\janeAdd{To effectively and sustainably advance protections and oversight for drivers, consumers must first be aware of }\del{policymakers and the public in general require a transformed perception of }the labor, logistics and costs involved with \del{rideshare }\janeAdd{ridehail} driving.
\del{In this study, }
\janeAdd{To motivate consumers to practice more socially responsible consumption behaviors and foster solidarity with drivers,}
we explore the potential for gamified in-ride interactions to \del{advance passengers'}\janeAdd{facilitate engagement with real (and lived) driver experiences}\del{ understanding, empathy and advocacy for underexposed rideshare driving conditions and driver vulnerabilities}.
Through \del{a series of}\janeAdd{nine} workshops with 19 drivers and 15 \del{riders}\janeAdd{passengers}, 
we surface how gamified in-ride interactions revealed passenger knowledge gaps around latent ridehail conditions, \janeAdd{prompt reflection and shifts in perception of their relative power and consumption behaviors, and highlight drivers' preferences for creating more immersive and contextualized service experiences, and identify opportunities to design safe and appropriate passenger-driver interactions that motivate solidarity with drivers. 
In sum, we advance conceptual understandings of in-ride social and managerial relations, demonstrate potential for future worker advocacy in algorithmically-managed labor, and offer design guidelines for more human-centered workplace technologies.}
 
\end{abstract}


\maketitle

\section{Introduction}
Platform-based gig labor occupies a growing proportion of the global workforce\del{, offering a wide variety of digitally-mediated services, ranging from driving and couriering to crowdwork and freelancing to home-based and personal care services}. Ridehail services in particular (e.g., Uber, Lyft), are crucial to advancing urban mobility \cite{anomalies}, work access \cite{ sannondisabilities}, and in stimulating local economies \cite{wealth, arrived}. 
But alongside increasing \janeAdd{consumer} adoption\del{ of gig services}, scholars raise concerns around the physical \cite{health, safety, consent, brush}, financial \cite{bargaining, beyond, reimagined} and psycho\del{logical}\janeAdd{social} \cite{hazards, making, psychosocial, buttons} working conditions\del{ of gig laborers}. 
Legal and critical scholars, for instance, challenge platforms' abuse of power through algorithmic management and control \cite{Jarrahiam, fte}, \del{as well as the absence of policy and regulations that require platform provisions of protections and}\janeAdd{which are left unchecked by labor regulations and their consequences remain unaddressed due to inadequate social} safety nets \cite{kreuger, dubal_ab5}. 

Besides platforms, consumers also shape worker conditions in significant ways\del{,}\janeAdd{: as \textit{managers} of individual interactions when they evaluate and expect service quality of workers} \del{through expectations of service quality and pricing} \cite{influence, immaterial}\del{, ratings \cite{immaterial}} and \janeAdd{as \textit{labor market regulators} who hold significant} \del{scaled}collective political power \cite{triangle, handbook}\janeAdd{. Platforms often intentionally campaign to amass such consumer support to bypass regulatory constraints imposed by legislatures, such as with Prop 22} \cite{thelen, surveillance}. 
But despite their influence, consumers remain largely unaware of laborers' harsh working realities -- Pew Research found that nearly half of Americans to have never heard of ongoing debates around the classification of ridehail drivers \cite{pew}. 

\del{Critically,}\janeAdd{In addition to cultivating political alignment with consumers \cite{thelen, legitimize}, platforms also intentionally design interfaces that suppress} 
\janeAdd{information-sharing among users \cite{asymmetries},} 
\del{algorithmic management practices that platforms employ are }\del{deliberately opaque and undocumented, obscured from consumer perception and}\janeAdd{which constrain the capacities of workers and consumers to engage in mutual support \cite{atom}}\del{ scrutiny}. 
\del{More inconspicuous tactics include (1) psychologically-controlling mechanisms like gamification \cite{making_out} and ratings \cite{rating, durlauf2019commodification}, which manipulatively promote prolonged engagement \cite{long_hours}\del{ and surveillance} (2) legal evasions of employer responsibilities including tax or tort liability \cite{corporate} and consumer misbehaviors \cite{schor} as well as (3) undisclosed and unpredictable wage adjustments that reduce and minimize worker earnings \cite{wager, vasudevangame}. }
\janeAdd{In attempts to alleviate the resulting information asymmetries,} a burgeoning body of HCI studies engage with workers \janeAdd{such as ridehail drivers} to expose the hidden and undocumented risks of gig labor \janeAdd{(\textit{e.g.}, pay rates \cite{fairfare}, deactivation \cite{fareshare}, tax or tort liability \cite{corporate}, consumer misbehaviors \cite{schor})}, offering \janeAdd{an array of} worker-centered tools to collectivize and resist \cite{gig2gether, zhang2023stakeholder, fairfare, bargaining, turkopticon}. \janeAdd{Workers also leverage a host of commercial tools to self-track data, as a means to manage their own accountabilities \cite{self_tracking}. These systems sought to empower laborers through interactions that directly interface with workers, but workers occupy relatively vulnerable positions vis-a-vis platforms, making them powerless}
\del{relying primarily on workers to push back }against \janeAdd{harmful} platform tactics and insufficient regulatory infrastructures\del{ can add to their vulnerabilities, financially, psychologically and career-wise}. Consumers, on the other hand, \del{have}\janeAdd{hold significantly} more capacity, resources and power to \janeAdd{and influence} worker rights and conditions \cite{fastdrink, handbook}. \janeAdd{But it remains unknown how technologies can surface driving labor in ways that \textit{motivate}, mobilize and catalyze consumers to advocate for and practice more socially-conscious consumption behaviors.}
\del{and insights from service design suggest that more empathetic understanding (and rapport in general) between consumers and workers create more meaningful, satisfied and pleasant interactions \cite{consumer_empathy, rapport}.}

\janeAdd{In many on-demand gig work settings, the potential to build solidarity between consumers and workers is limited by the lack of opportunities for social interaction between stakeholders, 
due to physical (\textit{e.g.,} crowdwork) or temporal distance (\textit{e.g.,} food delivery, petsitting).}
The casual\del{rideshare}\janeAdd{, co-located} setting\del{ where }\janeAdd{ of ridehail creates} \janeAdd{a unique space where} passenger-driver pairs  \janeAdd{can engage in conversations that foster mutual understanding, rapport and support. However, the temporal constraints and social boundaries of rides -- \textit{i.e.} their fleeting duration and awkwardness of engaging with strangers -- can still prevent both parties from initiating more meaningful and transformative exchanges. Raising the question: what forms of technological probes can motivate passengers to overcome these barriers and engage in more substantial and in-depth in-ride social interactions with drivers?}

\del{are generally strangers, social boundaries prevent consumers from broaching and contemplating these sensitive and uncertain topics.}

Gamification \janeAdd{represents} one approach for motivating an audience to engage \del{and empathize }with serious but sensitive prosocial causes -- \textit{e.g.,} gender-based violence \cite{standbyme}, interpersonal racism \cite{provotypes} and HIV prevention \cite{hiv} -- \janeAdd{particularly within time-constrained contexts such as platform-mediated rides}. While platforms leverage gamification for more managerial purposes \cite{drives}, transformational and persuasive games (achieved through mechanisms of ``\textit{embedded}'' messaging, interactive narratives, etc.) offer players immersive spaces to learn about or experience driving conditions \janeAdd{over a short time span} without being subjected to personally vulnerable positions.
This study explores the potentials of game-based interventions as boundary objects for \janeAdd{initiating, }mediating \janeAdd{and potentially transforming} consumer \del{education and }discourse around the obscured realities of ride\del{share}\janeAdd{hail} driving conditions. 
\del{Previous works of persuasive games revealed their potential to transform players' attitudes and perceptions on serious social issues, while creating psychological distance between the player and intended message \cite{transformational}. }
\del{Leveraging}\janeAdd{Following} relevant game design \janeAdd{techniques and frameworks for transforming player attitudes on serious social issues }\cite{transformational}, we worked with ride\del{share}\janeAdd{hail} drivers and passengers over a series of co-design sessions to explore whether gameplay interventions can \del{transform}\janeAdd{mobilize} passengers to \janeAdd{engage with,} understand\del{, empathize towards,}, care and advocate for latent labor issues of ride\del{share driving}\janeAdd{hail labor}.

\begin{enumerate}
\item[\textbf{RQ 1}] 
\del{Which gamified experiences allow effective embedding of ridesharing}\janeAdd{Which forms of gamification can effectively embed latent ridehail} concepts and experiences that drivers prioritize\del{d presenting to}\janeAdd{ sharing with} passengers?

\item[\textbf{RQ 2}] How can playable interventions motivate scalable passenger \janeAdd{engagement with, }understanding of and advocacy for the working conditions of ride\del{share}\janeAdd{hail} drivers?
\end{enumerate}

We took an iterative design process to approach these inquires, \del{starting off }\janeAdd{beginning }with goal delineation\del{ via literature review and}\janeAdd{ that combined broader insights from related ridehail studies and} formative interviews (\S\ref{goal}), followed by implementation of \del{four}\janeAdd{six} gamified prototypes (\S\ref{prototypes}) and a series of \janeAdd{nine} co-design workshops to gather driver and passenger feedback (\S\ref{iterative}).
\janeAdd{Our results show (1) passenger knowledge gaps around (latent) ridehail conditions (e.g., pay, logistics, rating pressures, long-term consequences) and early evidence for gamified interventions to mediate in-depth driver-passenger interactions and solidarity (2) driver preferences for more immersive and personalized in-ride passenger-facing play experiences and (3) design tradeoffs for shaping future passenger-driver interactions that motivate passenger-led advocacy for ridehail labor. We close by discussing prospective implications for design, advocacy and technology.}






\section{Related Work}
\janeAdd{We overview documented stressors of ridehail driving (\S\ref{stressors}), existing approaches to advocacy (\S\ref{advocacy_background}), potential for consumers to initiate influence individual service encounters (\S\ref{managers}) and regulations more collectively (\S\ref{collective}), as well as how in-ride games (\S\ref{state}) can transform (\S\ref{transform}) rather than coerce and manufacture driver labor (\S\ref{manufacture}).}
\vspace{-.75em}
\subsection{\janeAdd{App-based Ridehail Services}} \label{stressors}
App-based ridehail services have proliferated in the US market since their introduction more than a decade and half ago\del{, emerging as the largest sector of the on-demand economy \cite{making_out}}. With more than 36\% of the US adults having used ridehail services \cite{jiang2019more}\janeAdd{, Uber alone captured \$37.28 billion \cite{investoruber} in 2023 while Lyft netted \$4.4 billion}.
\janeAdd{Together with delivery work, NBER found transportation services such as ridehail to comprise the largest component of the gig workforce, as measured by the number of workers \cite{evolution}. Below, we summarize existing studies documenting subpar conditions of ridehail driving, as well as prior HCI and CSCW investigations that sought to counteract exploitative platform practices via interventions such as worker data-sharing.}
\vspace{.25em}
\subsubsection{\janeAdd{Latent Stressors,} Labor \janeAdd{and Power Asymmetry in}\del{ Vulnerabilities and Consumer Knowledge Gaps in} Ride\janeAdd{hail} Driving} 
\del{App-based rideshare services have proliferated in the US market since their introduction more than a decade and half ago, emerging as the largest sector of the on-demand economy \cite{making_out}, with more than 36\% of the US adults having used rideshare services \cite{jiang2019more}.}
\vspace{-0.95em}
Despite increasing \janeAdd{consumer adoption of app-based ridehailing services, drivers' earnings \cite{fairfare, fareshare}, job quality \cite{good, decent} and work conditions \cite{acute, pilot, crashes, safety, festering} in general have declined -- leaving them stressed \cite{stressfulride, happy, distress}, overworked \cite{fatigue, workaholism} and seeking strategies to resist platform control \cite{manufactures, laundering}.}
\janeAdd{Researchers extensively documented 
how algorithmic management and intense competition creates immense psychological stress for drivers \cite{stressfulride, distress} -- who also deal with material stressors such as low pay \cite{navigating, fairfare}, hidden health and safety risks from accidents on the road \cite{pilot, crashes, hazards, safety}, violence from passengers \cite{brush, festering}, fatigue \cite{fatigue}, job precarity \cite{understanding, instability} as well as more latent long-term consequences, including musculoskeletal and urinary disorders \cite{acute, stressfulride}. However, many harmful, latent and delayed effects \cite{acute} remain unobservable to passengers,}
\del{concern, there remains a knowledge gap between consumer perceptions of gig work such as rideshare driving and comprehensive understanding of the invisible risks, stressors, and vulnerabilities that drivers and other workers assume \cite{navigating, hazards, distress}, along with unseen immaterial, emotional and logistical labor \cite{immaterial, exploitation}. R}
\janeAdd{and computing scholars studying labor overlooked how consumers are critically elevated to positions of relative (but uninformed) managerial power, due to their objectives to center and impact worker priorities \cite{reimagined}. Specifically, platforms leverage strict 5-star rating thresholds to impose reputational pressures \cite{laundering} that exploitatively }\del{(and their accompanying deactivation thresholds) to} discipline drivers \cite{rating}.\del{
Such}
\janeAdd{ The rating mechanism} \del{coerce several forms of unpaid emotional labor from drivers, including }\janeAdd{is central to platform operations \cite{asymmetries,rosenblat_discriminating} and enable (unrealistic) consumer }expectations of \janeAdd{service quality, including requirements to maintain }\del{maintaining a} ``\textit{friendly}'', ``\textit{positive}'' and ``\textit{respectful}'' attitude\janeAdd{s}\del{to please the passenger}, regardless of how \del{riders}\janeAdd{passengers} themselves behave \cite{mediatization}. \janeAdd{Literature from organizational science \cite{org_justice, laundering} show how dominant ridehail platforms such as Uber design interactions that create greater power imbalance between consumers and workers to effectively ``\textit{launder}'' latent emotional \cite{making_out}, psychological \cite{psychological} and immaterial \cite{immaterial} labor from ridehail drivers \cite{exploitation}.
Such platform designs empower passengers without providing sufficient context on latent driver labor, which can cause uninformed and/or unconscious consumption behaviors.}
\del{But while drivers experience immense pressures to satisfy riders, passengers themselves may not be aware of the heavy implications that ratings carry \cite{immaterial, translating}. }

\vspace{-.75em}
\subsubsection{Technological Advocacy for \janeAdd{Ridehail}\del{\& Consumer Perceptions of Rideshare Driving}} \label{advocacy_background}
Scholars at the intersection of HCI and labor studies made several attempts to \del{leverage technological probes and interventions to surface and }curb the harmful impacts of algorithmic management \janeAdd{\cite{fte, Jarrahiam} and information asymmetries \cite{asymmetries} by actively involving workers to participate in the co-design and development of alternative technological (data) probes and interventions.}
\del{, as well as to advocate for and design alternative infrastructures to prioritize driver well-being.} \citet{stein2023you} imagined alternative data uses and more plural sociotechnical infrastructures with drivers\del{ to uncover key design objectives surrounding privacy, agency and utility}. \citet{zhang2023stakeholder} invited drivers to propose algorithmic imaginaries that \del{offer more worker-centered transparency, incentives and insights to drive}\janeAdd{center worker} well-being. \citet{alternative} worked with multiple stakeholder groups to reveal\del{ need for platform-based changes, technological innovations as well as civic advancements such as}\janeAdd{ (mis)aligned objectives to advance technology, service and public infrastructure, including} more accurate public perceptions of workers. \del{Recent studies stepped b}\janeAdd{B}eyond co-design\del{ to reveal potential for}\janeAdd{, recent studies showed the promise for} data probes \cite{zhang2023stakeholder}\del{ as well as}\janeAdd{,} data-sharing tools \cite{gig2gether, fairfare, bargaining} and collectives \cite{workshop} to\del{ advocate and} elevate worker priorities\janeAdd{ as well as labor regulations and protections \cite{fairfare}}.
While these proposed interactions demonstrated shared worker motivations, they require effortful data contributions from \del{workers}\janeAdd{drivers}, many of whom labor for long hours to balance financial needs \cite{good} with unstable job opportunities \cite{instability}, making it \janeAdd{impractical}\del{infeasible} for them to engage in additional (uncompensated) interactions. 



\subsection{\janeAdd{Role of Consumers in Ridehail Labor}}
\janeAdd{Given the capacity constraints of workers, we turn to the untapped potential of mobilizing consumers as users of worker-advocacy technologies. First, we examine the role of passengers in managing frontline ridehail service (quality) before showing how as a collective, consumers can offer workers \textit{\textbf{political leverage over platform decisions}}.}
\subsubsection{\janeAdd{Passengers as Co-managers of Ridehail Service}}\label{managers}
\janeAdd{Besides the harmful effects of algorithmic management, \textbf{consumers} also play a crucial role in rating systems \cite{triangle}, which are intentionally designed to provide safe, reliable and quality service to paying users \cite{quality}. 
Initially, maintaining quality service was necessary for platforms to amass a sizable consumer base and establish market position \cite{laundering}; but over time, strict deactivation thresholds \cite{rating_service} became  strategic political tools to forge political alliances with consumers, who can help defend against (\textit{e.g.,} Prop 22 \cite{thelen}) or even preempt \cite{disrupt} emerging regulations that threaten platform operations. 
Combined with platform nudges \cite{long_hours} and mechanisms (\textit{e.g.,} tipping \cite{tipping}), the resulting reputational pressures discipline drivers into performing unpaid service -- \textit{e.g.,} maintaining a friendly mood \cite{mediatization} or clean car \cite{quality}, playing comfortable music \cite{immaterial}, dressing professionally \cite{laundering}, offering water, snacks or phone chargers \cite{schor}. 
The relative empowerment of passengers also enables abuse and disrespect towards drivers \cite{laundering, org_justice, incivility}, not to mention legally questionable practices -- \textit{e.g.,} sneaking in too many people, or alcohol \cite{laundering}. 
In response, drivers must undertake yet more emotional labor to absorb or resist unjust treatment \cite{sabotage}, or to proactively educate passengers about consequences through in-car fliers or conversation \cite{asymmetries}. }

\janeAdd{By redistributing, decentralizing and delegating \textbf{managerial oversight} to individual consumers, platforms (1) sustain competitive pools of drivers \cite{rating_service, schor} and (2) sidestep employer obligations when they circumvent the need to establish explicit service rules \cite{laundering, immaterial, thelen, power_resources}. Prior work highlight the power of algorithmic management \cite{stein2023you, workshop}, but overlooked how \textit{consumer} data contributions (\textit{e.g., }ratings) constitute ultimate sources of this power \cite{power_resources}. Consequently, drivers are coerced into unpaid emotional labor, akin to traditional frontline services (\textit{e.g.,} call centers \cite{laundering}) and more face-to-face, yet increasingly algorithmically-managed sectors like hospitality \cite{spektor, f2f}.}

\del{Meanwhile, the ways rideshare passengers perceive and interact with driving conditions remain relatively underexplored.} 

\del{In food delivery, \citet{fastdrink} began probing this space by prototyping an interaction providing users with their courier's demographic information during waiting time, which shifted users away from affective empathy, but toward compassionate empathy -- an experience that incentivizes further prosocial actions to help others \cite{compassion}. 
But while technology-mediated interactions show promise to foster users' interpersonal empathy for individual workers, it remains unclear whether they hold the potential to cultivate consumer empathy in a way that motivates them to further care, take action and advocate for vulnerabilities that affect the broader, scaled ridesharing driving workforce -- objectives related to but opposing the intents of ``\textit{consumer empathization}'', which rideshare platforms adopt to establish legitimacy for their businesses \cite{legitimize}.}

\vspace{-.5em}
\subsubsection{\janeAdd{Collective Consumer Influence on Policy \& Regulation}}\label{collective}

\janeAdd{Crucially}, how \janeAdd{drivers and} consumers perceive a platform's \janeAdd{labor conditions can} influence their use, recommendation of \cite{role}\janeAdd{, or even retaliation against \cite{sabotage} it}
\janeAdd{. Competing companies (\textit{e.g.,} Lyft \cite{laundering}) with less coercive practices can attract  socially conscious consumers \cite{role}, threatening dominant platforms}\janeAdd{. Thus, consumers hold scalable} political power\del{ at scale } \cite{sceptics}\janeAdd{ that} platforms actively seek to influence. 

\janeAdd{\citet{thelen}'s comparative political analysis showed how Uber built coalitions with consumers to shield \cite{primed} against regulatory challenges. When London refused to renew their taxi service license in 2017 for its ``\textit{lack of corporate responsibillty}'' \cite{guardian}, Uber successfully mobilized half a million users within one day using its ``\textit{tried-and-tested tactic of asking customers for help when it locks horns with regulators}'' \cite{tested, million} -- resulting in license reinstatement. 
\citet{sceptics} explored} how workers might ``\textit{gain support from consumers they serve}'' \cite{triangle}\janeAdd{, to influence not only individual service interactions (managerially) but also broader platform accountability through collective political power \cite{handbook}.} 

\janeAdd{However, consumers currently lack awareness of the need to further regulate ridehail platforms: }while consumers with knowledge about driver classification were 20\% more likely to support additional regulation,
\janeAdd{over 40\% of American adults had never heard about the debate around the non-employee status of ridehail drivers in 2021 \cite{pew, kreuger} -- shortly after Prop 22 overturned Assembly Bill 5. }
Meanwhile, 2022 surveys show conflicted consumer opinions about platform-based labor, especially regarding hidden aspects of working conditions -- e.g., long-term consequences on career \cite{triangle}. 
\janeAdd{But \citet{triangle} also revealed a cooling trend in consumers' favorable perceptions of platforms, uncovering an opportunity to awaken their citizen identities \cite{primed} and mobilize consumer solidarity for workers through their concern for causes such as social responsibility \cite{power_resources}, environmentalism \cite{role} or philanthropy \cite{csr}. }

\subsection{\janeAdd{Gamification to Motivate Citizen Engagement}} 
\janeAdd{Prior work has attempted to design for consumer  awareness, compassion and action \cite{awareness} in the gig economy, but educating and motivating the geographically-dispersed population of ridehail passengers can be challenging, especially if constrained by the short time span of a single ride. Below, we summarize how platforms design algorithmic games to incentivize driver labor, the potential for gamification to rapidly mobilize passengers for driver advocacy, as well as the state of the art on games designed for, or related to ridehail.}

\subsubsection{\janeAdd{Gamification: Manufacturers of Consent vs Drivers of Transformation}} \label{manufacture}
\janeAdd{Platforms employ several gamification techniques to} psychologically trick and coerce labor \cite{making_out, prabowo2019does, tricks} -- which drivers resist \cite{vasudevangame}. Mechanisms include Quests and Challenges that yield monetary prizes, as well as more symbolic reward structures such as badges, points and status programs  \cite{vasudevangame, rating}. \janeAdd{But while such tactics motivate initial driver engagement, managerially-imposed gamification also take away their ``\textit{sense of consent}'' and agency to set practical boundaries with work \cite{terrain, mandatory}. Worse, the combination of opaque but rapid algorithmic rewards result in paradoxical senses of stressful and frustrating motivation that cannot be resolved by human managers \cite{terrain, asymmetries}. Such managerial gamification \cite{mandatory} leverage phenomena such as ``\textit{income targeting}'' from behavioral economics to trick drivers into ``\textit{grinding}'' away for long hours \cite{buttons}.}
\del{To compound, information asymmetries deprive drivers' individual agency when platforms choose to withhold key details of a trip -- e.g., exact destination and fare \cite{asymmetries}. }

\janeAdd{On the other side of managerially controlling workplace games is the possibility of more humanistic and transformative game play \cite{transformational, humanistic}, where players engage in autonomous, and growth-oriented games that guides them to grow, make meaning and learn -- such as when drivers practice oppositional play to resist workplace games \cite{vasudevangame, making_out}. Transformational game design \cite{tandem, transformational} adopts the latter, less managerial rhetoric and opens up immense opportunities to embed principles from various disciplines (\textit{e.g.,} asymmetric paternalism from behavioral economics \cite{paternalism}, incentive compatibility from mechanism design) to develop games that illustrate alternative choice architectures that motivate better player decisions and growth that can extend beyond the game environment into real-life choices.}

\subsubsection{\janeAdd{Games to Transform Passenger Engagement with Driver Experiences \& Labor}}\label{transform}
A key barrier to approaching \janeAdd{and discussing} the challenges and mechanisms of ridehail work is the sensitive and private nature of financial and emotional vulnerabilities that drivers experience \cite{sannondisabilities}, which \janeAdd{can} prevent consumers from \del{learning}\janeAdd{probing} about hidden driving labor and logistics (\S \ref{stressors}).
Gamified environments and gameful designs \cite{gameful} present an opportunity to build safe and inclusive spaces that can rapidly engage player awareness of sensitive \cite{kaufman2stealth, provotypes}, complex, and overlooked topics \cite{seat}. 

Games have functioned as a medium for promoting critical thinking and social consciousness around various pressing societal issues, ranging from racism (e.g., \textit{SimCity} \cite{consciousness}) to colonialism (e.g., \textit{Civilization} \cite{civ}) to capitalism (e.g., \textit{Animal Crossing} \cite{crossing}, \textit{World of Warcraft,  Second Life} \cite{negotiate}), including specific dimensions such as immaterial labor (e.g.,  \textit{Mario} \cite{wow}). Persuasive games intentionally leverage techniques like procedural rhetoric (the use of rules, mechanics and decisions) to model and portray social systems \cite{persuasive}, embedded approaches (e.g., distancing and intermixing) to address controversial topics, and immersive methods like narrative role-play and role reversal \cite{mr_empathy, equity} to affectively engage players in more marginalized and constrained perspectives
\cite{papers, narration}. \janeAdd{Temporal features such as immediate feedback and timed challenges/pressure \cite{self_brand} are shown to foster better and quicker user engagement \cite{enhancing}. In ridehail, gamification offer ideal opportunities to rapidly engage players by}
\begin{enumerate}
    \item \janeAdd{psychologically distancing the driver and passenger, who can then feel more }
    \begin{enumerate}
        \item \janeAdd{safe and motivated to explore driving conditions through immersive and (temporally) challenging spaces}
        \item \janeAdd{socially comfortable to discuss legitimate and lived driving experiences with drivers}
    \end{enumerate}
    \item \janeAdd{simulating gamification tactics that platforms impose to exert psychological control.}
\end{enumerate}

\subsubsection{\janeAdd{State of the Art for Ridehail Games}}\label{state}
\janeAdd{Presently, we are aware of only one system occupying the space of in-ride interactive games: the \href{https://playoctopus.com/}{Play Octopus Network}, which markets itself as the world's ``\textit{largest rideshare advertising network}'' and provides drivers in-car tablets with advergames such as trivia that primarily function to engage passenger attention for generating advertising revenue. 
\citet{cards} attempted to leverage gamified role-playing to mobilize drivers, which presupposes additional effort and capacity that most drivers lack. A few browser-based games also exist that intend to capture a broader audience, including the \textit{Uber Game} \cite{UberGame} (which won an Online Journalism Award) and \textit{Cherry Picker} \cite{Bitter}, but we remain unaware of any games designed for the situated context of a ride. 
\del{as the primary player audience not only requires extra effort from already-burdened workers (\S\ref{advocacy_background}), it also forfeits the opportunity to engage consumers, a population containing both potential driver advocates and future drivers, in gaining more knowledge around hidden risks and conditions of rideshare driving.}
}
\vspace{-0.5em}
\section{Design Criteria for Gamified In-ride Interactions} \label{connection}
To effectively convey hidden driving labor and conditions to passengers, we identified relevant criteria from game design and heuristics to support our study goals -- \janeAdd{described below and summarized in Table \ref{criteria_table}}.
Through pilot interviews and goal refinement (detailed in \S\ref{goal}), we also identified three additional ridehail-specific criteria in \S\ref{specific}. 

\subsection{Game Design Heuristics} \label{technique}

\subsubsection{Replayable} \label{replayable}
One metric for assessing a game's engagement potential is the player's desire to play again \cite{HEP}. Replay can enhance learning and understanding of game content \cite{edu_replay}, which is crucial first step to our goal of engaging passenger interest in ridehail labor conditions. Replays also promote social interaction among players (e.g., through discussion of its content) \cite{replayability}, which further supports our objective of mobilizing consumers.

\subsubsection{(Timed) Challenge} \label{challenge} Another standard  playability heuristic in both video and mobile games \cite{HEP, mobile} revolves around the presence of a meaningful challenge -- i.e. the level of difficulty that players must overcome to reach a winning condition -- which is central to creating enjoyable experiences.
\citet{malone} defined that challenging games must contain ``\textit{a goal whose outcome is uncertain}'', while \citet{play} also considered temporal factors: ``\textit{well-paced challenge(s) that makes the game worth playing}''. 
In ridehail games, timed challenges additionally simulate drivers' realistic time constraints \cite{stressfulride}. Despite incorporating challenges, we refrained from incentives such as leaderboards or challenge invites to contacts (\textit{e.g.,} friends or family) since they can trivialize driver vulnerabilities \cite{ensitive, standbyme}.

\subsection{Embedded Design Strategies}
\citet{kaufman2stealth} recommends ``\textit{stealthily}'' embedding persuasive messaging to ease players into receiving intended messages. We overview below ways to leverage embedded design strategies --- i.e., intermixing, obfuscating and distancing --- to effectively communicate latent ridehail driving conditions.

\subsubsection{Intermixing} \label{intermix}
Interspersing thematic on-message (\textit{e.g.,} ridehail content) and playful off-message (\textit{e.g.,} narrative content) material guides players into subconsciously internalizing a game's intended themes -- offsetting initial player discomfort and reservation when presented with emotionally-taxing topics \cite{chimeria}. 
Passengers who resist explicitly acknowledging the effects of their consumption behaviors may prefer interwoven content over overt content delivery.

\subsubsection{Obfuscating} \label{obfuscate}
To further bypass players' psychological defense, obfuscation conceals the persuasive intent of purposive games, reorienting player attention toward apparent game mechanics and objectives. 
The technique has been applied to topics such as bias against women in STEM \cite{Freedman}, complex social identities \cite{buffalo} and health advocacy \cite{kaufman2stealth}. 
By introducing serious and persuasive material messages covertly, obfuscation helps passengers avoid feeling pressures from their consumer identities (or associated managerial powers) while still provoking critical reflection from the user. 

\subsubsection{Psychological Distancing through Immersive Fictional Narratives} \label{distancing}
Narrative fiction is an effective medium for communicating complex and sensitive social experiences, including gender-based violence \cite{standbyme}, interpersonal racism \cite{provotypes}, healthcare \cite{fun}, and climate change \cite{climate}. 
Immersive narrative framings of social experiences also facilitate reflection \cite{close}, empathetic growth \cite{intent, growth} and prosocial behaviors \cite{video, prosocial}, allowing us to create safe spaces for players to explore sensitive driver topics in the first-person without directly experiencing harmful or disturbing work conditions.
\textit{Interactive} narrative fictions further enable designers to directly implement choice architectures (and principles such as asymmetric paternalism), allowing players to make firsthand decisions that influence in-game plots while building their sense of moral responsibility and self-efficacy \cite{moral}. 
In ridehail, interactive fiction (1) augments player understanding of driver labor while (2) revealing more growth-oriented and socially responsible choices.

\begin{table}[h!]
\centering
\resizebox{\textwidth}{!}{
\begin{tabular}{lcc|ccc|lcl}
                                                         & \multicolumn{2}{c|}{}                                                                         & \multicolumn{3}{c|}{\textbf{Embedded Design}}                  & \multicolumn{3}{c}{\textbf{Ridehail-Specific}}                                                                                                                                                                                                     \\ \cline{2-9} 
\multicolumn{1}{l|}{\textit{}}                           & \multicolumn{1}{c}{\textit{Replayability}} & \multicolumn{1}{c|}{\textit{\begin{tabular}[c]{@{}c@{}}Timed \\ Challenge\end{tabular}}} & \multicolumn{1}{c}{\textit{Obfuscating}} & \multicolumn{1}{c}{\textit{Intermixing}} & \multicolumn{1}{c|}{\textit{\begin{tabular}[c]{@{}c@{}}Fictional \\ Narrative\end{tabular}}}   & \multicolumn{1}{c}{\textit{\begin{tabular}[c]{@{}c@{}}Ground truth \\ answers\end{tabular}}} & \multicolumn{1}{c}{\textit{\begin{tabular}[c]{@{}c@{}}Playable \\ in-ride\end{tabular}}} & \multicolumn{1}{c}{\textit{\begin{tabular}[c]{@{}c@{}}Driver-Passenger \\ Interactions\end{tabular}}} \\ \hline
\multicolumn{1}{l|}{
\textit{\textbf{\elim{CrossRoads}  }}}        & \checkmark                    &                        & \multicolumn{1}{c}{\checkmark}                    & \multicolumn{1}{c}{} & \multicolumn{1}{c|}{}                                                                         & \multicolumn{1}{c}{\checkmark}                                                    & \checkmark                                                                                        & \multicolumn{1}{c}{\checkmark}                                                         \\
\multicolumn{1}{l|}{\textit{\textbf{\elim{Dilemmas @ Work}}}}   & \checkmark                      &                        & \multicolumn{1}{c}{} & \multicolumn{1}{c}{\checkmark}                    & \multicolumn{1}{c|}{}                                                                         &                                                                          & \multicolumn{1}{l}{}                                                                     &                                                                               \\
\multicolumn{1}{l|}{\textbf{\driven}}            & \checkmark                      &                        & \multicolumn{1}{c}{\checkmark}                    & \multicolumn{1}{c}{} & \multicolumn{1}{c|}{\checkmark}                                                  &                                                                          & \checkmark                                                                                        &                                                                               \\
\multicolumn{1}{l|}{\textbf{\trivia}}        & \checkmark                      & \checkmark & \multicolumn{1}{c}{} & \multicolumn{1}{c}{\checkmark}                    & \multicolumn{1}{c|}{}                                                                         & \multicolumn{1}{c}{\checkmark}                                                    & \checkmark                                                                                        & \multicolumn{1}{c}{\checkmark}                                                         \\
\multicolumn{1}{l|}{\textbf{\questions}} & \checkmark                      &                        & \multicolumn{1}{c}{} & \multicolumn{1}{c}{\checkmark}                    & \multicolumn{1}{c|}{}                                                                         &                                                                          & \checkmark                                                                                        & \multicolumn{1}{c}{\checkmark}                                                         \\
\multicolumn{1}{l|}{\textbf{\roads}}              & \checkmark                      & \checkmark & \multicolumn{1}{c}{\checkmark}                    & \multicolumn{1}{c}{} & \multicolumn{1}{c|}{\checkmark}                                                  &                                                                          & Mobile-Only                                                                              &                                                                              
\end{tabular}
}
\caption{\janeAdd{How game prototypes met game (including embedded) design and ridehail-specific criteria}}
\label{criteria_table}
\end{table} 
\FloatBarrier
\subsection{Ridehail-specific Criteria} \label{specific}
We conducted pilot interviews with 2 drivers (D1, D2 -- recruited from past studies) and 3 passengers (P1-P3) to elicit ridehail-specifc requirements to complement our own knowledge of app-based labor.

\subsubsection{In-Ride Compatibility} \label{compatibility}
Pilot passengers expressed a common preference for ``\textit{lightweight}'' and easy-to-pickup games to minimize chances of car sickness, although P2 and P3 also desired realistic simulations of ridehail driving. Pilot driver D2 suggested using the Octopus tablet currently in their car to integrate content in more natural ways while D1 cautioned how embedded content should not come across as a way for drivers ``\textit{vent your complaint}'' to passengers. 

\subsubsection{Driver-Passenger Interactions}
When discussing preferred genres, P2 indicated interest in simulations that shed light on how drivers interact with and ``\textit{talk to the person[/rider] in the backseat}'', since they're not a fan of actual driving whereas P3 suggested interactions where ``\textit{you have to talk to the driver, or engage with them}''.

\subsubsection{Ground Truth Answers}
To most effectively reconfigure and transform the role of passengers, concepts should convey accurate information regarding ridehail labor. Factual and apolitical content makes it easier for players to trust and learn, in addition to reaching larger audiences. 

\section{Playable Prototypes of Passenger-facing In-ride Games} \label{prototypes}

Drivers of the first two workshops made several concrete and actionable suggestions that informed our decisions to remove two prototypes -- \elim{CrossRoads} and \elim{Dilemmas @ Work}, which we describe more in the Appendix. Table \ref{timeline} offers a timeline that summarizes workshop order, and the presence or development of prototypes between sessions.
Early driver feedback also inspired the addition of \roads, a driver-facing settings page to indicate (1) topics to avoid in discussion and (2) currently preferred interaction levels with passengers, which range from ``\textit{not at all}'' to ``\textit{anytime}'', the selection of which dynamically reconfigures the passenger-facing menu screen -- \textit{e.g.,} more interactive games get grayed out when the driver chooses ``\textit{not at all}''. 

\begin{figure}[h!]
    \centering
\includegraphics[width=.9\linewidth]{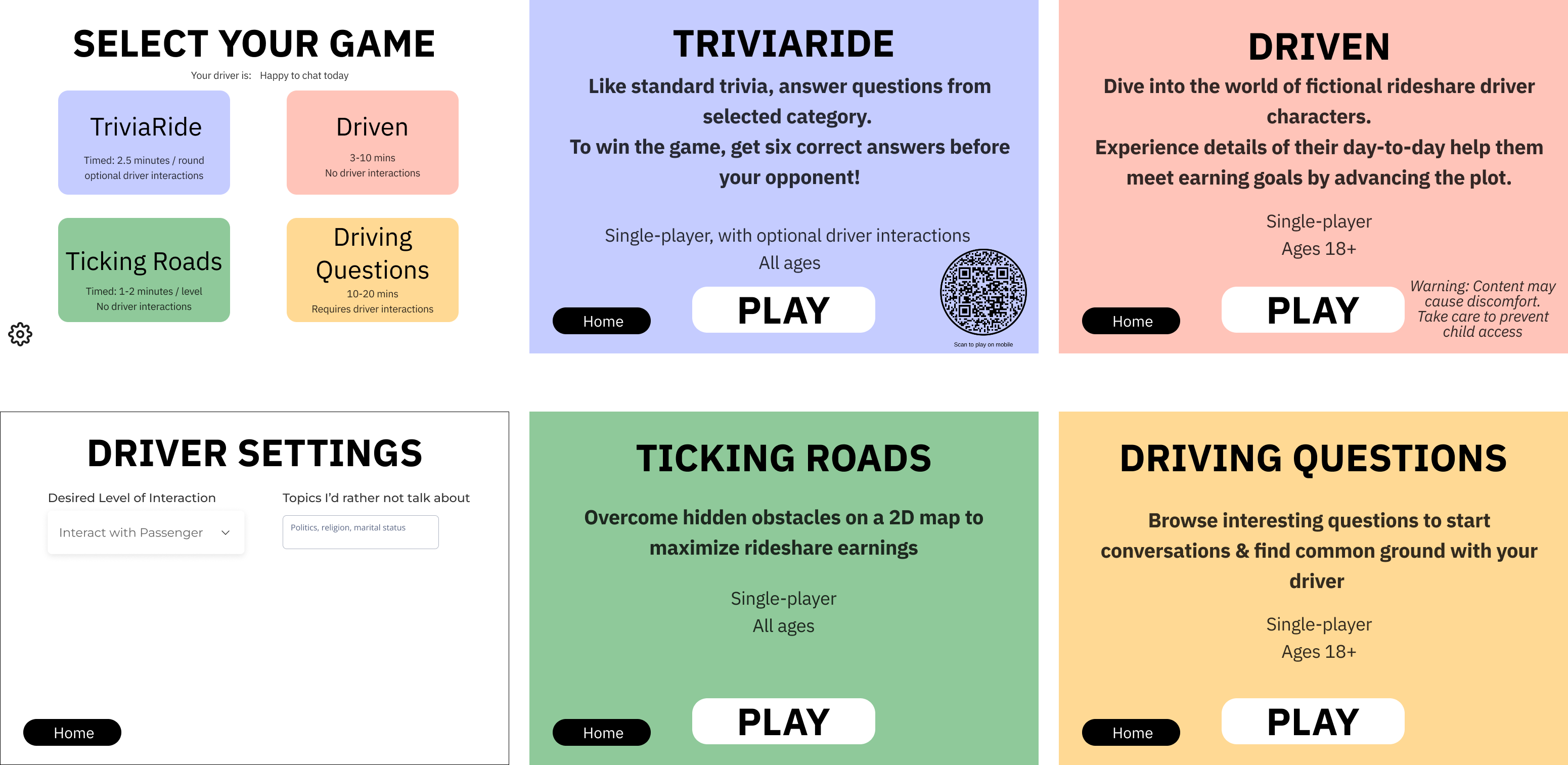}
    \caption{Menu informs passengers about expected content and interactions, as well as driver preferences when selecting a game}
    \vspace{-1em}
    \label{menu}
\end{figure}
\FloatBarrier

\subsection{Visual Novel: {\driven}} 
\driven is a visual novel (\textit{i.e.} a digital narrative) with interactive decision points that branch into multiple storylines -- creating spaces for players to experience ridehail conditions with psychological distance (\S\ref{distancing}). Pilot passenger P1 highlighted to us the value of more casual and broadly accessible experiences -- \textit{i.e.} ``\textit{games that are much more about the story \dots messaging \dots [to] make it easier, more accessible for everyone, even people who are not used to playing games}.'' Consequently, our choose-your-own-adventure\janeAdd{ offers} two driver characters whose plotlines unfold based on decisions that players make on their behalf to work towards income goals while balancing stressors from passengers and life. 

\begin{figure}[h]
    \centering
\includegraphics[width=.9\linewidth]{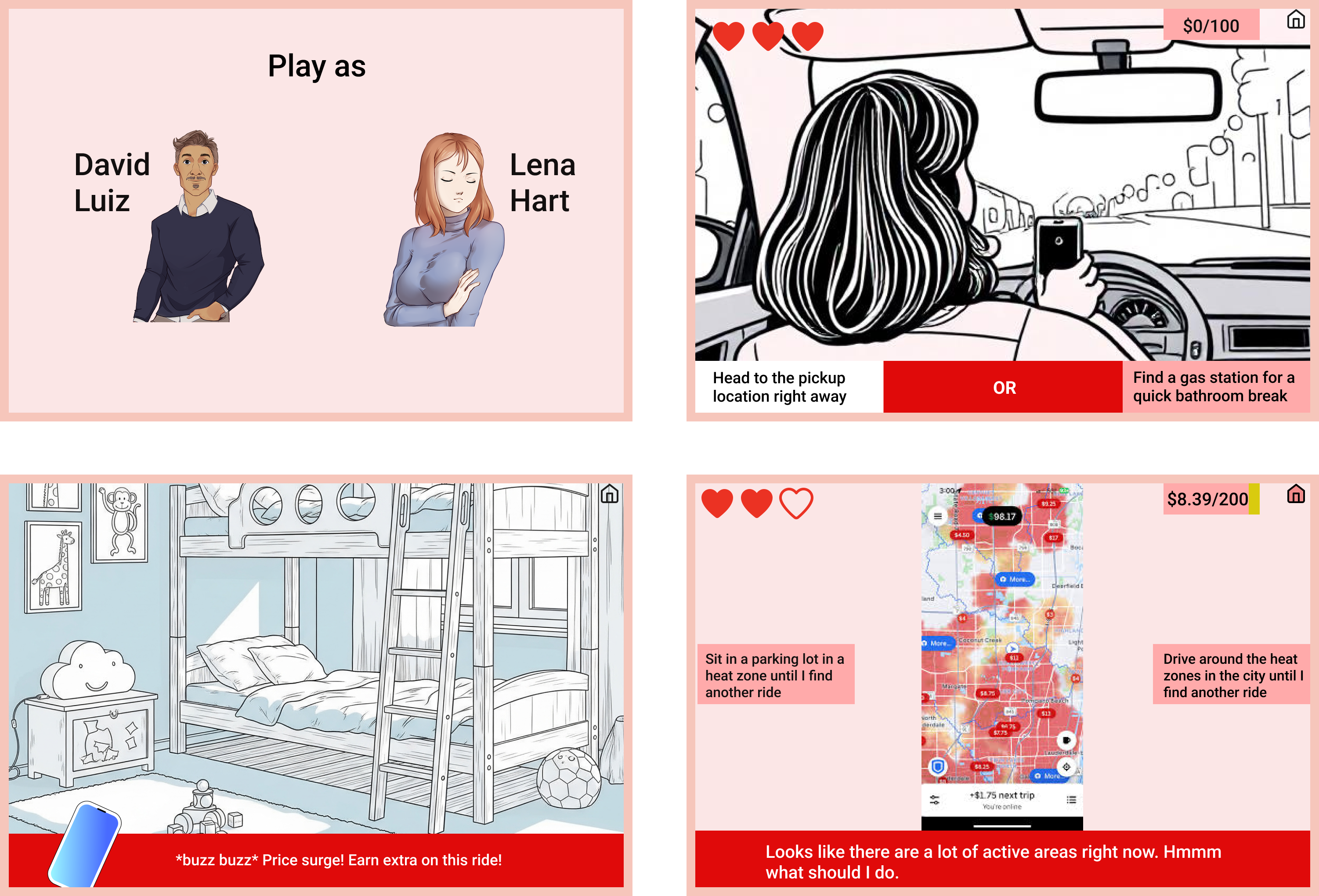}
    \caption{\driven is visual novel with point-and-click options that advance plotline of two NPC ridehail drivers}
    \label{crossroad}
\end{figure}

While \driven was heavily inspired by \textit{Cherry Picker} \cite{Bitter} and The \textit{Uber Game} \cite{UberGame}, it uniquely adopts a first-person perspective that leverages the persuasive power of procedural rhetoric (\S\ref{distancing}) to situate players within more complex issues -- \textit{e.g.,} work-life balance, algorithmic management, deadheading, driver-passenger (power) dynamics and interactions.
For instance, \driven foregrounds driver responsibilities -- \textit{e.g.,} those arising from family or primary occupations (main jobs) -- which prime players to consider the non-driving goals of drivers, which subtly intermixes narrative content while redirecting player attention away from the game's persuasive intent -- \textit{i.e. } obfuscation.
The divergent branching storylines provide high replayability, offering varied experiences across play sessions that drives player engagement and transformed understandings of driver-centric perspectives. 
Peripherally, passengers in-ride gameplay can also prompt further questions, conversations or reflections with the driver. 

\subsection{\trivia} 
Based off of the multiplayer game \textit{Trivia Crack}, \trivia challenges players to reach six correct answers before a hypothetical opponent. The game contained a mix of general trivia -- spanning categories like Social Studies, Science, Pop Culture, and the Arts (e.g., `\textit{What sport has been played on the moon?}') -- intermixed with ridehail-related policies or facts (e.g., `\textit{What law classified drivers as independent contractors in CA?}'). All questions contain a verifiable ground-truth answer, and
the ridehail questions are desigend to require minimal background knowledge, allowing the players to quickly learn from prompts like `\textit{Which location is the \textbf{most} lucrative for rideshare driving at 3am?}'. 

\FloatBarrier
\begin{figure}[h]
    \centering
\includegraphics[width=.65\linewidth]{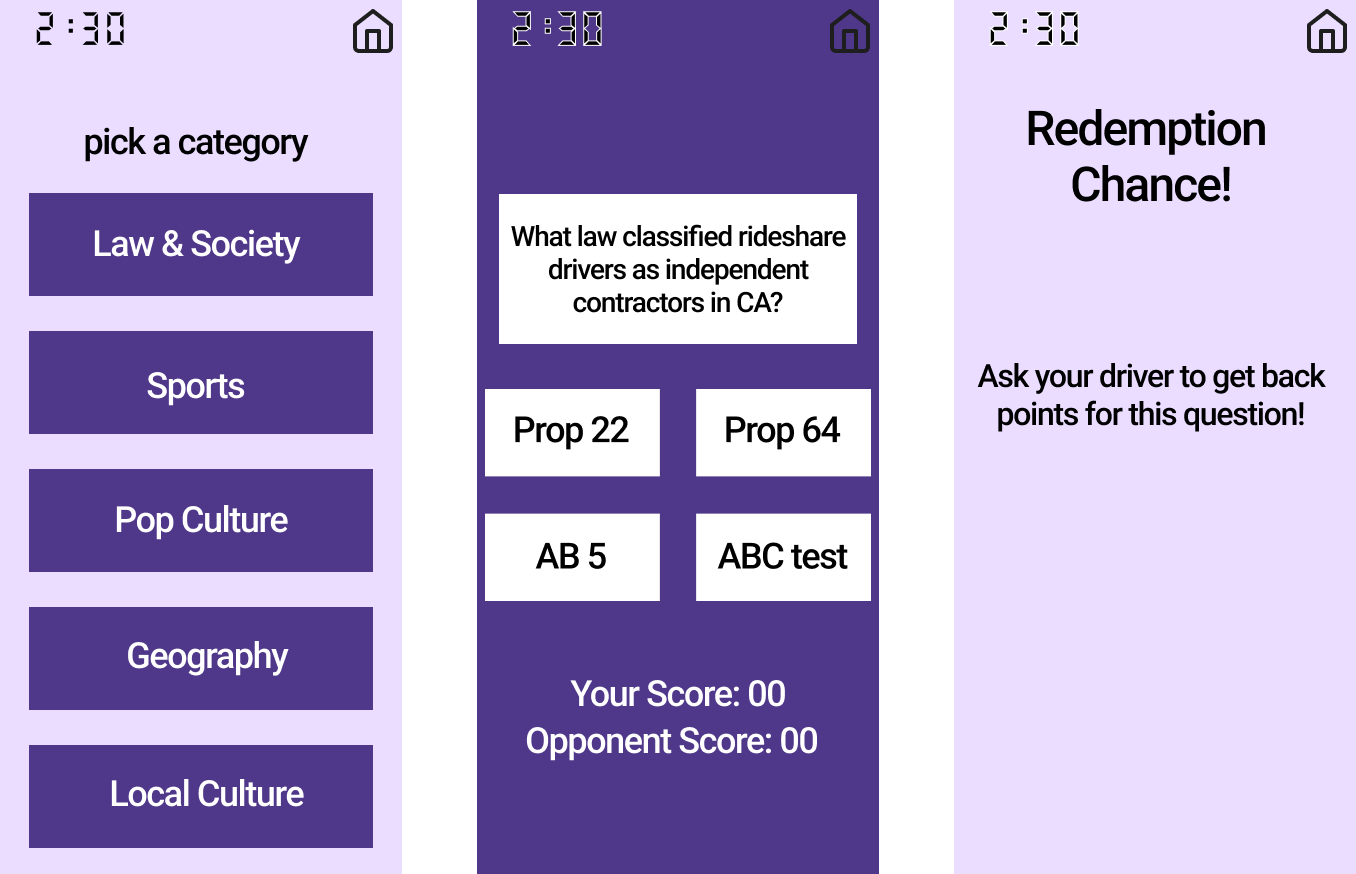}
    \caption{\trivia is a timed challenge with optional driver interactions and embedded ridehail concepts}
    \label{crossroad}
\end{figure}

\trivia also incentivizes driver-passenger interaction by giving ``\textit{redemption}'' chances after incorrect responses, delivered through an interface prompt that encourages the player to verbally ask their driver for help. \janeAdd{This design was motivated} by D2.2's suggestion of ``\textit{adding a driver-passenger collab mode would be super cool \dots A game where we solve a puzzle together}''. 
In later versions of the game, the first-to-six objective was replaced with a timed, point-based system, following D2.2's recommendation that ``\textit{putting a timer would be very good [to provide] urgency to answer the question}'' -- in alignment with the timed challenge criteria (\S\ref{challenge}).
\FloatBarrier
\vspace{-2em}
\subsection{\questions} \label{questions}

Inspired by the game \textit{We’re Not Really Strangers} (WRNS) and driver D1.2's suggestion for more driver-passenger interactions and emotional connections, \questions is a conversation prompting game that serves as boundary object to \janeAdd{initiate and }mediate conversations during a ride -- helping keep discussions topically related to driving labor while allowing the driver-passenger pair to get to know each other as people. We repeatedly updated the content to minimize intrusive content (concepts embedded include mental health impact, logistics, take rate) while intermixing more locally-grounded questions. Since driver-passenger pairings are almost always unique, replayability of \questions is strong and its selection of 18 conversational questions also increase response variation with every new driver-passenger combination. 
\begin{figure}[h]
    \centering
\includegraphics[width=.8\linewidth]{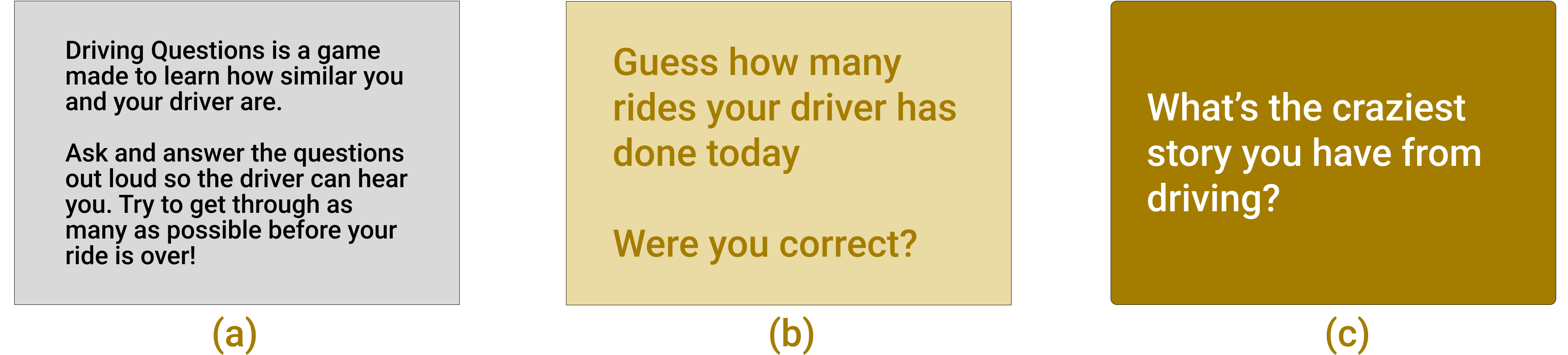}
    \caption{\questions bridges the driver-passenger social gap with conversation prompts for passengers (b) and drivers (c)}
    \label{crossroad}
\end{figure}
\vspace{-2em}
\subsection{\roads} 
Borne from the recommendation of drivers from the first session, \roads is a time management game \janeAdd{(similar to \textit{Diner Dash} and \textit{Overcooked})} that surfaces stressors of the road to players by simulating ridehail obstacles on a map. Players undertake the task of picking up passengers at designated locations and receive immediate feedback when performing actions (e.g., move around, speed up/down, wait at pick up location). By framing stressors as obstacles and introducing time pressure, \roads diffuses (or obfuscates) the tensions of exposing dangerous road conditions. It also creates psychological distance between the player and realities of driving on the road by presenting a fictional, abridged simulation of logistical burdens that drivers regularly encounter.
\begin{figure}[h]
    \centering
\includegraphics[width=.9\linewidth]{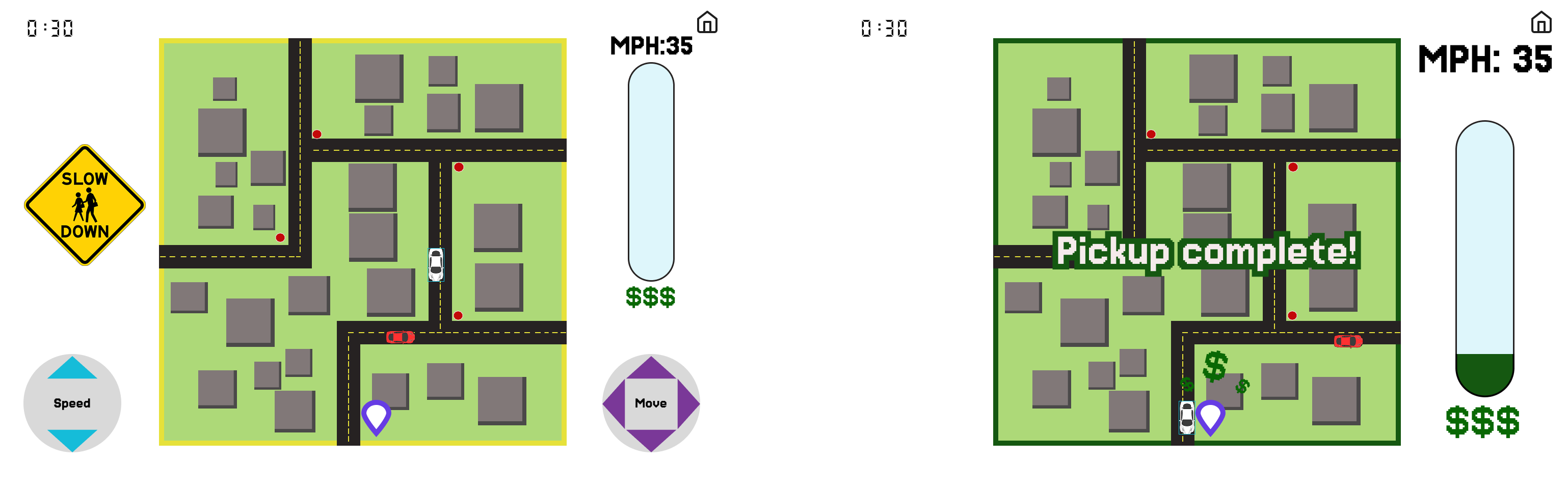}
    \caption{\roads simulates passenger pickup (logistics) and immediate feedback from ``\textit{driving}'' a car on the map}
    \label{ticking}
\end{figure}
\vspace{-2em}
\section{Methods} \label{method}
Given the unconventional and multi-disciplinary problem space (i.e. \janeAdd{raise awareness around realistic ridehail driving conditions}), intended audience (i.e. passengers) and goal (i.e. \janeAdd{motivate and mobilize consumers to care and advocate for driver labor}), we followed the (Tandem) Transformational Game Design process \cite{tandem, transformational}. All interviews
\subsection{Phase 1: Goal Delineation} \label{goal}
Starting with a cycle goal delineation, we synthesized relevant and recent \textbf{ridehail literature} to surface key but potentially underexposed conditions -- summarized in \S\ref{stressors}. In parallel, we identified potential techniques and genres from scholarship on transformational, serious and persuasive \textbf{game design} \janeAdd{that may support our goal of motivating passengers' perception change around ridehail driving conditions (\S\ref{technique})}. Next, we conducted \textbf{formative interviews} with 2 drivers and 3 passengers (as mentioned in \S\ref{specific}) to garner initial ideas and understanding around latent topics that drivers desire to communicate to passengers, levels of comfort and concern for a passenger-facing game addressing such issues, as well as preliminary reactions around (and suggestions for) potential game genres.
\subsection{Phase 2: Iterative Game Prototyping} 
\label{iterative}
\janeAdd{After the initial round of goal delineation, we presented three prototypes to drivers across two workshops, followed by a second round of goal refinement before moving on to passenger feedback. Table \ref{timeline} shows when prototypes were conceived, developed, presented and eliminated while} Table \ref{receptions} shows how people perceived prototypes across sessions.

\subsubsection{Driver Workshops}
\vspace{-.25em}
\janeAdd{Based on feedback from the formative study in Phase 1, we began implementing three game prototypes (\S\ref{prototypes}) and presented these to drivers across DW1 and DW2. In driver sessions, we inquired about issues they prioritized to share with passengers, probed for initial reactions and hesitation to prototypes and embedded concepts, and elicited ideas for alternative game designs and concepts to embed that align with the overarching goal -- see supplementary materials for detailed driver workshop protocols. }Then, we presented three initial prototypes -- \roads was not yet conceived/developed --  through two initial workshops with drivers (DW1, DW2). Afterwards, we completed another round of goal delineation \cite{tandem} to
\begin{enumerate}
    \item conceptually map (see supplementary materials) relevant concepts in ridehail driving to refine \S\ref{stressors} 
    \item highlight concepts at the intersection of
    \begin{enumerate}
        \item what drivers prioritize to communicate with passengers and
        \item conditions and vulnerabilities that are underexposed to passengers.
    \end{enumerate}
\end{enumerate}
Incorporating insights from goal refinement and initial driver feedback, we iterated on game mechanisms and content, and implemented \roads. To continuously adapt prototypes based on feedback from both stakeholder groups, we held three more driver workshops (DW3-DW5), interspersed among passenger sessions.
\begin{table}[h!]
\begin{tabular}{cccccccccc} & \textbf{DW1}                    & \textbf{DW2}                    & \textbf{DW3} & \cellcolor[HTML]{C0C0C0}\textbf{PW1} & \cellcolor[HTML]{C0C0C0}\textbf{PW2} & \textbf{DW4} & \cellcolor[HTML]{C0C0C0}\textbf{PW3} & \cellcolor[HTML]{C0C0C0}\textbf{PW4} & \textbf{DW5} \\ \cline{2-10} 
\multicolumn{1}{c|}{\elim{CrossRoads}}                                                   & \multicolumn{2}{c|}{\cmark
}                          & \multicolumn{7}{c}{\xmark}                                                    \\ \hline
\multicolumn{1}{c|}{\elim{Dilemmas @ Work}}   & \multicolumn{2}{c|}{\cmark
}                          & \multicolumn{7}{c}{\xmark}                                                                                                                   \\ \hline
\multicolumn{1}{c|}{\driven}                                                       & \multicolumn{9}{c}{\cmark
}                                                                                                                                                                     \\ \hline
\multicolumn{1}{c|}{\trivia}                                                   & \multicolumn{9}{c}{\cmark
}                                                                                                                                                                     \\ \hline
\multicolumn{1}{c|}{\questions} & \multicolumn{9}{c}{\cmark
}                                                                                                                                                                     \\ \hline
\multicolumn{1}{c|}{\roads}      & \multicolumn{1}{c|}{} & \multicolumn{1}{c|}{} & \multicolumn{7}{c}{\cmark
}                                                                                                                   \\ \hline
\multicolumn{1}{c|}{\textit{*Menu}}                                               & \multicolumn{1}{c|}{} & \multicolumn{3}{c|}{}                                     & \multicolumn{5}{c}{\cmark
}                             
                       \\ \hline
\end{tabular}

\begin{tabular}{c@{}c @{\quad} c@{}c @{\quad} c@{}c @{\quad} c@{}c}
 \cmark
 = & ~ Presented &
 \xmark = & ~ Eliminated &
  = & ~ Idea Conception &
  = & ~ Under Development
\end{tabular}
\caption{\janeAdd{Timeline of prototypes presented across workshops. DW1-5 were driver-facing workshops while PW1-4 (in gray) represent passenger-facing ones. Progression of the menu is also included (*).}}
\label{timeline}
\vspace{-2em}
\end{table}
\FloatBarrier
\begin{table}[]
\caption{\janeAdd{Passenger demographics across four workshops (labeled PW1-4), with individual passengers labeled ``\textit{PW1.1}'' etc.}}
\begin{tabular}{|l|r|l|l|l|l|l|l|l|}
\hline
                     & \multicolumn{1}{l|}{\textbf{ID}} & \textbf{Age} & \textbf{State} & \textbf{Gender} & \textbf{Income} & \textbf{Platforms}     & \textbf{Drives to Commute} & \textbf{Employment} \\ \hline
\multirow{4}{*}{PW1} & P1.1                             & 18-29        & MA             & NB              & \$12k - \$50k   & Uber (4.94), Lyft      & \textless Once a year      & 5-10 years          \\ \cline{2-9} 
                     & P1.2                             & 18-29        & PA             & M               & \$12k - \$50k   & Uber (4.81)            & Never                      & \textless 3 years   \\ \cline{2-9} 
                     & P1.3                             & 18-29        & CA             & M               & \$100k - \$200k & Uber, Lyft (5)         & Everyday                   & \textless 3 years   \\ \cline{2-9} 
                     & \multicolumn{1}{l|}{P1.4}        & 18-29        & TN             & M               & \$12k - \$50k   & Uber, Lyft (5)         & A few times a year         & 3-5 years           \\ \hline
\multirow{4}{*}{PW2}                  & P2.1                             & 18-29        & CT             & M               & \$50k - \$100k  & Uber (5)               & Everyday                   & 3-5 years           \\ \cline{2-9} 
                     & P2.2                             & 65+          & NV             & F               & \$50k - \$100k  & Uber (5)               & Never                      & 10+ years           \\ \cline{2-9} 
                     & P2.3                             & 18-29        & IL             & NA              & \$12k - \$50k   & Uber: (4.68)           & Never                      & 3-5 years           \\ \cline{2-9} 
                     & P2.4                             & 30-44        & NY             & M               & \$100k - \$200k & Uber (4.9), Lyft       & Never                      & 10+ years           \\ \hline
\multirow{4}{*}{PW3} & P3.1                             & 18-29        & NJ             & F               & \$50k - \$100k  & Uber (4.98), Lyft (5)  & Few times / year           & \textless 3 years   \\ \cline{2-9} 
                     & P3.2                             & 18-29        & MA             & M               & \$0 - \$12k     & Lyft (5)               & Everyday                   & 3-5 years           \\ \cline{2-9} 
                     & P3.3                             & 45-64        & TX             & F               & \$50k - \$100k  & Uber, Lyft (5)         & \textless Once a year      & 10+ years           \\ \cline{2-9} 
                     & P3.4                             & 30-44        & PA             & M               & \$12k - \$50k   & Uber (4.9), Lyft (4.9) & Never                      & 10+ years           \\ \hline
\multirow{3}{*}{PW4} & P4.1                             & 18-29        & MA             & M               & \$50k - \$100k  & Uber (5)               & Few times / year           & \textless 3 years   \\ \cline{2-9} 
                     & P4.2                             & 18-29        & NJ             & F               & \$12k - \$50k   & Uber (4.99)            & Never                      & 3-5 years           \\ \cline{2-9} 
                     & P4.3                             & 45-64        & FL             & M               & \$100k - \$200k & Uber (4.90) Lyft (5.0) & \textgreater Once a week   & 10+ years           \\ \hline
\end{tabular}
\label{riders}
\vspace{-2em}
\end{table}
\subsubsection{Passenger Workshops} 
Next, we probed passengers to gather initial understanding and concerns around ridehail driving, evaluations of prototypes, and hesitations or ideas for alternative interactions that align with our study goal. 
\label{pwp} Each session started with ``\textit{Character Card}'' introductions where passengers took turns sharing their name/location/experiences on sticky notes of specific colors, which we used \janeAdd{to collect responses to} questionnaire-style prompts about their (1) knowledge of ridehail labor (2) level of empathy with drivers (based on questions adapted from the QCAE \cite{qcae} and IRI \cite{ma}) and (3) willingness advocate for drivers. \janeAdd{Then, each passenger took turns playing a game of their choice (presented via Figma prototypes) and rated their interaction experiences} along seven dimensions: fun, replayability, sneakiness (at embedding ridehail concepts), ride-friendliness, lightweight vs taxing, recommendability and how thought-provoking it was. \janeAdd{At the end of each session, we re-administered the questionnaire to survey perception changes around, knowledge of, empathy with and motivation to advocate for ridehail labor, which serve as proxy measures to help us gain initial understanding of whether game design can transform users to engage in more solidarity with drivers and socially conscious consumption behaviors.}
Supplementary materials contain full ranking results and detailed workshop protocols. 

\begin{table}[h!]
\begin{tabular}{|
>{\columncolor[HTML]{FFFFFF}}l |
>{\columncolor[HTML]{FFFFFF}}r |
>{\columncolor[HTML]{FFFFFF}}l |
>{\columncolor[HTML]{FFFFFF}}l |
>{\columncolor[HTML]{FFFFFF}}l |
>{\columncolor[HTML]{FFFFFF}}l |
>{\columncolor[HTML]{FFFFFF}}l |
>{\columncolor[HTML]{FFFFFF}}l |
>{\columncolor[HTML]{FFFFFF}}l |
>{\columncolor[HTML]{FFFFFF}}l |}
\hline
{\color[HTML]{333333} }                                                       & \multicolumn{1}{l|}{\cellcolor[HTML]{FFFFFF}{\color[HTML]{333333} \textbf{ID}}} & {\color[HTML]{333333} \textbf{Age}}   & {\color[HTML]{333333} \textbf{City}} & \cellcolor[HTML]{FFFFFF}{\color[HTML]{333333} \textbf{Gender}} & {\color[HTML]{333333} \textbf{Education}}                          & {\color[HTML]{333333} \textbf{Status}} & {\color[HTML]{333333} \textbf{Ridehail Income}} & {\color[HTML]{333333} \textbf{Hrs/week}} & {\color[HTML]{333333} \textbf{\begin{tabular}[c]{@{}l@{}}Ridehail \\ Experience\end{tabular}}} \\ \hline
\cellcolor[HTML]{FFFFFF}{\color[HTML]{333333} }                               & {\color[HTML]{333333} \text{1.1}}                                              & {\color[HTML]{333333} 45-54}          & {\color[HTML]{333333} PA}            & \cellcolor[HTML]{FFFFFF}{\color[HTML]{333333} \textit{M}}      & {\color[HTML]{333333} \textit{High School}}                        & {\color[HTML]{333333} \textit{FT}}     & {\color[HTML]{333333} Essential for basic needs} & {\color[HTML]{333333} N/A}               & {\color[HTML]{333333} 6 years}                                                                  \\ \cline{2-10} 
\cellcolor[HTML]{FFFFFF}{\color[HTML]{333333} }                               & {\color[HTML]{333333} 1.2}                                                       & {\color[HTML]{333333} \textit{45-54}} & {\color[HTML]{333333} \textit{IL}}   & {\color[HTML]{333333} \textit{M}}                              & {\color[HTML]{333333} \textit{High School}}                        & {\color[HTML]{333333} \textit{FT}}     & {\color[HTML]{333333} Essential for basic needs} & {\color[HTML]{333333} N/A}               & {\color[HTML]{333333} 4 years}                                                                  \\ \cline{2-10} 
\multirow{-3}{*}{\cellcolor[HTML]{FFFFFF}{\color[HTML]{333333} \textit{DW1}}} & {\color[HTML]{333333} 1.3}                                                       & {\color[HTML]{333333} \textit{18-29}} & {\color[HTML]{333333} \textit{CA}}   & {\color[HTML]{333333} \textit{M}}                              & {\color[HTML]{333333} \textit{Bachelor's}}                         & {\color[HTML]{333333} \textit{FT}}     & {\color[HTML]{333333} Essential for basic needs} & {\color[HTML]{333333} 25-40}             & {\color[HTML]{333333} 3 years}                                                                  \\ \hline
\cellcolor[HTML]{FFFFFF}{\color[HTML]{333333} }                               & {\color[HTML]{333333} 2.1}                                                       & {\color[HTML]{333333} \textit{45-64}} & {\color[HTML]{333333} \textit{NC}}   & \cellcolor[HTML]{FFFFFF}{\color[HTML]{333333} \textit{F}}      & {\color[HTML]{333333} \textit{Some college}}                       & {\color[HTML]{333333} \textit{FT}}     & {\color[HTML]{333333} Essential for basic needs} & {\color[HTML]{333333} 25-40}             & {\color[HTML]{333333} One month}                                                                \\ \cline{2-10} 
\cellcolor[HTML]{FFFFFF}{\color[HTML]{333333} }                               & {\color[HTML]{333333} 2.2}                                                       & {\color[HTML]{333333} \textit{30-44}} & {\color[HTML]{333333} \textit{NY}}   & {\color[HTML]{333333} \textit{M}}                              & {\color[HTML]{333333} \textit{Associate's}}                        & {\color[HTML]{333333} \textit{PT}}     & {\color[HTML]{333333} Nice but not essential}    & {\color[HTML]{333333} 10-25}             & {\color[HTML]{333333} 2 years}                                                                  \\ \cline{2-10} 
\multirow{-3}{*}{\cellcolor[HTML]{FFFFFF}{\color[HTML]{333333} \textit{DW2}}} & {\color[HTML]{333333} 2.3}                                                       & {\color[HTML]{333333} \textit{18-29}} & {\color[HTML]{333333} \textit{GA}}   & \cellcolor[HTML]{FFFFFF}{\color[HTML]{333333} \textit{F}}      & {\color[HTML]{333333} \textit{Bachelor's}}                         & {\color[HTML]{333333} \textit{PT}}     & {\color[HTML]{333333} Essential for basic needs} & {\color[HTML]{333333} 25-40}             & {\color[HTML]{333333} 3 years}                                                                  \\ \hline
\cellcolor[HTML]{FFFFFF}{\color[HTML]{333333} }                               & {\color[HTML]{333333} 3.1}                                                       & {\color[HTML]{333333} \textit{30-44}} & {\color[HTML]{333333} \textit{TX}}   & {\color[HTML]{333333} \textit{M}}                              & {\color[HTML]{333333} \textit{Bachelor's}}                         & {\color[HTML]{333333} \textit{PT}}     & {\color[HTML]{333333} Essential for basic needs} & {\color[HTML]{333333} 25-40}             & {\color[HTML]{333333} 5 years}                                                                  \\ \cline{2-10} 
\cellcolor[HTML]{FFFFFF}{\color[HTML]{333333} }                               & {\color[HTML]{333333} 3.2}                                                       & {\color[HTML]{333333} \textit{30-44}} & {\color[HTML]{333333} \textit{WA}}   & {\color[HTML]{333333} \textit{M}}                              & {\color[HTML]{333333} \textit{Bachelor's}}                         & {\color[HTML]{333333} \textit{FT}}     & {\color[HTML]{333333} Essential for basic needs} & {\color[HTML]{333333} 40+}               & {\color[HTML]{333333} 6 years}                                                                  \\ \cline{2-10} 
\cellcolor[HTML]{FFFFFF}{\color[HTML]{333333} }                               & {\color[HTML]{333333} 3.3}                                                       & {\color[HTML]{333333} \textit{18-29}} & {\color[HTML]{333333} \textit{NY}}   & {\color[HTML]{333333} \textit{NB}}                             & {\color[HTML]{333333} \textit{Associate's}}                        & {\color[HTML]{333333} \textit{FT}}     & {\color[HTML]{333333} Essential for basic needs} & {\color[HTML]{333333} 25-40}             & {\color[HTML]{333333} 3 years}                                                                  \\ \cline{2-10} 
\multirow{-4}{*}{\cellcolor[HTML]{FFFFFF}{\color[HTML]{333333} \textit{DW3}}} & {\color[HTML]{333333} 3.4}                                                       & {\color[HTML]{333333} \textit{18-29}} & {\color[HTML]{333333} \textit{NY}}   & \cellcolor[HTML]{FFFFFF}{\color[HTML]{333333} \textit{F}}      & {\color[HTML]{333333} \textit{Post-Graduate}}                      & {\color[HTML]{333333} \textit{N/A}}    & {\color[HTML]{333333} N/A}                       & {\color[HTML]{333333} N/A}               & {\color[HTML]{333333} 2 years}                                                                  \\ \hline
\cellcolor[HTML]{FFFFFF}{\color[HTML]{333333} }                               & {\color[HTML]{333333} 4.1}                                                       & {\color[HTML]{333333} \textit{30-44}} & {\color[HTML]{333333} \textit{CA}}   & \cellcolor[HTML]{FFFFFF}{\color[HTML]{333333} \textit{N/A}}    & {\color[HTML]{333333} \textit{Associate's}}                        & {\color[HTML]{333333} \textit{PT}}     & {\color[HTML]{333333} Nice but not essential}    & {\color[HTML]{333333} 25-40}             & {\color[HTML]{333333} 5 years}                                                                  \\ \cline{2-10} 
\cellcolor[HTML]{FFFFFF}{\color[HTML]{333333} }                               & {\color[HTML]{333333} 4.2}                                                       & {\color[HTML]{333333} \textit{30-44}} & {\color[HTML]{333333} \textit{FL}}   & {\color[HTML]{333333} \textit{M}}                              & {\color[HTML]{333333} \textit{Associate's}}                        & {\color[HTML]{333333} \textit{FT}}     & {\color[HTML]{333333} Nice but not essential}    & {\color[HTML]{333333} 25-40}             & {\color[HTML]{333333} 7 years}                                                                  \\ \cline{2-10} 
\cellcolor[HTML]{FFFFFF}{\color[HTML]{333333} }                               & {\color[HTML]{333333} 4.3}                                                       & {\color[HTML]{333333} \textit{30-44}} & {\color[HTML]{333333} \textit{CO}}   & {\color[HTML]{333333} \textit{NB}}                             & {\color[HTML]{333333} \textit{Bachelor's}}                         & {\color[HTML]{333333} \textit{PT}}     & {\color[HTML]{333333} Nice but not essential}    & {\color[HTML]{333333} 25-40}             & {\color[HTML]{333333} 4 years}                                                                  \\ \cline{2-10} 
\multirow{-4}{*}{\cellcolor[HTML]{FFFFFF}{\color[HTML]{333333} \textit{DW4}}} & {\color[HTML]{333333} 4.4}                                                       & {\color[HTML]{333333} \textit{18-29}} & {\color[HTML]{333333} \textit{TX}}   & {\color[HTML]{333333} \textit{NB}}                             & {\color[HTML]{333333} \textit{Associate's}}                        & {\color[HTML]{333333} \textit{FT}}     & {\color[HTML]{333333} Nice but not essential}    & {\color[HTML]{333333} 25-40}             & {\color[HTML]{333333} 3 years}                                                                  \\ \hline
\cellcolor[HTML]{FFFFFF}{\color[HTML]{333333} }                               & {\color[HTML]{333333} 5.1}                                                       & {\color[HTML]{333333} \textit{30-44}} & {\color[HTML]{333333} \textit{PA}}   & {\color[HTML]{333333} \textit{M}}                              & {\color[HTML]{333333} \textit{Bachelor's}}                         & {\color[HTML]{333333} \textit{PT}}     & {\color[HTML]{333333} Essential for basic needs} & {\color[HTML]{333333} 10-25}             & {\color[HTML]{333333} 7.5 years}                                                                \\ \cline{2-10} 
\cellcolor[HTML]{FFFFFF}{\color[HTML]{333333} }                               & {\color[HTML]{333333} 5.2}                                                       & {\color[HTML]{333333} \textit{30-44}} & {\color[HTML]{333333} \textit{IL}}   & {\color[HTML]{333333} \textit{M}}                              & {\color[HTML]{333333} \textit{Bachelor's}}                         & {\color[HTML]{333333} \textit{FT}}     & {\color[HTML]{333333} Essential for basic needs} & {\color[HTML]{333333} 25-40}             & {\color[HTML]{333333} 5 years}                                                                  \\ \cline{2-10} 
\cellcolor[HTML]{FFFFFF}{\color[HTML]{333333} }                               & {\color[HTML]{333333} 5.3}                                                       & {\color[HTML]{333333} \textit{18-29}} & {\color[HTML]{333333} \textit{PA}}   & \cellcolor[HTML]{FFFFFF}{\color[HTML]{333333} \textit{NB}}     & \cellcolor[HTML]{FFFFFF}{\color[HTML]{333333} \textit{Bachelor's}} & {\color[HTML]{333333} \textit{FT}}     & {\color[HTML]{333333} Essential for basic needs} & {\color[HTML]{333333} 25-40}             & {\color[HTML]{333333} 8 years}                                                                  \\ \cline{2-10} 
\cellcolor[HTML]{FFFFFF}{\color[HTML]{333333} }                               & {\color[HTML]{333333} 5.4}                                                       & {\color[HTML]{333333} \textit{45-54}} & {\color[HTML]{333333} \textit{PA}}   & {\color[HTML]{333333} \textit{M}}                              & {\color[HTML]{333333} \textit{Some college}}                       & {\color[HTML]{333333} \textit{N/A}}    & {\color[HTML]{333333} Nice but not essential}    & {\color[HTML]{333333} N/A}               & {\color[HTML]{333333} 2.75 years}                                                               \\ \cline{2-10} 
\multirow{-5}{*}{\cellcolor[HTML]{FFFFFF}{\color[HTML]{333333} \textit{DW5}}} & {\color[HTML]{333333} 5.5}                                                       & {\color[HTML]{333333} \textit{45-64}} & {\color[HTML]{333333} \textit{GA}}   & {\color[HTML]{333333} \textit{M}}                              & {\color[HTML]{333333} \textit{Bachelor's}}                         & {\color[HTML]{333333} \textit{FT}}     & {\color[HTML]{333333} Essential for basic needs} & {\color[HTML]{333333} 25-40}             & {\color[HTML]{333333} 6 years}                                                                  \\ \hline
\end{tabular}
\caption{Driver demographics \janeAdd{across five workshops (DW1-5), individual workers are labeled ``\textit{DW1.1}'' etc.}} 
\label{drivers}
\end{table} 
\vspace{-2em}
\subsection{Recruitment}
During formative interviews, we recruited 2 drivers (D1, D2) based on contacts from prior studies, as well as 3 passengers (P1 --- P3) based on convenience sampling from our home universities. 
For workshops, we reached out to \janeAdd{(1) drivers from} past studies, \janeAdd{(2)} subreddits, Craiglist posts and physical flyers in local professional communities. 
All formative interviews and workshops were conducted via Zoom.
Both participants groups were compensated at a rate of \$60/hour and selected based on eligibility, location and experience levels, indicated by pre-study screening forms. 
Tables \ref{riders} and \ref{drivers} summarize passenger and driver demographics \footnote{Due to an oversight to record the meeting, participant data from DW4 are based primarily on internal notes}.

\subsection{Thematic Analysis}
After our workshops, three researchers open coded all 12 hours of workshop transcripts (transcribed by Otter.ai) to extract relevant themes and opinions on improvements for each prototype. Then, we combined all individual driver codes in an affinity diagram to map out common ideas, extracting 8 main categories: existing practices/strategies, frustrations, reactions to prototypes, current consumer perceptions, underexposed knowledge, design objectives and alternative gamified interactions/interventions. The first five categories helped us understand how well the prototypes capture realistic driver experiences, while the remainder guided our next iterations of prototype adjustments -- these codes also helped eliminate less effective prototypes and introduce new features. \janeAdd{In particular, codes in the category of \textit{reactions to prototypes} informed the construction of Table \ref{receptions}. We considered a prototype ``\textit{preferred}'' in a workshop if more than half of participants expressed affinity to it. If a session was divided on their sentiments -- roughly half positive and half negative -- we categorized it as ''\textit{neutral}''. Remaining prototypes that received mainly negative reactions were considered ``\textit{resisted}''. We halted recruitment and data collection when new themes no longer emerged.}

\subsection{Positionality}
\janeAdd{Our study activities (from recruitment, consent and data collection) followed procedures of an IRB-approved protocol. However, working with sensitive topics across two stakeholder groups carries unique dangers that our internal team continuously deliberate on to reach further understanding of our relative positions \cite{positionality}.}
We reflect on ways to center driver experiences and reduce replacing their voices and opinions with our own values and epistemologies, paying particular attention discussing vulnerabilities in ways that uplift and empower, rather than silence, suppress or overshadow worker experiences.
\janeAdd{Towards passengers, we disclosed our intent to raise awareness around ridehail driving conditions at the start of each session. Yet we still acknowledge that even upfront notices cannot fully eliminate risks of imposing our own values and positions onto passenger participants, which may cause discomfort -- especially when using techniques such as obfuscation. Thus we also explicitly reminded participants about their right to opt out at any time throughout the session. }
Our team members receive training in Computer Science, Media Arts \& Sciences, Software Engineering and Human Computer Interaction, where two authors have experience researching and working with ridehail drivers.
One author has part-time experience working for a food delivery platform, while two authors have extensive experience laboring across service occupations.

\subsection{Limitations}
While our \textit{Character Card} activity during passenger workshops aimed capture early evidence of how passengers perspectives changed before and after experiencing gamifiied interactions, these results are limited by the size of our participant sample.  \janeAdd{Future works might consider more scalable evaluations of similar tools to more rigorously examine the impact of such interventions on player understanding and advocacy for driving conditions. Our US-based investigation can also limit external validity and generalizability -- both geographically and culturally. For instance, how do more collectivist (\textit{e.g.,} Asian-Pacific) societies or nations in the Global South (\textit{e.g.,} Chile \cite{south}) receive or resist such mechanisms of control? How would a post-colonial lens of analysis offer different results between migrant workers of the US vs Europe \cite{morality}? We hope this study offer a case study that might inspire compare analysis with other national, cultural and geographic contexts.} 

\section{Results}
\renewcommand{\arraystretch}{1.5}
\begin{table}[h!]
\begin{tabular}{l|lllllllll}
\cline{2-10}
                           & DW1                                             & DW2                                             & DW3                                             & PW1                                             & PW2                                             & DW4                                             & PW3                                             & PW4                                             & DW5                                             \\ \hline
\textbf{\elim{CrossRoads}}        & \cellcolor[HTML]{7A4960}                        & \cellcolor[HTML]{7A4960}                        & \cellcolor[HTML]{C7C7C7}{\color[HTML]{000000} } & \cellcolor[HTML]{C7C7C7}{\color[HTML]{000000} } & \cellcolor[HTML]{C7C7C7}{\color[HTML]{000000} } & \cellcolor[HTML]{C7C7C7}{\color[HTML]{000000} } & \cellcolor[HTML]{C7C7C7}{\color[HTML]{000000} } & \cellcolor[HTML]{C7C7C7}{\color[HTML]{000000} } & \cellcolor[HTML]{C7C7C7}{\color[HTML]{000000} } \\ \cline{1-1}
\textbf{\elim{Dilemmas @ Work}}   & \cellcolor[HTML]{7A4960}                        & \cellcolor[HTML]{DCC75A}                        & \cellcolor[HTML]{C7C7C7}{\color[HTML]{000000} } & \cellcolor[HTML]{C7C7C7}{\color[HTML]{000000} } & \cellcolor[HTML]{C7C7C7}{\color[HTML]{000000} } & \cellcolor[HTML]{C7C7C7}{\color[HTML]{000000} } & \cellcolor[HTML]{C7C7C7}{\color[HTML]{000000} } & \cellcolor[HTML]{C7C7C7}{\color[HTML]{000000} } & \cellcolor[HTML]{C7C7C7}{\color[HTML]{000000} } \\ \cline{1-1}
\textbf{\driven}            & \cellcolor[HTML]{446644}                        & \cellcolor[HTML]{446644}                        & \cellcolor[HTML]{446644}                        & \cellcolor[HTML]{DCC75A}                        & \cellcolor[HTML]{DCC75A}                        & \cellcolor[HTML]{DCC75A}                        & \cellcolor[HTML]{DCC75A}                        & \cellcolor[HTML]{DCC75A}                        & \cellcolor[HTML]{446644}                        \\ \cline{1-1}
\textbf{\trivia}        & \cellcolor[HTML]{DCC75A}                        & \cellcolor[HTML]{DCC75A}                        & \cellcolor[HTML]{446644}                        & \cellcolor[HTML]{446644}                        & \cellcolor[HTML]{446644}                        & \cellcolor[HTML]{446644}                        & \cellcolor[HTML]{DCC75A}                        & \cellcolor[HTML]{446644}                        & \cellcolor[HTML]{DCC75A}                        \\ \cline{1-1}
\textbf{\questions} & \cellcolor[HTML]{446644}                        & \cellcolor[HTML]{446644}                        & \cellcolor[HTML]{DCC75A}                        & \cellcolor[HTML]{446644}                        & \cellcolor[HTML]{7A4960}                        & \cellcolor[HTML]{446644}                        & \cellcolor[HTML]{446644}                        & \cellcolor[HTML]{446644}                        & \cellcolor[HTML]{446644}                        \\ \cline{1-1}
\textbf{\roads}     & \cellcolor[HTML]{1F3650}{\color[HTML]{000000} } & \cellcolor[HTML]{C7C7C7}{\color[HTML]{000000} } & \cellcolor[HTML]{446644}{\color[HTML]{000000} } & \cellcolor[HTML]{446644}                        & \cellcolor[HTML]{7A4960}                        & \cellcolor[HTML]{446644}                        & \cellcolor[HTML]{DCC75A}                        & \cellcolor[HTML]{7A4960}                        & \cellcolor[HTML]{446644}                        \\ \cline{1-1}
\textit{Menu*}   & \cellcolor[HTML]{1F3650}{\color[HTML]{000000} } & \cellcolor[HTML]{C7C7C7}                        & \cellcolor[HTML]{C7C7C7}                        & \cellcolor[HTML]{C7C7C7}                        & \cellcolor[HTML]{7A4960}                        & \cellcolor[HTML]{DCC75A}                        & \cellcolor[HTML]{446644}                        & \cellcolor[HTML]{DCC75A}                        & \cellcolor[HTML]{DCC75A}                       
\end{tabular}

\vspace{0.5em}
\begin{tabular}{*{5}{c c}} 
\textcolor[HTML]{446644}{\rule{1em}{1em}} & Preferred &
\textcolor[HTML]{DCC75A}{\rule{1em}{1em}} & Neutral &
\textcolor[HTML]{7A4960}{\rule{1em}{1em}} & Resisted &
\textcolor[HTML]{1F3650}{\rule{1em}{1em}} & Origin of conception &
\textcolor[HTML]{C7C7C7}{\rule{1em}{1em}} & Not Presented 
\end{tabular}

\caption{Heatmap shows participant perceptions of prototypes across workshops, which are ordered chronologically. DW1-5 represent driver-facing workshops and PW1-4 shows passenger-facing ones. Feedback for the menu selection is also included (*).}
\label{receptions}
\vspace{-2.5em}
\end{table} 
\janeAdd{Below we first report how passengers received the game prototypes, early evidence of how they shifted their perceptions of ridehail, and their knowledge gaps that the probes revealed. Next, we discuss driver preferences for purposeful and personalized prototypes that initiate unforced and fulfilling interactions with passengers, as well as driver perspectives on and experiences with common passenger knowledge gaps. Finally, we share how both stakeholder groups reacted to  proposed driver-passenger interactions, including reasons for hesitations and ways to better motivate engagement.}

\subsection{Passenger Receptions to Gamified Interventions, Early Perceptual Changes \& Knowledge Gaps} \label{gaps}
During formative interviews, all passengers reflected on gaps in their understanding of ridehail work, noting that incorporating content related to ridehail conditions would motivate engagement with the game. For instance, P1 related how ``\textit{[he]'ll be more inclined to try it out}'' while P3 shared that ``\textit{[she] would definitely play a rideshare driver simulator \dots where your goal is to get from one place to another}''. 
\janeAdd{Building on such potential, we next report how passengers received our game prototypes during workshop sessions (\S\ref{engagement}), early evidence of how the play experience changed their perceptions towards ridehail conditions (\S\ref{evolved}), and prominent knowledge gaps that the game probes revealed (\S\ref{surfaced}). }

\subsubsection{Passenger Feedback on In-Ride Game Prototypes}
\janeAdd{We overview how drivers ranked prototypes along (1) playability heuristics and (2) their abilities to provoke reflective thinking around ridehail conditions.}

\begin{figure}[h]
    \centering
\includegraphics[width=.6\linewidth]{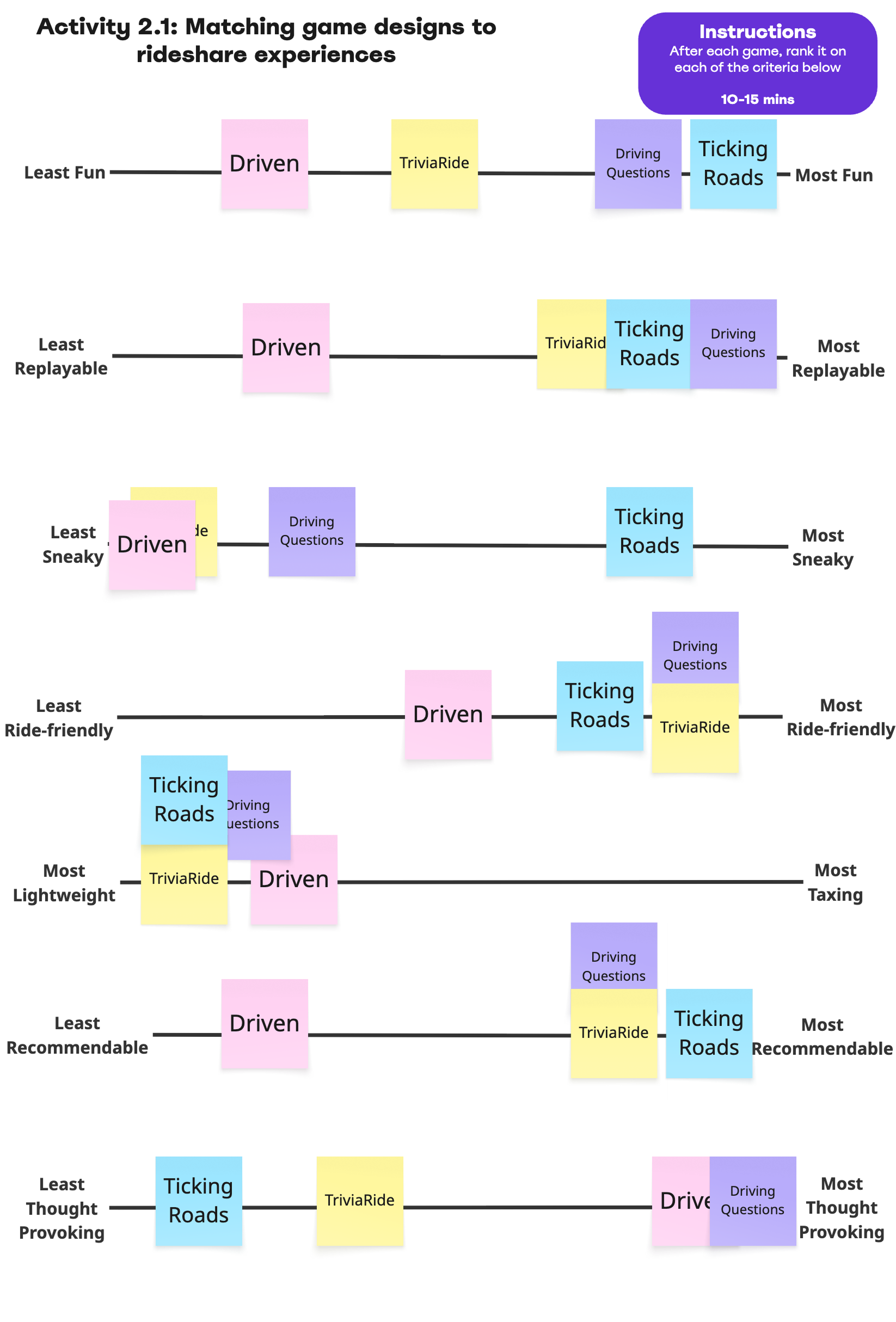}
    \caption{\janeAdd{After playthroughs, passengers ranked each prototype along game heuristics, an embedded design criterion (``\textit{sneakiness}'') and ridehail-specific criteria (ride-friendliness, ability to provoke) -- Figure shows example from PW1.}}
    \vspace{-1em}
    \label{menu}
\end{figure}

\paragraph{Fun vs Provocation, Replayability \& Recommendability} \label{engagement} 
Passengers found \questions the most fun (PW3, PW4) and replayable (PW1, PW3, PW4). \trivia came in second for fun, preferred by passengers like P2.1, who found it ``\textit{a little bit fun --- the interface was very simple, and so that made it, not super thrilling. But I enjoy trivia \dots and I really like the idea of having questions that are specific to the area where you're riding, I found that very inventive}''.

In contrast, \driven was rated as least fun but most provocative (PW1, PW3, PW4) and second most replayable.
Largely due to technical issues, \roads consistently placed last for replayabiltiy -- P4.3 noted how ``\textit{controlling issues were a pain, it was kind of wonky}'', surfacing a tension between immersion and implementation effort. 
Initially, PW1 ranked \roads as least recommendable. \janeAdd{After we implemented improvements based on initial feedback from PW1, passengers in PW2-4 ranked it as most recommmendable. }

\paragraph{Ridehail-specific Rankings} \label{rideshare_rankings}
\janeAdd{Regarding} ridehail content, we asked passengers to rank each prototype by how (1) ``\textit{sneaky}'' it was at embedding driving-related content (a proxy measure for success of obfuscation) (2) ride-friendly it was and (3) lightweight or taxing it was to navigate game mechanisms and content. \roads was most successful at obfuscation, with P3.4 admitting how ``\textit{that was pretty sneaky. I'm not gonna lie. I didn't even realize it. Thought I was [just] driving around}'' while \driven was most ineffective at obfuscation. 
The conversation-prompting prototype \questions was determined to be most ride-friendly, while \trivia came in close second. The lightweight-taxing ranking measurement was interpreted by passengers along both dimensions of gameplay mechanics and content: \driven was deemed most emotionally-taxing since it is ``\textit{frighteningly realistic \dots and depressing, and for once it seems like accurate for a lot of people's [driver's] situations}'' (P2.4). 
By contrast, \trivia was considered most lightweight, both in terms of content and mechanics, although its ridehail content was most exposed, obvious and not well obfuscated: ``\textit{they were very apparent}'' - P2.3.

\FloatBarrier

\subsubsection{\janeAdd{How Games Evolved Passenger Perceptions \& Solidarity with Ridehail Drivers} }\label{evolved}
Leveraging the \textit{Character Cards} (\S\ref{pwp}) activity, we surveyed how passengers perceived, empathized with, and desired to advocate for driving conditions both before and after introducing prototypes. Supplementary materials show how passengers placed themselves on these scales across all four workshops. Below, we report changes in their (1) knowledge around ridehail conditions as well as (2) empathy with and willingness to advocate for drivers -- which we consider as proxy measures for solidarity.
\begin{figure}[]
    \centering
\includegraphics[width=.9\linewidth]{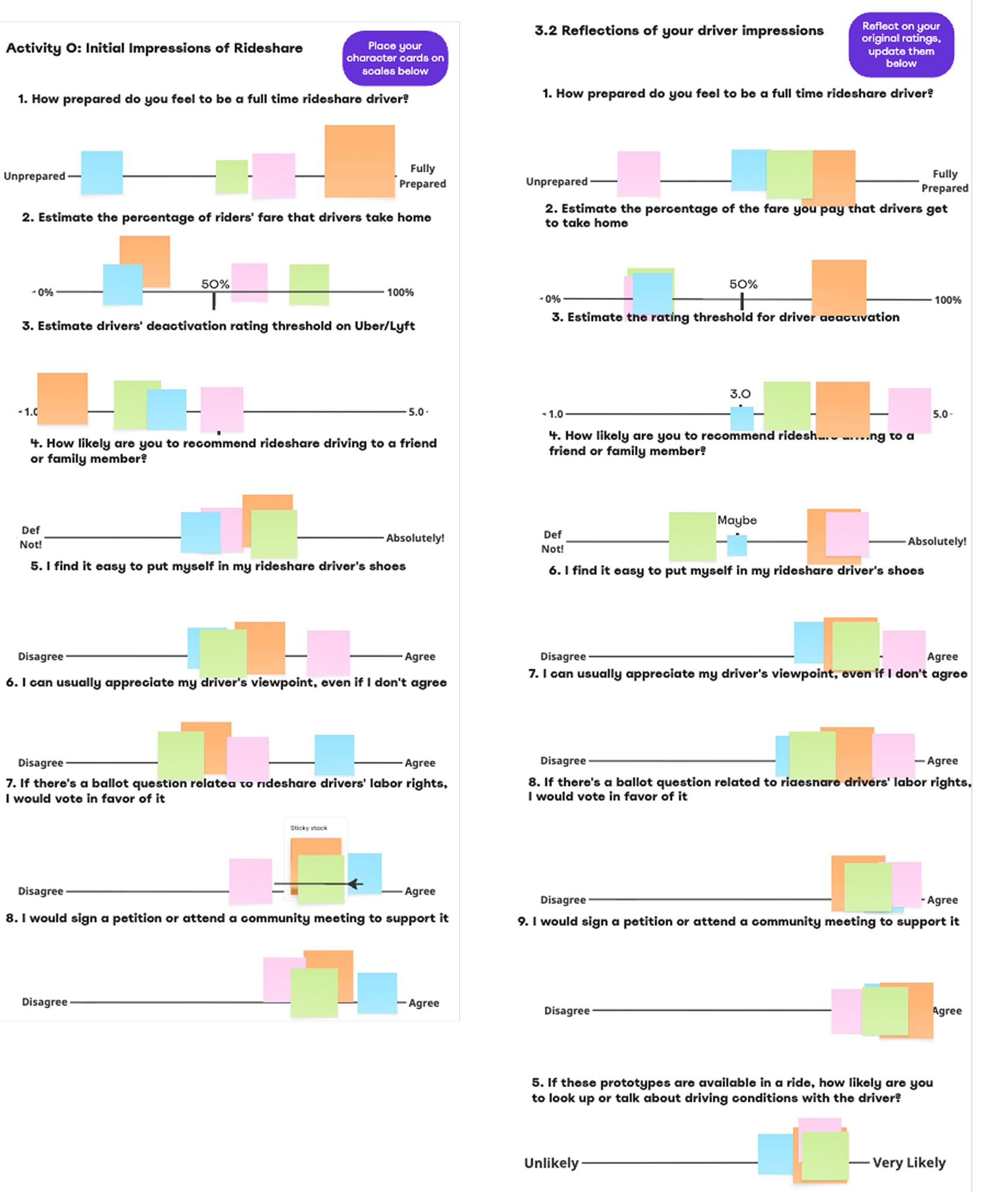}
    \caption{\janeAdd{\textit{Character Cards} (with each color representing a passenger) helped us survey passenger understanding, empathy and willingness to advocate before and after each workshop. Example above show (changes in) responses from PW3.}}
    \vspace{-1em}
    \label{PW3}
\end{figure}
\paragraph{\janeAdd{Shifts in Passenger Knowledge of Ridehail Driving}}
When probed about their readiness to drive for ridehail themselves (\textit{i.e. How prepared do you feel to be a full time rideshare driver?}), confidence increased slightly in PW1 and PW4, whereas all passengers in PW2 felt unprepared before and after. 
Estimations of driver take-home rate was unanimously lowered across all four workshops, \janeAdd{suggesting that prototypes helped passengers reach} \janeAdd{\textbf{more accurate understandings of}} \janeAdd{\textbf{platform take rates}}.
Meanwhile, passenger estimations of deactivation rates surged -- especially in PW3, where ratings flipped entirely -- see Figure \ref{PW3} -- indicating \textbf{a significant knowledge gap around how heavily ratings impacted the job stability of drivers}. Passenger willingness to recommend ridehail driving remained largely unchanged.

\paragraph{\janeAdd{Early Indicators of Change in Passenger Empathy, Solidary \& Advocacy for Ridehail Labor}} 
Empathy-related questions (\textit{i.e., ``I find it easy to put myself in my rideshare driver’s shoes”; “I can usually appreciate my driver’s viewpoint, even if I don’t agree''}) were introduced after the first passenger workshop (\textit{i.e.} PW2–PW4). Across these sessions, \janeAdd{passengers showed slightly higher sensitivity to driver perspectives}, with effects most prominent in PW3.
In the later three workshops, we also added two items to probe passenger willingness to advocate for ridehail drivers. Across all sessions, \textbf{passengers became more motivated to vote in favor of ``\textit{a ballot question related to rideshare drivers’ labor rights}''} \janeAdd{with those in PW2 and PW3 expressing absolute agreement by the end of their sessions. In response to the follow-up question (whether they would \textit{``sign a petition or attend a community meeting to support it''}) all passengers showed slightly \textit{higher inclinations to advocate}.}

\subsubsection{\janeAdd{Passenger Knowledge Gaps Surfaced by Game Prototypes}}
\label{surfaced}
\janeAdd{Based on how passenger reacted to embedded ridehail content, we next present prominent knowledge gaps that our prototypes helped surface.
}
For instance, P3.1 ``\textit{feel uncomfortable asking them how much they're getting paid, or their take-home pay}'', let alone more personal inquiries about what drivers miss (\S\ref{questions}), since ``\textit{it can be painful when you leave your home country}'' (P2.2). 
However, the optional yet inviting nature of games like \questions serve as boundary objects for mediating such conversations, even among more quiet passengers ---
P3.3 identified as ``\textit{an introvert [\dots] and I have trouble starting conversations [\dots] this game would help me be more talkative, because it gives me things to say}''. 
\FloatBarrier
\paragraph{Pay rates} \label{pay}

All passengers of PW1 \& PW3 were ``\textit{surprised about the 30\% [take-home rate our drivers estimated.] I didn't realize that it could be that low. That's bad.}'' P2.2 (a more sympathetic passenger who regularly asked drivers about rising take rates) observed how ``\textit{all the drivers are anxious to answer [\dots to ensure] riders know that even though [passengers] are paying more, [drivers] are not making more}''. \textbf{\driven was most effective at bringing out reactions towards low pay rates}, leaving P3.2 outraged: driving ``\textit{80 trips for \$65 is crazy!}'' 
\trivia was also clear at communicating low pay rates, with P2.3 and P3.1 noticing a question about most profitable hours of operation. While \questions carried potential to spark driver-passenger discussion around pay (P1.2), many passengers hesitated to discuss the financial and potentially sensitive topic -- see \S\ref{balance}.
Although \roads was designed to show how obstacles influence a driver’s earnings, players generally did not notice these financial effects — indicating the concept may need to be presented more clearly or reinforced through replays.

\paragraph{Pickup Logistics} \label{pickup}
The time that drivers spent waiting for passengers during pickup is a commonly overlooked factor that affects driver pay.
P3.3 exhibited concern about how their location may impact driver earnings  ``\textit{I live in a suburb \dots and \dots I don't think they're getting paid between the different ones. \dots I always wonder if they're making a lot less money and just driving basically for free between the places.}''
\janeAdd{The time pressure in }\roads \textbf{guided passengers like P4.3 to empathize about wait times at pick up}: ``\textit{that must be frustrating for the the driver to wait for the person to come up}.''
Similarly, P4.2 noticed through the plot of \driven how drivers must decide on an exact pickup location when directions from the app are not clear: ``\textit{when drivers are going to somewhere that has an entrance and an exit, and knowing where to pick up someone.}''
With \questions, P2.2 was the only one who noticed a question centering wait times during pickups, even though all players went through the same set of questions. Given its need of ground-truth answers, we struggled to embed factual content on pick up logistics within \trivia.

\paragraph{Rating Pressures \& Passenger Expectations of Service} \label{ratings}
In terms of mechanisms, \roads was most effective at \textbf{eliciting passenger reactions to behaviors of impatient riders}, with P2.3 rating the prototype as ``\textit{provoking, because you have a person texting you angrily when you're trying your best \dots it'll teach them to put themselves in the drivers' shoes}.'' 
Similar to frustrations that drivers often feel about road conditions, P4.3 also grew annoyed at the simulated passenger in the time management game who was ``\textit{Texting, `}where are you?\textit{' But there's obviously a car right there, lots of traffic}''.
\questions reminded P1.2 of ``\textit{one driver \dots he was telling me about \dots all the passengers that he picked up during the night, which are usually all like drunk college kids}'', suggesting that the conversational nature of the interaction can help drivers more comfortably share experiences of poor passenger behaviors.
Despite how \driven embedded consequences of \textit{receiving} bad ratings as a result of interactions with passenger, the prototype did not actually demonstrate tangible consequences (e.g., deactivation, reduced work availability) beyond small pay differences --- which may have caused passengers to not notice the embedded concept. 
\trivia also did not embed concepts on rating pressures.

\paragraph{Platformic Management \& Long-term Consequences} \label{longer_term}

While passengers may rarely observe the longer-term tolls and costs that ridehail cause for drivers, psychologically-distanced narrative elements in \driven (e.g., pop-up app notifications at home) helped surface a key tension that last over time --- family-life balance --- which was noticed by every passenger who played through (P1.4, P2.4, P3.3, P4.2). In their second playthrough, P2.4 started to \janeAdd{\textbf{anticipate extractive platform tactics}}: ``\textit{I feel like they're gonna ping me and say}, `Why are you taking so long use the bathroom?' '' The modular and timed nature of \driven again made it unsuitable for addressing psychological control, even though D1.2 saw potential for simulations like \roads to give character to ``\textit{the big evil ride sharing company}''. While the conversational nature of \questions could prompt discussion around long term consequences and control, passenger discomfort with potential intrusion may hinder in-depth probing (\S\ref{balance}).


\subsection{Towards Driver-Centered \& Integrated In-ride Gaming Interactions} \label{driver_reactions}
\janeAdd{Next, we overview driver receptions of game prototypes, including their preferences for immersive and personalized environments that help passengers feel more at ease and willing to engage in interactions with the driver.
Then, we describe how drivers observed passenger knowledge gaps}, including \textbf{\textit{covert logistical, emotional and immaterial labor}} that range from general car maintenance (e.g., gas, oil changes) to in-ride services -- \textit{e.g.,} keeping comfortable temperature and mood while ensuring safety. 

\subsubsection{Drivers Prefer Memorable, \janeAdd{Contextualized} but Unobtrusive Games} \label{alignments}
Drivers expressed strong enthusiasm for the potential of games to create \textbf{purposeful, immersive and memorable} gamified experiences around ridehail knowledge. 
For instance, D2.2 emphasized how ``\textit{the story game: it actually rocks --- taps into emotions, memories and creativity, and then it's relatable and often openly hilarious}'' and imagined cases of ``\textit{making [passengers] go} `yeah, man, I never thought about it like that' '', causing many (D1.1, D2.1, D2.2, D2.3, D3.4) to rank \driven highest: ``\textit{love the story games because [\dots] it builds empathy in in a more subtle way}'' (D3.4). 

An immersive narrative experience also help drivers create an environment that \textbf{distracts passengers from time pressures} of reaching their destination. D3.2 imagines how ``\textit{When a rider is into a story game \dots the whole mood in the car becomes more quiet, more relaxed, no pressure, no rush. They're just absorbing something soft and engaging}.'' \janeAdd{But alongside immersion, D2.2 also cautioned against cognitively demanding interfaces: ``\textit{style choices that spark a bit of fun without demanding too much attention}'', while D3.1 warned about the ``\textit{risk of getting halfway through something, and then, boom, you are there [the ride] stop[s], and it actually ends awkwardly. So for me, the sweet spot would be not keeping the stories short, flexible and rider-led}''.}
Through this relaxed and non-confrontational approach, drivers shared desires to communicate latent knowledge including pickup logistics (D1.2, D3,1, D3.2), long hours (D2.1, D2.3), unreasonable passenger expectations (D2.1) and behaviors (D2.3, D1.3), traffic (D1.3, DW4) -- some of which drivers would've been difficult to broach since they ``\textit{don't want to make [passengers] scared to be in the ride with you}'' (D1.1). 

Drivers who have experience with Octopus tablets noted that its existing trivia content was not designed for ridehail contexts, leading them to prefer alternatives like \trivia (ranked highest by D3.1, D3.3, D2.3) that feel more \textbf{personalized, local} and \textbf{mood-aware}. 
For instance, D2.2 ``\textit{want games that feel like they belong in the rideshare world, not like they were copied from somewhere else and shoved into into my car}'', suggesting instead interactions with ``\textit{more personalization, more mood awareness, more empathy, fun}''. Numerous participants (D2.2, D3.1, P2.1, P2.4, DW4 members) also suggested location-specific content, to \textit{``add a local city flavor, trivia about Seattle [\dots] we have landmarks nearby, or [\dots] which coffee shop this quote is from -- it makes the ride feel connected to where we are, and breaks the ice faster than [games on] a plane''} (D2.2).

With a lighter interaction like \questions, drivers saw potential to start off \textbf{unforced conversations} around topics of mutual interest. For instance, D3.2 discusses how this prototype ``\textit{changes the mood \dots spark[s] some light interaction [between drivers and passengers and] also between the people onboard}.'', in addition to engaging both quiet and extroverted passengers:
\begin{quote}
    \textit{It's a game, so people are not forced to to play it. So this is just one way in which you will create the conversation. If the rider feels like he/she is not compatible to play the game, then it's okay. But if one is ready for that, then absolutely, there are some people who are extroverted \dots and I think this would match quite well with my riders.} -(D2.2)
\end{quote}
For drivers like D5.5, such connection with (introverted) passengers generates an \textbf{intrinsic sense of fulfillment}:
\textit{``I really love seeing where they say that they prefer to be quiet, and you could still get a conversation out of them --- makes me feel like I'm doing my job, plus more''} while simultaneously helping passengers ``\textit{keeps them off their mind of traffic}''.

Since \roads was conceived at workshop DW1, it was only presented in DW3-5 (see Table \ref{receptions}), but it was a favorite for drivers like D5.2, who found it relatively ``\textit{more interactive}.'' 
Drivers in DW4 also saw its mechanics and graphics as helpful simulations of driver experiences, especially for populations under (time pressure) such as busy parents managing small children onboard. 
\janeAdd{However, D2.2 stresses that the interactions should engage passengers without disrupting their ability to focus on driving: ``\textit{as someone who is behind the wheel all day, I'm gonna think about mood, safety and how natural it all feels. If a game fits seamlessly into the ride, awesome. If not, then it just adds noise, pressure or openness.}''}
In general, drivers saw potential for \roads to quickly capture attention, which \textbf{ease passenger anxieties and pressures}, with D3.2 observing how it can ``\textit{reduce rider frustration during delays [and] \dots construction. Time[d] games give the rider something else to think about, so it lowers complaints and makes the ride feel more like a break than a hassle.}''
\vspace{-.5em}
\subsubsection{\janeAdd{Driver Reflections on Latent Labor Conditions}}
\janeAdd{Across workshops, drivers leveraged game prototypes to consider and reflect on experiences and labor that typically remain concealed from, or gets overlooked by passengers. While these are ordered by drivers' expressed priorities, we note how they also \textbf{show more surface-level, salient and cognitively-obvious symptoms of algorithmic control first, followed by more nuanced, subtle and psychological reasons that cause harmful conditions}, demonstrating the ability of game probes to elicit more underlying fears and rationales.}
In their experiences talking to passengers, pilot drivers like D1 shared how ``\textit{Very few [passengers] -- maybe one or two -- out of the couple thousands of rides I’ve done have asked me what my pay for that ride versus what they were paying. So I think [they] probably don't know [or] don't care}''. 
D2 similarly thought that only folks who worked on ``\textit{a gig app \dots or if there's somebody in their immediate circle of life (friends or family) that does it}'' are likely to know anything about it, suggesting that passengers lack motivation, spaces and occasions to learn about latent ridehail conditions. 
\vspace{-.5em}
\paragraph{Pay Rates} \janeAdd{Drivers face a wide variety of stressors in daily operations that plague their well-being, but consistently prioritized pay as the most important misconception to clarify for passengers.}
For instance, D3.2 shared how they would ``\textit{love [for passengers] to know that I don't really get the full fare they are paying. [\dots] Uber, first of all, take their cut, then I cover fuel, car maintenance, time -- all these are swallowed by [what's broadly considered] service.}''
A direct result of this gap is lack of motivation to tip: ``\textit{[passengers] assume a lot of times that they don't have to tip [\dots] they feel like their charge all goes to the driver}'' (D2.1). 
In sum, D1.1 explains \textbf{how pay dominates all other concerns} since adequate compensation can alter drivers perceptions of all other stressors: 

\begin{quote}
    ``\textit{If we are investing our energy, our time, our efforts, frankly, we just don't [want] a feeling of being taken advantage of [\dots] of being manipulated [\dots] of being not cared about. I want to feel like somebody gives a [expletive] about me for the energy I'm putting in, [like] I'm making the money I'm making.}'' -- (D1.1)
\end{quote}

\janeAdd{As a result of low pay (``\textit{they take these risks simply because they're not being paid}''), D1.1 further describe how some drivers protest by ``\textit{flip[ping passengers] to cash rides}''\footnote{strategy where drivers and passengers mutually agree to pay in cash to avoid platform fees}, even though doing carries serious insurance and deactivation risks, as well as} ``\textit{a mandatory impounding of the[ir] vehicle [for] the charge of providing commercial transportation in a non-commercial vehicle}'' (D1.2). 

\paragraph{Pickup Logistics}
Drivers such as D1.3 detail how \textbf{platform mechanisms fall short to ensure timely passenger arrival during pick up}: 
``\textit{the customer doesn't really pay that much more in wait time [\dots] it's not enough to be prohibitive [\dots] It's almost like \dots [platforms] want to encourage that}''. Consequently, D3.3 explained how pickup delays lead to wasted gas and lost time that drivers could have used for additional trips: ``\textit{the waiting time is really expensive, especially when gas prices go up [\dots] it could just eat into our profits [so bad], and I feel like only us drivers really understand that.}'' 

\janeAdd{By contrast, D1.2 described their prior experience working for multiple cab companies -- where dispatchers would tell drivers to ``\textit{get out of there}'' and ''\textit{get another [passenger]}'' if the current one doesn't arrive within ``\textit{literally, in 30 seconds}''. Meanwhile, various road conditions (\textit{e.g.,} parking availability) can delay pickup when driving for ridehail:}
\begin{quote}
    ``\textit{I had to pick up someone, it was center of the city, and there's literally is no parking, it's red lanes on both sides -- bus only lanes. You can't drive in those lanes. You can't stop, you can't park [\dots] So I was sitting in that lane waiting [and] of course, the bus came through. I got a ticket.}'' -- (D1.2)
\end{quote}

\paragraph{\janeAdd{Rating Pressures Coerce Unpaid Services \& Labor}}
Another significant factor exerting psychological control in ridehail labor are passenger ratings of drivers, since strict thresholds can put many drivers at risk of deactivation by platforms \cite{free} -- making ''\textit{ratings feel personal}'' (D3.1) and thus coerced into performing a variety of \textbf{unpaid labor in service of ratings}. 
Inside of a ride, drivers frequently described how passengers expect ``\textit{a premium vibe}'' (D3.4), ''\textit{control over traffic}'' (D3.1), whereas drivers grapple with a variety of unpredictable factors, stressors and costs. Outside the car, D3.1 also point out external factors: ``\textit{construction, detours, sixth street chaos and airport gridlock}''.

\begin{quote}
    ``\textit{Ratings can drop for stuff I can't control: sometimes I'm reading the vibe and giving folks their space, not just ignoring them. We are not just driving, we are juggling through navigation, personalities and keeping things safe. The app itself is not perfect, it doesn't show everything we are dealing with [\dots] I'm paying attention to your body language, your tone, your energy, from the moment you get [in] [\dots] If you're if you're chatty, I'll match that. If you're quiet or stressed. I'll try to keep things calm and give you space so it's little things -- e.g., adjusting the music so it fits your mood, making sure the temperature is comfortable, choosing the smoothest route so you're not bouncing around in the back}.'' -- (D3.1)
\end{quote}

\paragraph{\janeAdd{Long-term Consequences of Algorithmic Control}}
Corroborating related work on psychological and algorithmic control, D1.2 described how platforms intimidate drivers with warning notifications (\textit{e.g.,} \textit{''You should only use last ride when you're when on your last ride''}) to discourage use of the feature since it would make driver supply less predictable. Unfortunately, D1.1 also discuss how many financially vulnerable drivers who primarily depend on ridehail earnings for a living cannot afford to challenge such tactics: ``\textit{most drivers they don't see the s***, they don't understand it \dots a lot of foreign nationals \dots [because] when you're here just trying to put food on the plate for your family, when you see a warning like that come up, you're not going to f*** around with your only source of income}''.

Beyond app mechanisms, \textbf{platforms also incentivize drivers to take short-term risks for small rewards} (e.g., bonuses or pay boosts) without regard for longer-term consequences. For instance, D3.3 described how drivers sometimes sacrifice bathroom breaks to maximize time toward bonuses:

\begin{quote}
    \textit{Those few seconds, even though I pee fast, it can make the difference between capturing it and not capturing [the ride]. We shouldn't be facing those choices. Most people, it's within reason. They can go to the bathroom on their job without facing, bonus losses [\dots] The pay boost isn't always worth the effort. [\dots] What if for that I have to go to a doctor, having to pay more money than I would've even got with a little bonus?}.''
\end{quote}

\subsection{\janeAdd{Designing Interactions that Motivate Driver-Passenger Solidarity }} \label{tradeoffs}
\janeAdd{To motivate and approach healthier interactions between passengers and drivers, participants shared several barriers, tradeoffs and incentives that can inform future designs of in-ride engagements between the two stakeholder groups.}

\subsubsection{\janeAdd{Respectfully Navigating Driver-Passenger Social Boundaries}}
While \questions presents an opportunity \janeAdd{for further social interactions between} drivers and passengers, both \janeAdd{stakeholder}s expressed \janeAdd{reservations} to initiate conversation, surfacing how interactions trade off with safety.

\paragraph{\janeAdd{Driver Vulnerability to Harassment}} Reflecting recent reports of harassment problems for both drivers and passengers \cite{festering}, participants such as D2.2 \janeAdd{expressed concern around} suggestive topics: ``\textit{sexually suggestive or flirty questions? No, no, not, not the place [\dots] It's gross, it's unsafe, and it creates a very dangerous precedent, especially for for women drivers and riders.}'' Even male-identifying drivers like D5.2 shared common experiences of passenger-imposed ``\textit{sexual harassment. I've gotten that a lot. Like: `}hey, I'll give you this money and you do this to me, or let me do this to you.\textit{' }'' In fact, D5.2 has even been rematched to a perpetrating passenger: ``\textit{I've reported several times to the Uber people about behaviors, and once I got this offender [that I once reported] again as my rider -- we were matched again}''

\paragraph{Safe \& Socially Appropriate Spaces} Up front, drivers feared \janeAdd{to cause \textbf{social awkwardness} or instigate potential \textbf{compromises to safety}:} ``\textit{not every rider wants to interact. And if the game feels too personal or like a distraction from driving, it can backfire}'' (D3.1). \janeAdd{Safety resonated with several} participants (P1.1-3, D2.2, D3.2-4,) -- D1.2 for instance, expressed approval with the \janeAdd{varied} interaction mode\janeAdd{s available in} \questions because ``\textit{it's safe[r] when [the passenger] talk versus the driver [\dots] We don't want to drive [while] interacting with anything, for safety.}''
In the backseat, passengers like P2.3 \janeAdd{reciprocated driver needs to keep the conversation comfortable and topically appropriate}, noting how ``\textit{in a car there is that dynamic of: they don't want to make you feel uncomfortable, and if I just bring up a topic that makes them feel uncomfortable, I don't think that [helps]}''. \janeAdd{The highly socially-conscious passenger P2.2 acknowledged:}

\begin{quote}
    ``\textit{a fine line between having a genuine interest and prying [\dots] as passengers, we have to be conscious of the fact that these drivers don't want to offend us, because after all, their rating is at stake, so questions should not be intrusive}''
\end{quote}
\janeAdd{In addition to fears of crossing personal boundaries of drivers,} passengers \janeAdd{also} took care to \textbf{minimize chances of belittling or criticizing }\janeAdd{\textbf{drivers}, out of awareness for their vulnerabilities:}
P1.2 ``\textit{don't really feel that it's like right for me to criticize the way they're doing their job, because I don't really know anything about what it's like being a rideshare driver}''. \janeAdd{P2.2 reflected on their own relative power as a consumer more explicitly:} ``\textit{it is important that we recognize what our position is in the rideshare thing, and we don't offend people or put them on spots}''.

To help them approach heavier topics, passengers (P1.4, P2.4, P3.2) found \driven effective at presenting thought-provoking content (e.g., harassment or family obligations) in creative and memorable ways -- ``\textit{fun to play in the backseat [\dots] I was pretty invested}'' (D2.2, D3.1) -- but it can also be too long, heavy or ``\textit{emotionally taxing}'' for a single ride (P2.4). For instance, D2.2 ``\textit{worry about it being too much for some passengers, [since] not everybody [is] in the mood for deep [content] or or even hearing stories}''.
However, passengers felt more comfortable discussing \textbf{lighter, ridehail-specific logistics}, including the number of rides the driver completed and reasons for cancellations (P3.3), time spent between rides (P2.2) or how ratings compare across service platforms (P2.3). When combined captivating game mechanics, more casual topics also help alleviate driver responsibilities to lighten the mood, by drawing passenger attention away from pressing factors (traffic, pressures of arriving on-time to destination):  ``\textit{it passes time fast, also especially helpful on short trips or when we are stuck in traffic, riders stop checking how long it's taking}.'' - (D3.2). 
However, \janeAdd{highly engaging or competitive gaming environments might} also risk upsetting or stressing out passengers who desire a more relaxed environment --- D3.1 and D2.2 consider how
``\textit{not everyone wants to think hard during a ride, some folks just wanna zone out or scroll their phone and if someone gets questions wrong over and over, it might actually stress them out}'', highlighting priorities to \textbf{avoid overly stressful or challenging game mechanisms} that are ``\textit{too competitive, too loud, or makes someone feel dumb}''.

\subsubsection{\janeAdd{Motivating Socially Responsible Passenger Engagement}} \label{incentives}
\janeAdd{Instead of expecting increasingly higher qualities of service, we discuss alternative ways that passengers and drivers desire to engage. Topically, these include conversations that discuss personal interests, local cultures or logistical labor. Other incentives that motivate passenger engagement include monetary contributions to drivers or charitable causes, as well as more punitive rating mechanisms that admonish passengers for irresponsible consumption behaviors.}

\paragraph{Connecting via Driver Interests, Labor \& Local Culture} Passengers yearned to connect with drivers and local events, pointing to in-ride content that could surface such knowledge. For instance, P1.1 recalled how a photo once sparked a conversation with a driver, and proposed physical cards featuring \textit{``little things that maybe a rider could connect to and be like, `}Oh, cool, you also watch Love Island.\textit{' ''} Drivers were generally enthused to share ridehail-specific insights (e.g., ``\textit{Uber drivers make playlists just for open silences}'') and personal information (e.g., \textit{``Did you know your driver once drove a pet pig to a party?''}) as well as music playlists (D5.2, D3.1); a few also suggested tailoring ``\textit{to rider interest or local culture}'' (D2.2)

\begin{quote}
    ``\textit{Seattle has the most coffee shops in the US, [so it's] not only about drivers, but also about local city parts -- weird, but true [facts]. Or we can have brainy or historical thing here, ridiculous laws}'' (D3.1)
\end{quote}

While \roads simulates the challenges of navigating to pick up passengers, drivers suggested other aspects of their labor as gameplay content. For instance, D1.2 suggested
gamifying micro-decisions such as \textit{``declining rides that aren't good''}, which can improve player competency at identifying exploitative rides.
D2.2 envisioned \janeAdd{games where the player acts as the driver to guess a passenger's mood}, to foreground emotional labor that drivers perform: \textit{``guess the rider's mood game based on small clues: the way they greeted you, or a storytelling round where the driver shares one situation and the rider has to react [\dots] so it helps highlight how much reading the room is part of what we do}''. 

\paragraph{Incentive Mechanisms} In terms of motivating further engagement and play, passengers expressed a desire for more tangible ways to act and help alleviate driver labor and working conditions. 
For instance, even though \href{https://playoctopus.com/}{Octopus} tablets can display driver profile information, P1.1 and P1.4 suggested additional ways of presenting surface-level information about drivers to spark conversation. 
Beyond knowledge-sharing, passengers like P2.2 proposed that directing their winnings to drivers (or charitable causes) would incentivize their engagement: ``\textit{I would think there should be an option in there where you could play and win something for your driver.}'' Drivers like D5.3 suggested mechanisms that make ``\textit{the games more interesting, by adding a leaderboard}'' which could even be extended across platforms: ``\textit{a leaderboard for the scores -- it would even be more challenging if you could do Uber against Lyft}'' (D2.1). In addition to positive rewards, P1.1 also considered the use of penalties to dissuade poor and rude passenger behaviors, such as giving greater weight to instances when drivers rate passengers poorly:

\begin{quote}
    \textit{If you're consistently reported as a rude a rider by drivers maybe you have to pay a fine [\dots so riders think]: Oh, I should probably not be rude to this driver, because then my next ride is gonna cost me more} (P1.1)
\end{quote}

\section{Discussion}
More than a decade ago, \citet{kitturfuture} posed the question of whether we can ``\textit{foresee a future a crowd workplace in which we would want our children to participate?}'' 
Borrowing this lens of healthier (crowd) workplaces for posterity, \janeAdd{we discuss implications for future  designers of algorithmically-mediated services, renewed approaches to advocacy and directions for how future technology interventions can support workers and their allies.}

\subsection{\janeAdd{Design Implications for Algorithmically-Mediated Production and Creation}
}
Our study surfaced opportunities for new \janeAdd{passenger-driver} interactions and incentives that can promote more aligned and mutually beneficial \janeAdd{behaviors} between drivers and passengers (\S\ref{incentives}). \janeAdd{However, advancing and realizing worker-consumer solidarity in practice -- \textit{e.g.,} socially responsible consumption, multi-stakeholder unity and altered platform design decisions -- will require core changes to how we all perceive and manage these services, at multiple scales. At the level of individual service interactions, designers might consider tacit ways of reminding consumers to engage in more aware, intentional and responsible micro-decisions that carry managerial function \cite{role} -- \textit{e.g., whether to tip drivers before or after a ride} -- we suggest in \S\ref{tech} future directions for training drivers to resist harmful micro-management by algorithms. More collectively, consumers can begin looking for more local means of advocating and influencing protective driver regulations \cite{handbook} -- described further in \S\ref{regulate}. }

\paragraph{\janeAdd{Examining Consumer Interactions to Visibilize Labor}} \janeAdd{\citet{morality} showed how platform designers experience internal conflict that arise from value differences between their moral identities and platforms' business objectives, corroborating the claim by \citet{legitimize} that platforms intentionally cultivate expectation-setting consumer behaviors such as tipping. Here we observe a triadic power relationship between algorithmic managers (or AMs \cite{f2f}), consumers and workers (ordered from high to low) \cite{triangle}, which can inform the design of an entire ecosystem of algorithmically-mediated services. With the exception of \citet{laundering}, we are unaware of previous lines of research within the US that conducted grounded, qualitative analysis on the consumer-AM side of this triangle, which can inform future designs of human-AI co-management or co-creation in general. We encourage future work to analyze more closely the sociomaterial working relations of algorithmically-mediated labor to better document and manage (in)visible accountabilities in labor \cite{located}.
}

\paragraph{\janeAdd{Towards Dignified, Autonomy-Preserving \& Fulfilling Work}}
\janeAdd{Platforms were heralded for their affordances of flexible and independent work. Much prior research attempted to measure and improve workers' degraded senses of job satisfaction \cite{quality} and well-being \cite{psychological, reimagined} in the gig economy. But underlying these surface level objectives are missing but central motivations of autonomy, dignity \cite{dignity} and agency \cite{mandatory} -- the lack of which can cause enormous psychological harm. How to design more purposive and negotiable platform mechanisms that satisfy intrinsic worker fulfillment while creating meaning at work? DW5 members mentioned fears of displacement by autonomous vehicles \cite{traffic}: what are underlying human motivations for resisting them \cite{resistance}? Can AVs fulfill core service needs of flexibility or convenience?}
\janeAdd{We took care to use the term \textit{ridehail} rather than \textit{rideshare} to (1) emphasize the lack of symmetric power-sharing and (2) bridge the gap between US-based and global scholars \cite{emails, triangle, power_resources, thelen}.
Historically, taxi services evolved from for-hire horse-drawn carriages in the 1600's \cite{taxi_hire}. But while traditionally licensed taxi operators owned fleets of taxicabs, does this driver-carriage-rider metaphor fit ridehail services? 
}

\subsection{\janeAdd{Mobilizing Public Discourse and Consumer-led Labor Advocacy through End-User Auditing}} \label{regulate}

\begin{quote}
    \emph{\janeAdd{``Consumers’ sympathies have not yet been decisively won by platforms or yet lost to the cause of improving gig work conditions'' -- \citet{handbook}}}
\end{quote}

\janeAdd{Crucially, ridehail offered a unique context of study where the consumer and service-provider are physically and temporally co-located to engage in interpersonal interactions, and our study offer the first attempt within HCI and CSCW to explore possibilities of reshaping the passenger-consumer interaction within a ride \cite{triangle, schor}. 
Since consumers as a collective carry significant political power in influencing platform decisions and policy \cite{triangle}, we expand the locus of resistance \cite{locus} to ``\textit{activate}'' a method ``\textit{soft action}'' -- gamified interaction \cite{considerations}, with an objective to \janeAdd{motivate behavioral change in} how consumers respond to, use and resist disruptive technology. By engaging consumers in labor advocacy, we open up opportunities for bottom-up end-user audits \cite{eu_audits, karahalios, sousveillance, bu_eu} (see \S\ref{lighter} for technical possibilities) or designs (\textit{e.g.,} nudges, asymmetric paternalism) that encourage behaviors of collective resistance \janeAdd{more intentional} choices (\textit{e.g.,} relatively socially conscious, sustainable competitor alternatives \cite{csr, soft_hard}) or non-use \cite{non_use}. }

\janeAdd{Despite the presence of competitors, platform choices in collude can still inhibit regulatory progress (such as in the case of Prop 22 \cite{thelen}) -- this is the juncture where consumer-led refusal is critical \cite{non_use}.}
Compared to the revenue-oriented interactions presented by Play Octopus, our prototypes carry additional purpose of raising awareness around driving conditions. 
Instead of relying on external ads as a source of support --- which harm passenger engagement (\textit{e.g.,} P1.1) --- many passengers (\textit{e.g.,} P2.2) inquired about potential prizes and rewards. \janeAdd{During workshops}, we informally introduced the idea of embedding and promoting local businesses, who can offer direct prizes (\textit{e.g.,} giftcards) and incentives (\textit{e.g.,} coupons) that in turn benefit their operations -- \janeAdd{enabling more local (and thus scalable) resistance} \cite{local}.

\janeAdd{Legally, tablets like Play Octopus cannot be formally banned from ridehail vehicles \cite{join}, since they provide drivers additional earnings when enough passengers engage. Depending on the region, regulations of their presence could put platforms in danger of misclassification lawsuits. 
However, platforms’ intentional emphasis on treating workers as ``\textit{independent contractor}'' users (or prosumers \cite{prosumer}) also means that regulatory attention can turn towards consumer protections to advance driver rights \cite{cpr}.
As ridehail internationalizes \cite{emails, triangle, power_resources, thelen}, how should wide-spanning academic (\textit{e.g.,} computing, labor, organizational management) and geographic communities remain in conversation with one another? As labor increasingly modularize \cite{flash}, what are more opportunities for more democratic policy creation \cite{policycraft} that allow workers to liberally voice concerns and share insights of working conditions \cite{gig2gether} across generations, without fears of platform retaliation?}


\subsection{\janeAdd{Connect Local Communities \& Build Meta-Cognitive Abilities with Power-Balanced, Immersive \& Creative Technologies}} \label{tech}
\janeAdd{In addition to potentials for engaging collective influence, we discuss how lightweight future technologies can help build more sustainable and connected gig communities, as well as ways to augment more meta-cognitive human skills and intuition so drivers (and workers more broadly) can better detect, monitor and conquer algorithmic management.}

\vspace{-.25em}
\paragraph{\janeAdd{Lighter, More Immersive Interactions to Surface Labor \& Connect Communities}} \label{lighter}
Our study uncovered driver desires for more in-ride interactions that better integrate passengers with ridehail-specific knowledge while connecting them to local facts and events (\S\ref{alignments}). Beyond connecting passengers to immediate physical realities (\textit{e.g.,} traffic, weather), drivers also appreciated how immersive simulations (e.g., \trivia or \roads) surfaced (exploitative) labor practices, which persist despite empirical evidence from more than a decade back \cite{immaterial} as well as rising labor consciousness and tech wariness \cite{back}. But while we intended to follow the Tandem Transformational process, our prototypes did not materialize physically, limiting the degree to which we can simulate realistic experiences. 
Follow up studies might consider how \janeAdd{emerging techniques can surface additional driver (and gig) labor at different scales. However, we must preface these ideas by cautioning the reader to first consider potential dangers of applying emerging simulation technologies, what guardrails are necessary?
How can mixed or virtual realities \cite{mr_empathy} simulate individual worker narratives? 
How can emerging video- and world-generation capabilities capture \cite{video_capture} or re-enact \cite{replay} labor accountabilities \cite{self_tracking}?
What insights can social simulations reveal on collective behaviors \cite{simulated_CA}, or policy formulations \cite{policycraft} of driver communities?}

\janeAdd{Beyond tablets, passengers expressed preferences for mobile software so they can engage on their own devices. Mobile interactions offer user affordances including social media-like interactions (\textit{e.g.,} discussions \cite{quallm}, chat rooms\footnote{\href{https://x.com/driversc2c?t=UbCM_Ag5ysfAqaUEKLzQow}{\textit{E.g.,} the weekly X space hosted by Drivers Coast 2 Coast}}) as well as more collection and organization of multimodal (\textit{e.g.,} spatial \cite{fairness}, audio, video) and multi-media data (\textit{e.g.,} new articles, videos, discussions and blogs \cite{locus}).}

\vspace{-.25em}
\paragraph{\janeAdd{Creatively Engaging Drivers' Meta-Cognitive Capabilities to Promote Driver Dignity}} 
We hope that exposure through \janeAdd{immersive} gamification can help players (including potential future drivers) to develop behaviors and strategies for resisting manipulative platform tactics \cite{wow}. 
\janeAdd{But what are sustainable ways to help drivers regain autonomy, agency and dignity over presently algorithmically-controlled actions? \citet{making_out} showed how more resourceful drivers engage in playful acts of resistance, which she termed relational games. Example strategies that drivers employed to win this game included psychological distancing to resist managerial nudges (\textit{e.g.,} notifications, badges) and the accepting of only profitable rides to ``\textit{make out}'' with net positive earnings. As tactics of algorithmic management evolve, what are other creative interaction probes and interventions that we can build to train workers' metacognitive abilities (\textit{e.g.,} planning \cite{planning}, monitoring \cite{monitoring}, reflecting \cite{metacognitions}) so they can more actively reflect on, track and resist mechanisms of algorithmic control?}

\vspace{-.25em}
\section{Conclusion}
\janeAdd{This work demonstrated the possibility for playable in-ride interventions to motivate consumer-led advocacy for labor conditions of ridehail drivers. We contribute a novel direction for worker advocacy that bridges the computing and labor communities. Future work can draw upon its insights to (1) gain more in-depth understanding of algorithmically-managed labor relations and organizations (2) foster and mobilize scalable consumer solidarity and advocacy (3) build creative interventions that train human workers to resist harmful algorithmic control.}

\bibliographystyle{ACM-Reference-Format}
\bibliography{references}

@article{wealth,
  title={Ride-Sharing the Wealth: Effects of Uber and Lyft on Jobs, Wages and Economic Growth},
  author={Koling, Adam and Armanios, Daniel Erian and Michalek, Jeremy J and Forsythe, Connor and Jha, Akshaya},
  note={SSRN working paper},
  doi={10.2139/ssrn.4865183},
  year={2024}
}

@article{traffic,
title = {Can sharing a ride make for less traffic? Evidence from Uber and Lyft and implications for cities},
journal = {Transport Policy},
volume = {102},
pages = {1-10},
year = {2021},
issn = {0967-070X},
doi = {https://doi.org/10.1016/j.tranpol.2020.12.015},
url = {https://www.sciencedirect.com/science/article/pii/S0967070X20309525},
author = {Bruce Schaller}
}

@techreport{arrived,
  author       = {Caitlin Gorback},
  title        = {Your Uber Has Arrived: Ridesharing and the Redistribution of Economic Activity},
  institution  = {Retail, Real Estate, and the Built Environment Initiative (RERI), Wharton School, University of Pennsylvania},
  type         = {Working Paper},
  number       = {RERI WP},
  year         = {2020},
  month        = {January},
  url          = {https://www.reri.org/research/files/2020_gorback_jmp.pdf}
}

@article{social,
  title={Understanding and modeling the social preferences for riders in rideshare matching},
  author={Cui, Yu and Makhija, Ramandeep Singh Manjeet Singh and Chen, Roger B and He, Qing and Khani, Alireza},
  journal={Transportation},
  volume={48},
  pages={1809--1835},
  year={2021},
  publisher={Springer},
  doi={10.1007/s11116-020-10112-0}
}

@inproceedings{emails,
author = {Kooti, Farshad and Grbovic, Mihajlo and Aiello, Luca Maria and Djuric, Nemanja and Radosavljevic, Vladan and Lerman, Kristina},
title = {Analyzing Uber's Ride-sharing Economy},
year = {2017},
isbn = {9781450349147},
publisher = {International World Wide Web Conferences Steering Committee},
address = {Republic and Canton of Geneva, CHE},
url = {https://doi.org/10.1145/3041021.3054194},
doi = {10.1145/3041021.3054194},
booktitle = {Proceedings of the 26th International Conference on World Wide Web Companion},
pages = {574–582},
numpages = {9},
location = {Perth, Australia},
series = {WWW '17 Companion}
}

@inproceedings{narration,
author = {Curran, Max T. and Gordon, Jeremy Raboff and Lin, Lily and Sridhar, Priyashri Kamlesh and Chuang, John},
title = {Understanding Digitally-Mediated Empathy: An Exploration of Visual, Narrative, and Biosensory Informational Cues},
year = {2019},
isbn = {9781450359702},
publisher = {Association for Computing Machinery},
address = {New York, NY, USA},
url = {https://doi.org/10.1145/3290605.3300844},
doi = {10.1145/3290605.3300844},
booktitle = {Proceedings of the 2019 CHI Conference on Human Factors in Computing Systems},
pages = {1–13},
numpages = {13},
location = {Glasgow, Scotland Uk},
series = {CHI '19}
}

@book{pew,
  title={The state of gig work in 2021},
  author={Anderson, Monica and McClain, Colleen and Faverio, Michelle and Gelles-Watnick, Risa},
  year={2021},
  publisher={Pew Research Center Washington, DC}
}

@article{distress, title={Casually cynical or trapped? Exploring gig workers’ reactions to psychological contract violation}, DOI={10.1108/jmp-10-2023-0624}, 
 journal={Journal of Managerial Psychology}, author={Saksida, Tina and Maffie, Michael and Mihelič, Katarina Katja and Culiberg, Barbara and Merkuž, Ajda}, year={2024} }

@article{triangle, title={The triangular relationship in platform gig work: Consumers, platform beneficence and worker vulnerability}, volume={40}, ISSN={0268-1072}, DOI={10.1111/ntwe.12310}, abstractNote={Platform gig work is created and contested in dynamic, triangular relationships between platforms, workers, and consumers. Compared with the first two groups, however, evidence about the role of the third—consumers—is sparse. In this paper, we investigate consumers’ changing perceptions of work in the platform gig economy and argue that their perspective warrants greater attention in sociological analyses. Using data from two Australian public opinion surveys conducted 5 years apart (2017 and 2022), we explore how consumers’ views of platform gig work evolved during a period of rapid change that includes the first 2 years of the COVID-19 pandemic. We find that while overall platform use increased markedly, many consumers felt conflicted about gig workers’ conditions and key features of platforms’ typical labour practices. There is a pronounced, enduring, and consequential tension in consumers’ views of the merits and drawbacks of this work; between, on the one hand, an acceptance that platforms do benefit workers to some extent and, on the other hand, misgivings about workers’ vulnerability to harm. In centring consumers, our paper empirically enriches current triangular conceptions of labour relations in the platform gig economy, by showing how consumers mediate the interests of platforms and workers, to shape how gig work manifests and who benefits from it. We also contribute useful new practical knowledge, by elucidating the prevailing concerns of consumers that could be developed into resonant themes for campaigns aimed at improving platform gig workers’ rights.}, number={2}, journal={New Technology, Work and Employment}, author={Healy, Joshua and Pekarek, Andreas}, year={2025}, pages={265–284} }

@article{influence,
  title={Influence of service excellence on consumer satisfaction of ridesharing industry},
  author={Ziyad, Ahmed and Rehman, ZU and Batool, Zahara and Khan, Ammad Hassan},
  journal={International Journal for Traffic and Transport Engineering},
  volume={10},
  number={4},
  pages={468--481},
  year={2020},
  publisher={Scientific Research Journal Ltd.}
}

@inproceedings{kitturfuture,
  title={The future of crowd work},
  author={Kittur, Aniket and Nickerson, Jeffrey V and Bernstein, Michael and Gerber, Elizabeth and Shaw, Aaron and Zimmerman, John and Lease, Matt and Horton, John},
  booktitle={Proceedings of the 2013 conference on Computer supported cooperative work},
  pages={1301--1318},
  year={2013}
}

@inproceedings{prabowo2019does,
  title={Does gamification motivate gig workers? A critical issue in ride-sharing industries},
  author={Prabowo, Rahmanto and Sucahyo, Yudho Giri and Gandhi, Arfive and Ruldeviyani, Yova},
  booktitle={2019 International Conference on Advanced Computer Science and information Systems (ICACSIS)},
  pages={343--348},
  year={2019},
  organization={IEEE}
}

@article{vasudevangame,
  title={Gamification and work games: Examining consent and resistance among Uber drivers},
  author={Vasudevan, Krishnan and Chan, Ngai Keung},
  journal={new media \& society},
  volume={24},
  number={4},
  pages={866--886},
  year={2022},
  publisher={SAGE Publications Sage UK: London, England}
}

@article{long_hours,
  title={Online manipulation: Hidden influences in a digital world},
  author={Susser, Daniel and Roessler, Beate and Nissenbaum, Helen},
  journal={Geo. L. Tech. Rev.},
  volume={4},
  pages={1},
  year={2019},
  publisher={HeinOnline}
}

@article{kaufman2stealth,
  title={Creating stealth game interventions for attitude and behavior change: An ‘embedded design’model},
  author={Kaufman, Geoff and Flanagan, Mary and Seidman, Max},
  journal={Persuasive gaming in context},
  pages={73},
  year={2021},
  publisher={Amsterdam University Press}
}

@book{transformational,
  title     = {The Transformational Framework: A Process Tool for the Development of Transformational Games},
  author    = {Culyba, Sabrina H.},
  year      = {2018},
  publisher = {ETC Press},
  address   = {Pittsburgh, PA},
  url       = {https://press.etc.cmu.edu/books/transformational-framework}
}

@article{provotypes, title={Interactive Fiction Provotypes for Coping with Interpersonal Racism}, DOI={10.1145/3491102.3502044}, abstractNote={Reducing uncertainty around the nature of racist interactions is one of the key motivations driving individual behaviors for coping with those incidents. However, there are few appropriate technologies to support BIPOC (Black, Indigenous, People of Color) in engaging in social uncertainty reduction around this vulnerable, sensitive topic. This paper reports on an exploratory design study investigating how social technology might facilitate uncertainty reduction through three “provotypes” - provocative prototypes of user-generated speculative design concepts. U.S.-based participants engaged with the provotypes through an interactive fiction to explore their usefulness in the context of a racist microaggression. Results showed that engaging the provotypes through interactive fiction facilitated complex and productive interactions and critiques. This work contributes a novel method for conducting exploratory design, remote user studies using interactive fiction as well as priorities, tensions, and further information what role, if any, technology might play in managing racist interactions.}, journal={CHI Conference on Human Factors in Computing Systems}, author={To, Alexandra and Carey, Hillary and Shrivastava, Riya and Hammer, Jessica and Kaufman, Geoff}, year={2022}, pages={1–14} }

@article{hiv,
  title={A qualitative study to inform the development of a videogame for adolescent human immunodeficiency virus prevention},
  author={Hieftje, Kimberly and Rosenthal, Marjorie S and Camenga, Deepa R and Edelman, E Jennifer and Fiellin, Lynn E},
  journal={GAMES FOR HEALTH: Research, Development, and Clinical Applications},
  volume={1},
  number={4},
  pages={294--298},
  year={2012},
  publisher={Mary Ann Liebert, Inc. 140 Huguenot Street, 3rd Floor New Rochelle, NY 10801 USA}
}

@inproceedings{standbyme,
  title={Standbyme: a gamified educational platform to raise awareness on gender-based violence},
  author={Roumelioti, Eftychia and Gini, Federica and Jakobi, Antonia Laura Philipa and Marconi, Annapaola and Ny{\'u}l, Bogl{\'a}rka and Paladino, Maria Paola and Schiavo, Gianluca and Zancanaro, Massimo},
  booktitle={Companion proceedings of the annual symposium on computer-human interaction in play},
  pages={108--113},
  year={2023}
}

@article{sannondisabilities,
  title={Toward a more inclusive gig economy: Risks and opportunities for workers with disabilities},
  author={Sannon, Shruti and Cosley, Dan},
  journal={Proceedings of the ACM on Human-Computer Interaction},
  volume={6},
  number={CSCW2},
  pages={1--31},
  year={2022},
  publisher={ACM New York, NY, USA}
}

@article{anomalies,
  title={Ridesharing and digital resilience for urban anomalies: Evidence from the New York City taxi market},
  author={Zhang, Yingjie and Li, Beibei and Qian, Sean},
  journal={Information Systems Research},
  volume={34},
  number={4},
  pages={1775--1790},
  year={2023},
  publisher={INFORMS}
}

@inproceedings{sousveillance,
author = {Do, Kimberly and De Los Santos, Maya and Muller, Michael and Savage, Saiph},
title = {Designing Gig Worker Sousveillance Tools},
year = {2024},
isbn = {9798400703300},
publisher = {Association for Computing Machinery},
address = {New York, NY, USA},
url = {https://doi.org/10.1145/3613904.3642614},
doi = {10.1145/3613904.3642614},
booktitle = {Proceedings of the CHI Conference on Human Factors in Computing Systems},
articleno = {384},
numpages = {19},
location = {Honolulu, HI, USA},
series = {CHI '24}
}

@article{making,
author = {Blaising, Allie and Kotturi, Yasmine and Kulkarni, Chinmay and Dabbish, Laura},
title = {Making it Work, or Not: A Longitudinal Study of Career Trajectories Among Online Freelancers},
year = {2021},
issue_date = {December 2020},
publisher = {Association for Computing Machinery},
address = {New York, NY, USA},
volume = {4},
number = {CSCW3},
url = {https://doi.org/10.1145/3432925},
doi = {10.1145/3432925},
journal = {Proc. ACM Hum.-Comput. Interact.},
month = {jan},
articleno = {226},
numpages = {29}
}

@article{health,
  title={Work-life balance and gig work: ‘Where are we now’and ‘where to next’with the work--life balance agenda?},
  author={Warren, Tracey},
  journal={Journal of Industrial Relations},
  volume={63},
  number={4},
  pages={522--545},
  year={2021},
  publisher={SAGE Publications Sage UK: London, England},
  doi = {10.1177/00221856211007161}
}

@article{safety,
  title={The health and safety risks for people who drive for work in the gig economy},
  author={Christie, Nicola and Ward, Heather},
  journal={Journal of Transport \& Health},
  volume={13},
  pages={115--127},
  year={2019},
  publisher={Elsevier},
  doi = {10.1016/j.jth.2019.02.007}
}

@article{consent,
author = {Juliet B Schor and Christopher Tirrell and Steven Peter Vallas},
title ={Consent and Contestation: How Platform Workers Reckon with the Risks of Gig Labor},
journal = {Work, Employment and Society},
volume = {38},
number = {5},
pages = {1423-1444},
year = {2024},
doi = {10.1177/09500170231199404},
URL = { 
https://doi.org/10.1177/09500170231199404
},
eprint = {https://doi.org/10.1177/09500170231199404}
}

@inproceedings{brush,
author = {Ma, Ning F. and Rivera, Veronica A. and Yao, Zheng and Yoon, Dongwook},
title = {“Brush it Off”: How Women Workers Manage and Cope with Bias and Harassment in Gender-agnostic Gig Platforms},
year = {2022},
isbn = {9781450391573},
publisher = {Association for Computing Machinery},
address = {New York, NY, USA},
url = {https://doi.org/10.1145/3491102.3517524}, 
doi = {10.1145/3491102.3517524},
booktitle = {Proceedings of the 2022 CHI Conference on Human Factors in Computing Systems},
articleno = {397},
numpages = {13},
location = {New Orleans, LA, USA},
series = {CHI '22}
}

@article{bargaining,
author = {Calacci, Dan and Pentland, Alex},
title = {Bargaining with the Black-Box: Designing and Deploying Worker-Centric Tools to Audit Algorithmic Management},
year = {2022},
issue_date = {November 2022},
publisher = {Association for Computing Machinery},
address = {New York, NY, USA},
volume = {6},
number = {CSCW2},
url = {https://doi.org/10.1145/3570601},
doi = {10.1145/3570601},
journal = {Proc. ACM Hum.-Comput. Interact.},
month = {nov},
articleno = {428},
numpages = {24}
}

@misc{beyond,
  title={Beyond disruption: How tech shapes labor across domestic work and ridehailing},
  author={Ticona, Julia and Mateescu, Alexandra and Rosenblat, Alex},
  journal={Data \& Society},
  year={2018},
  publisher={APO: Analysis \& Policy Observatory}
}

@article{hazards,
  title={Physical and psychological hazards in the gig economy system: A systematic review},
  author={Taylor, Kelvin and Van Dijk, Pieter and Newnam, Sharon and Sheppard, Dianne},
  journal={Safety science},
  volume={166},
  pages={106234},
  year={2023},
  publisher={Elsevier}
}

@article{fte,
  title={Understanding perception of algorithmic decisions: Fairness, trust, and emotion in response to algorithmic management},
  author={Lee, Min Kyung},
  journal={Big data \& society},
  volume={5},
  number={1},
  pages={2053951718756684},
  year={2018},
  publisher={SAGE Publications Sage UK: London, England}
}

@INPROCEEDINGS{Jarrahiam,
  title     = "Algorithmic Management and Algorithmic Competencies:
               Understanding and Appropriating Algorithms in Gig Work",
  booktitle = "Information in Contemporary Society",
  author    = "Jarrahi, Mohammad Hossein and Sutherland, Will",
  publisher = "Springer International Publishing",
  pages     = "578--589",
  year      =  2019
}

@article{kreuger,
  title={Modernizing labor laws for twenty-first-century work: the" independent worker"},
  author={Harris, Seth and Krueger, Alan},
  year={2015},
  publisher={The Hamilton Project}
}

@article{dubal_ab5,
  title={Economic security \& the regulation of gig work in California: From AB5 to Proposition 22},
  author={Dubal, Veena B},
  journal={European labour law journal},
  volume={13},
  number={1},
  pages={51--65},
  year={2022},
  publisher={SAGE Publications Sage UK: London, England}
}

@article{fastdrink, title={Fast Drink: Mediating Empathy for Gig Workers}, DOI={10.1145/3588967.3588975}, journal={Proceedings of the 2nd Empathy-Centric Design Workshop}, author={Meijer, Wo and Verhoeff, Bent and Verma, Himanshu and Bourgeois, Jacky}, year={2023}, pages={1–6} }

@inproceedings{gig2gether,
author = {Hsieh, Jane and Zhang, Angie and Surati, Sajel and Xie, Sijia and Ayala, Yeshua and Sathiya, Nithila and Kuo, Tzu-Sheng and Lee, Min Kyung and Zhu, Haiyi},
title = {Gig2Gether: Datasharing to Empower, Unify and Demystify Gig Work},
year = {2025},
isbn = {9798400713941},
publisher = {Association for Computing Machinery},
address = {New York, NY, USA},
url = {https://doi.org/10.1145/3706598.3714398},
doi = {10.1145/3706598.3714398},
booktitle = {Proceedings of the 2025 CHI Conference on Human Factors in Computing Systems},
articleno = {99},
numpages = {25},
location = {
},
series = {CHI '25}
}

@inproceedings{zhang2023stakeholder,
author = {Zhang, Angie and Boltz, Alexander and Lynn, Jonathan and Wang, Chun-Wei and Lee, Min Kyung},
title = {Stakeholder-Centered AI Design: Co-Designing Worker Tools with Gig Workers through Data Probes},
year = {2023},
isbn = {9781450394215},
publisher = {Association for Computing Machinery},
url = {https://doi.org/10.1145/3544548.3581354},
doi = {10.1145/3544548.3581354},
booktitle = {Proceedings of the 2023 CHI Conference on Human Factors in Computing Systems},
articleno = {859},
numpages = {19},
location = {Hamburg, Germany},
series = {CHI '23}
}

@article{durlauf2019commodification,
  title={The commodification of digital labor in the gig economy: Online outsourcing, insecure employment, and platform-based rating and ranking systems},
  author={Durlauf, Maria},
  journal={Psychosociological Issues in Human Resource Management},
  volume={7},
  number={1},
  pages={54--59},
  year={2019},
  publisher={Addleton Academic Publishers}
}

@article{stressfulride,
  title={Stressful by design: Exploring health risks of ride-share work},
  author={Bartel, Emma and MacEachen, Ellen and Reid-Musson, Emily and Meyer, Samantha B and Saunders, Ron and Bigelow, Philip and Kosny, Agnieszka and Varatharajan, Sharanya},
  journal={Journal of Transport \& Health},
  volume={14},
  pages={100571},
  year={2019},
  publisher={Elsevier}
}

@inproceedings{immaterial,
author = {Raval, Noopur and Dourish, Paul},
title = {Standing Out from the Crowd: Emotional Labor, Body Labor, and Temporal Labor in Ridesharing},
year = {2016},
isbn = {9781450335928},
publisher = {Association for Computing Machinery},
address = {New York, NY, USA},
url = {https://doi.org/10.1145/2818048.2820026},
doi = {10.1145/2818048.2820026},
booktitle = {Proceedings of the 19th ACM Conference on Computer-Supported Cooperative Work \& Social Computing},
pages = {97–107},
numpages = {11},
location = {San Francisco, California, USA},
series = {CSCW '16}
}

@inproceedings{considerations,
author = {Raval, Noopur and Qadri, Rida and Wong, Richmond Y. and Kneese, Tamara and Hanna, Alex},
title = {Considerations for Building Solidarity among Academic and Tech Workers: Thinking through access, positionality and limits to collective action},
year = {2022},
isbn = {9781450391566},
publisher = {Association for Computing Machinery},
address = {New York, NY, USA},
url = {https://doi.org/10.1145/3491101.3516511},
doi = {10.1145/3491101.3516511},
booktitle = {Extended Abstracts of the 2022 CHI Conference on Human Factors in Computing Systems},
articleno = {153},
numpages = {3},
location = {New Orleans, LA, USA},
series = {CHI EA '22}
}

@article{exploitation,
author = {Niels van Doorn},
title = {Platform labor: on the gendered and racialized exploitation of low-income service work in the ‘on-demand’ economy},
journal = {Information, Communication \& Society},
volume = {20},
number = {6},
pages = {898--914},
year = {2017},
publisher = {Routledge},
doi = {10.1080/1369118X.2017.1294194},
URL = { 
        https://doi.org/10.1080/1369118X.2017.1294194
},
eprint = { 
        https://doi.org/10.1080/1369118X.2017.1294194
}
}

@article{eu_audits,
author = {Lam, Michelle S. and Gordon, Mitchell L. and Metaxa, Dana\"{e} and Hancock, Jeffrey T. and Landay, James A. and Bernstein, Michael S.},
title = {End-User Audits: A System Empowering Communities to Lead Large-Scale Investigations of Harmful Algorithmic Behavior},
year = {2022},
issue_date = {November 2022},
publisher = {Association for Computing Machinery},
address = {New York, NY, USA},
volume = {6},
number = {CSCW2},
url = {https://doi.org/10.1145/3555625},
doi = {10.1145/3555625},
journal = {Proc. ACM Hum.-Comput. Interact.},
month = nov,
articleno = {512},
numpages = {34}
}

@article{prosocial,
  title={Effects of immersive stories on prosocial attitudes and willingness to help: testing psychological mechanisms},
  author={Ma, Zexin},
  journal={Media Psychology},
  volume={23},
  number={6},
  pages={865--890},
  year={2020},
  publisher={Taylor \& Francis}
}

@inproceedings{tandem,
  title={Tandem transformational game design: A game design process case study},
  author={To, Alexandra and Fath, Elaine and Zhang, Eda and Ali, Safinah and Kildunne, Catherine and Fan, Anny and Hammer, Jessica and Kaufman, Geoff},
  booktitle={Proceedings of the International Academic Conference on Meaningful Play},
  year={2016}
}

@article{cards,
  title={Cards against gamification: Using a role-playing game to tell alternative futures in the gig economy},
  author={Popan, Cosmin and Perez, David and Woodcock, Jamie},
  journal={The Sociological Review},
  volume={71},
  number={5},
  pages={1058--1074},
  year={2023},
  publisher={SAGE Publications Sage UK: London, England}
}

@article{making_out,
  title={“Making out” while driving: Relational and efficiency games in the gig economy},
  author={Cameron, Lindsey D},
  journal={Organization Science},
  volume={33},
  number={1},
  pages={231--252},
  year={2022},
  publisher={INFORMS}
}

@article{jiang2019more,
  title={More Americans are using ride-hailing apps},
  author={Jiang, Jingjing},
  year={2019},
  publisher={Pew Research Center}
}

@article{consciousness,
author = {Gonzalo Frasca},
title = {Rethinking agency and immersion: video games as a means of consciousness-raising},
journal = {Digital Creativity},
volume = {12},
number = {3},
pages = {167--174},
year = {2001},
publisher = {CAA Website},
doi = {10.1076/digc.12.3.167.3225},
URL = { 
        https://doi.org/10.1076/digc.12.3.167.3225
},
eprint = { 
        https://doi.org/10.1076/digc.12.3.167.3225
}
}

@article{understanding, title={What Do Platforms Do? Understanding the Gig Economy}, volume={46}, ISSN={0360-0572}, DOI={10.1146/annurev-soc-121919-054857}, abstractNote={The rapid growth of the platform economy has provoked scholarly discussion of its consequences for the nature of work and employment. We identify four major themes in the literature on platform work and the underlying metaphors associated with each. Platforms are seen as entrepreneurial incubators, digital cages, accelerants of precarity, and chameleons adapting to their environments. Each of these devices has limitations, which leads us to introduce an alternative image of platforms: as permissive potentates that externalize responsibility and control over economic transactions while still exercising concentrated power. As a consequence, platforms represent a distinct type of governance mechanism, different from markets, hierarchies, or networks, and therefore pose a unique set of problems for regulators, workers, and their competitors in the conventional economy. Reflecting the instability of the platform structure, struggles over regulatory regimes are dynamic and difficult to predict, but they are sure to gain in prominence as the platform economy grows. Expected final online publication date for the Annual Review of Sociology, Volume 46 is July 30, 2020. Please see http://www.annualreviews.org/page/journal/pubdates for revised estimates.}, number={1}, journal={Annual Review of Sociology}, author={Vallas, Steven and Schor, Juliet B.}, year={2020}, pages={1–22} }

@article{role, title={The role of customers in the gig economy: how perceptions of working conditions and service quality influence the use and recommendation of food delivery services}, volume={15}, ISSN={1862-8516}, DOI={10.1007/s11628-020-00432-7}, abstractNote={This research examines how customers’ perceptions about controversial labor practices of food delivery platforms may affect their intentions to use and recommend these services. Three studies reveal that customers’ behavioral intentions depend on their perceptions of the working conditions for the delivery workers, as well as service quality. This influence is higher among customers with a high level of social conscious consumption. Our research also explores the costs that customers would be willing to assume to be served by a food delivery service that offers better working conditions. These insights reveal several relevant managerial implications for gig economy firms.}, number={1}, journal={Service Business}, author={Belanche, Daniel and Casaló, Luis V. and Flavián, Carlos and Pérez-Rueda, Alfredo}, year={2021}, pages={45–75} }

@inproceedings{navigating,
author = {Hsieh, Jane and Karger, Miranda and Zagal, Lucas and Zhu, Haiyi},
title = {Co-Designing Alternatives for the Future of Gig Worker Well-Being: Navigating Multi-Stakeholder Incentives and Preferences},
year = {2023},
isbn = {9781450398930},
publisher = {Association for Computing Machinery},
address = {New York, NY, USA},
url = {https://doi.org/10.1145/3563657.3595982},
doi = {10.1145/3563657.3595982},
booktitle = {Proceedings of the 2023 ACM Designing Interactive Systems Conference},
pages = {664–687},
numpages = {24},
location = {Pittsburgh, PA, USA},
series = {DIS '23}
}

@article{rating,
  title={The rating game: The discipline of Uber’s user-generated ratings},
  author={Chan, Ngai Keung},
  journal={Surveillance \& Society},
  volume={17},
  number={1/2},
  pages={183--190},
  year={2019}
}

@article{asymmetries,
  title={Algorithmic labor and information asymmetries: A case study of Uber’s drivers},
  author={Rosenblat, Alex and Stark, Luke},
  journal={International journal of communication},
  volume={10},
  pages={27},
  year={2016}
}

@article{fatigue,
  title={Driver fatigue in taxi, ride-hailing, and ridesharing services: a systematic review},
  author={Jaydarifard, Saeed and Behara, Krishna and Baker, Douglas and Paz, Alexander},
  journal={Transport Reviews},
  volume={44},
  number={3},
  pages={572--590},
  year={2024},
  publisher={Taylor \& Francis}
}

@incollection{tricks,
  title={How Uber uses psychological tricks to push its drivers' buttons},
  author={Scheiber, Noam},
  booktitle={Ethics of data and analytics},
  pages={362--371},
  year={2022},
  publisher={Auerbach Publications}
}

@article{festering,
  author       = {Emily Steel},
  title        = {Uber’s Festering Sexual Assault Problem},
  journal      = {The New York Times},
  date         = {2025-08-06},
  note         = {Updated 2025-08-07},
  url          = {https://www.nytimes.com/2025/08/06/business/uber-sexual-assault.html},
}

@article{mediatization,
  title={Mediatization of social space and the case of Uber drivers},
  author={Chan, Ngai Keung and Humphreys, Lee},
  journal={Media and Communication},
  volume={6},
  number={2},
  pages={29--38},
  year={2018}
}

@article{translating,
  title={Rideshare transparency: Translating gig worker insights on ai platform design to policy},
  author={Nagaraj Rao, Varun and Dalal, Samantha and Agarwal, Eesha and Calacci, Dana and Monroy-Hern{\'a}ndez, Andr{\'e}s},
  journal={Proceedings of the ACM on Human-Computer Interaction},
  volume={9},
  number={2},
  pages={1--49},
  year={2025},
  publisher={ACM New York, NY, USA}
}

@inproceedings{alternative,
  title={Co-designing alternatives for the future of gig worker well-being: Navigating multi-stakeholder incentives and preferences},
  author={Hsieh, Jane and Karger, Miranda and Zagal, Lucas and Zhu, Haiyi},
  booktitle={Proceedings of the 2023 ACM Designing Interactive Systems Conference},
  pages={664--687},
  year={2023}
}

@inproceedings{non_use,
  title={Data leverage: A framework for empowering the public in its relationship with technology companies},
  author={Vincent, Nicholas and Li, Hanlin and Tilly, Nicole and Chancellor, Stevie and Hecht, Brent},
  booktitle={Proceedings of the 2021 ACM Conference on Fairness, Accountability, and Transparency},
  pages={215--227},
  year={2021}
}

@inproceedings{stein2023you,
author = {Stein, Jake M L and Vizgirda, Vidminas and Van Kleek, Max and Binns, Reuben and Zhao, Jun and Zhao, Rui and Goel, Naman and Chalhoub, George and Albayaydh, Wael S and Shadbolt, Nigel},
title = {‘You are you and the app. There’s nobody else.’: Building Worker-Designed Data Institutions within Platform Hegemony},
year = {2023},
isbn = {9781450394215},
publisher = {Association for Computing Machinery},
address = {New York, NY, USA},
url = {https://doi.org/10.1145/3544548.3581114},
doi = {10.1145/3544548.3581114},
booktitle = {Proceedings of the 2023 CHI Conference on Human Factors in Computing Systems},
articleno = {281},
numpages = {26},
location = {Hamburg, Germany},
series = {CHI '23}
}

@article{fairfare,
author = {Calacci, Dana and Nagaraj Rao, Varun and Dalal, Samantha and Di, Catherine and Pua, Kok-Wei and Schwartz, Andrew and Spitzberg, Danny and Monroy-Hern\'{a}ndez, Andr\'{e}s},
title = {FairFare: A Tool for Crowdsourcing Rideshare Data to Empower Labor Organizers},
year = {2025},
publisher = {Association for Computing Machinery},
address = {New York, NY, USA},
issn = {1073-0516},
url = {https://doi.org/10.1145/3769680},
doi = {10.1145/3769680},
note = {Just Accepted},
journal = {ACM Trans. Comput.-Hum. Interact.},
month = oct
}

@inproceedings{workshop,
  title={Worker Data Collectives as a means to Improve Accountability, Combat Surveillance and Reduce Inequalities},
  author={Hsieh, Jane and Zhang, Angie and Kim, Seyun and Rao, Varun Nagaraj and Dalal, Samantha and Mateescu, Alexandra and Grohmann, Rafael Do Nascimento and Eslami, Motahhare and Lee, Min Kyung and Zhu, Haiyi},
  note={To Appear},
  url = {https://arxiv.org/abs/2409.00737},
  booktitle={Proceedings of the ACM 2012 Conference on Computer Supported Cooperative Work},
  year={2024},
  doi={3678884.3681829},
}

@article{rapport, title={Customer-Employee Rapport in Service Relationships}, volume={3}, ISSN={1094-6705}, DOI={10.1177/109467050031006}, abstractNote={Relationships are an important aspect of doing business, and few businesses can survive without establishing solid relationships with their customers. Although the marketing literature suggests that personal relationships can be important to service firms, little specificity has been provided as to which relational aspects should receive attention. In this study, the authors examine one specific aspect of customer-employee relationships, rapport, that they believe may be particularly salient in service businesses characterized by a high amount of interpersonal interactions. Rapport has received relatively little attention in the marketing literature; the goal of this study is to fill this gap in the literature. In two different service contexts, the authors find support for two empirically distinct dimensions of rapport. They also find a positive relationship between these dimensions and satisfaction, loyalty intent, and word-of-mouth communication. They conclude by suggesting future research directions for further academic inquiry of rapport in service contexts.}, number={1}, journal={Journal of Service Research}, author={Gremler, Dwayne D. and Gwinner, Kevin P.}, year={2000}, pages={82–104} }

@article{consumer_empathy, title={On the Role of Empathy in Customer-Employee Interactions}, volume={15}, ISSN={1094-6705}, DOI={10.1177/1094670512439743}, abstractNote={While the service literature repeatedly emphasizes the role of empathy in service interactions, studies on empathy in customer-employee interactions are nearly absent. This study defines and conceptualizes employee and customer empathy as multidimensional constructs and empirically investigates their impact on customer satisfaction and customer loyalty. A quantitative study based on dyadic data and a multilevel modeling approach finds support for two effects of empathy in service interactions. The study reveals that customer empathy strengthens the positive effect of employee empathy on customer satisfaction, leading to more “symbiotic interactions.” The findings also indicate that empathic customers are more likely to respond to a dissatisfying encounter with “forgiveness,” in the sense that customer empathy is able to mitigate negative effects of customer dissatisfaction on customer loyalty. From these empirical results, the authors derive several implications for service research and the management of service encounters. In particular, the present study provides a valuable basis for strategies of “interaction routing,” that is, matching customers and employees on the basis of their psychological profiles to create smooth and satisfying service interactions. The authors elaborate on approaches to implement this strategy in service organizations.}, number={3}, journal={Journal of Service Research}, author={Wieseke, Jan and Geigenmüller, Anja and Kraus, Florian}, year={2012}, pages={316–331} }

@inproceedings{compassion,
author = {Lee, Ken Jen and Davila, Adrian and Cheng, Hanlin and Goh, Joslin and Nilsen, Elizabeth and Law, Edith},
title = {“We need to do more... I need to do more”: Augmenting Digital Media Consumption via Critical Reflection to Increase Compassion and Promote Prosocial Attitudes and Behaviors},
year = {2023},
isbn = {9781450394215},
publisher = {Association for Computing Machinery},
address = {New York, NY, USA},
url = {https://doi.org/10.1145/3544548.3581355},
doi = {10.1145/3544548.3581355},
booktitle = {Proceedings of the 2023 CHI Conference on Human Factors in Computing Systems},
articleno = {66},
numpages = {20},
location = {Hamburg, Germany},
series = {CHI '23}
}

@article{sceptics,
  title={Sceptics or supporters? Consumers’ views of work in the gig economy},
  author={Healy, Joshua and Pekarek, Andreas and Vromen, Ariadne},
  journal={New Technology, Work and Employment},
  volume={35},
  number={1},
  pages={1--19},
  year={2020},
  publisher={Wiley Online Library}
}

@article{mr_empathy,
  title={6. A Breathtaking Journey. Appealing to Empathy in a Persuasive Mixed-Reality Game},
  author={Kors, Martijn and Ferri, Gabriele and van der Spek, Erik D and Ketel, Cas and Schouten, Ben},
  journal={Persuasive gaming in context},
  pages={95},
  year={2021},
  publisher={Amsterdam University Press Amsterdam}
}

@article{legitimize,
  title={How Consumer Empathy Drives Platform Success},
  author={Giesler, Markus and Veresiu, Ela and Humphreys, Ashlee},
  journal={Marketing Science Institute Working Paper Series},
  year={2019}
}

@article{seat,
author = {Rankin, Yolanda A. and Irish, India},
title = {A Seat at the Table: Black Feminist Thought as a Critical Framework for Inclusive Game Design},
year = {2020},
issue_date = {October 2020},
publisher = {Association for Computing Machinery},
address = {New York, NY, USA},
volume = {4},
number = {CSCW2},
url = {https://doi.org/10.1145/3415188},
doi = {10.1145/3415188},
journal = {Proc. ACM Hum.-Comput. Interact.},
month = oct,
articleno = {117},
numpages = {26}
}

@book{crossing,
  title={The rhetoric of video games},
  author={Bogost, Ian},
  year={2008},
  publisher={MacArthur Foundation Digital Media and Learning Initiative}
}

@article{wow,
  title={Corporate ideology in World of Warcraft},
  author={Rettberg, Scott and Corneliussen, Hilde G and Rettberg, Jill Walker},
  journal={Digital culture, play, and identity: A World of Warcraft reader},
  pages={19--38},
  year={2008},
  publisher={MIT Press Cambridge, MA}
}

@article{negotiate,
  title={Game over? Negotiating modern capitalism in virtual game worlds},
  author={Harambam, Jaron and Aupers, Stef and Houtman, Dick},
  journal={European Journal of Cultural Studies},
  volume={14},
  number={3},
  pages={299--319},
  year={2011},
  publisher={Sage Publications Sage UK: London, England}
}

@book{civ,
  title={Videogames and postcolonialism: Empire plays back},
  author={Mukherjee, Souvik},
  year={2017},
  publisher={Springer}
}

@book{persuasive,
  title={Persuasive games},
  author={Bogost, Ian},
  year={2007},
  publisher={MIT press}
}

@article{papers,
  title={Papers, Please and the systemic approach to engaging ethical expertise in videogames},
  author={Formosa, Paul and Ryan, Malcolm and Staines, Dan},
  journal={Ethics and Information Technology},
  volume={18},
  number={3},
  pages={211--225},
  year={2016},
  publisher={Springer}
}

@inproceedings{fun,
author = {Iacovides, Ioanna and Cox, Anna L.},
title = {Moving Beyond Fun: Evaluating Serious Experience in Digital Games},
year = {2015},
isbn = {9781450331456},
publisher = {Association for Computing Machinery},
address = {New York, NY, USA},
url = {https://doi.org/10.1145/2702123.2702204},
doi = {10.1145/2702123.2702204},
booktitle = {Proceedings of the 33rd Annual ACM Conference on Human Factors in Computing Systems},
pages = {2245–2254},
numpages = {10},
location = {Seoul, Republic of Korea},
series = {CHI '15}
}

@article{growth,
  title={The function of fiction is the abstraction and simulation of social experience},
  author={Mar, Raymond A and Oatley, Keith},
  journal={Perspectives on psychological science},
  volume={3},
  number={3},
  pages={173--192},
  year={2008},
  publisher={SAGE Publications Sage CA: Los Angeles, CA}
}

@inproceedings{intent,
  title={The INTENT Game: An Interactive Tool for Empathy in Neurotypicals},
  author={Girdhar, Varun and Tseng, Chao-Yang and Wang, Shiyu and Yang, Ruoxi and Ye, Zibo and Christel, Michael G and Stevens, Scott M and Evans, Morgan},
  booktitle={Joint International Conference on Serious Games},
  pages={433--439},
  year={2024},
  organization={Springer}
}

@inproceedings{video,
  title={Video games and their correlation to empathy: How to teach and experience empathic emotion},
  author={Wulansari, Ossy Dwi Endah and Pirker, Johanna and Kopf, Johannes and Guetl, Christian},
  booktitle={International Conference on Interactive Collaborative Learning},
  pages={151--163},
  year={2019},
  organization={Springer}
}

@inproceedings{HEP,
  title={Using heuristics to evaluate the playability of games},
  author={Desurvire, Heather and Caplan, Martin and Toth, Jozsef A},
  booktitle={CHI'04 extended abstracts on Human factors in computing systems},
  pages={1509--1512},
  year={2004}
}

@article{edu_replay,
  title={The Antecedents of and Associations with Elective Replay in an Educational Game: Is Replay Worth It?.},
  author={Liu, Zhongxiu and Cody, Christa and Barnes, Tiffany and Lynch, Collin and Rutherford, Teomara},
  journal={International Educational Data Mining Society},
  year={2017},
  publisher={ERIC}
}

@article{replayability,
  title={Replayability of video games},
  author={Frattesi, Timothy and Griesbach, Douglas and Leith, Jonathan and Shaffer, Timothy and DeWinter, Jennifer},
  journal={IQP, Worcester Polytechnic Institute, Worcester},
  year={2011}
}

@inproceedings{malone,
  title={Heuristics for designing enjoyable user interfaces: Lessons from computer games},
  author={Malone, Thomas W},
  booktitle={Proceedings of the 1982 conference on Human factors in computing systems},
  pages={63--68},
  year={1982}
}

@inproceedings{play,
author = {Desurvire, Heather and Wiberg, Charlotte},
title = {Game Usability Heuristics (PLAY) for Evaluating and Designing Better Games: The Next Iteration},
year = {2009},
isbn = {9783642027734},
publisher = {Springer-Verlag},
address = {Berlin, Heidelberg},
url = {https://doi.org/10.1007/978-3-642-02774-1_60},
doi = {10.1007/978-3-642-02774-1_60},
booktitle = {Proceedings of the 3d International Conference on Online Communities and Social Computing: Held as Part of HCI International 2009},
pages = {557–566},
numpages = {10},
location = {San Diego, CA},
series = {OCSC '09}
}

@inproceedings{mobile,
author = {Korhonen, Hannu and Koivisto, Elina M. I.},
title = {Playability heuristics for mobile games},
year = {2006},
isbn = {1595933905},
publisher = {Association for Computing Machinery},
address = {New York, NY, USA},
url = {https://doi.org/10.1145/1152215.1152218},
doi = {10.1145/1152215.1152218},
booktitle = {Proceedings of the 8th Conference on Human-Computer Interaction with Mobile Devices and Services},
pages = {9–16},
numpages = {8},
location = {Helsinki, Finland},
series = {MobileHCI '06}
}

@inproceedings{ensitive,
  title={The Challenge of Technology Research in Sensitive Settings: Case Studies in'ensitive HCI'},
  author={Waycott, Jenny and Wadley, Greg and Schutt, Stefan and Stabolidis, Arthur and Lederman, Reeva},
  booktitle={Proceedings of the Annual Meeting of the Australian Special Interest Group for Computer Human Interaction},
  pages={240--249},
  year={2015}
}

@article{acute,
  title={Acute musculoskeletal pain reported among rideshare drivers in the health/safety investigation among non-standard workers in the gig economy (HINGE) pilot study},
  author={Caban-Martinez, Alberto J and Santiago, Katerina M and Feliciano, Paola Louzado and Ogunsina, Kemi and Kling, Hannah and Griffin, Kevin and Solle, Natasha Schaefer},
  journal={Journal of occupational and environmental medicine},
  volume={62},
  number={5},
  pages={e236--e239},
  year={2020},
  publisher={LWW}
}

@article{pilot,
  title={Characterizing the health and safety concerns of US rideshare drivers: A qualitative pilot study},
  author={Louzado-Feliciano, Paola and Santiago, Katerina M and Ogunsina, Kemi and Kling, Hannah E and Murphy, Lauren A and Schaefer Solle, Natasha and Caban-Martinez, Alberto J},
  journal={Workplace health \& safety},
  volume={70},
  number={7},
  pages={310--318},
  year={2022},
  publisher={SAGE Publications Sage CA: Los Angeles, CA}
}

@article{crashes,
  title={Work-related crashes in rideshare drivers in the United States},
  author={Shannon, Brett and Friedman, Lee S and Hellinger, Andrew and Almberg, Kirsten and Ehsani, Johnathon},
  journal={Journal of safety research},
  volume={89},
  pages={13--18},
  year={2024},
  publisher={Elsevier}
}

@article{good,
  title={Good gig, bad gig: autonomy and algorithmic control in the global gig economy},
  author={Wood, Alex J and Graham, Mark and Lehdonvirta, Vili and Hjorth, Isis},
  journal={Work, employment and society},
  volume={33},
  number={1},
  pages={56--75},
  year={2019},
  publisher={Sage Publications Sage UK: London, England}
}

@article{license,
author = {Widder, David Gray and Dabbish, Laura and Herbsleb, James D. and Martelaro, Nikolas},
title = {Power and Play: Investigating "License to Critique" in Teams' AI Ethics Discussions},
year = {2024},
issue_date = {November 2024},
publisher = {Association for Computing Machinery},
address = {New York, NY, USA},
volume = {8},
number = {CSCW2},
url = {https://doi.org/10.1145/3686938},
doi = {10.1145/3686938},
journal = {Proc. ACM Hum.-Comput. Interact.},
month = nov,
articleno = {399},
numpages = {23}
}

@article{Freedman,
	author = {Freedman, Gili and Green, Melanie C. and Seidman, Max and Flanagan, Mary},
	journal = {Technology, Mind, and Behavior},
	number = {4},
	year = {2021},
	month = {nov 8},
	note = {https://tmb.apaopen.org/pub/6yzxhfgt},
	publisher = {},
	title = {The {Effect} of {Embodying} a {Woman} {Scientist} in {Virtual} {Reality} on {Men}\textquoteright{}s {Gender} {Biases}},
	volume = {2},
}

@article{free,
  title={A Free Market Approach to the Rideshare Industry and Worker Classification: The Consequences of Employee Status and a Proposed Alternative},
  author={Saltsman, Easton},
  journal={JL Econ. \& Pol'y},
  volume={13},
  pages={209},
  year={2017},
  publisher={HeinOnline}
}

@article{close,
author = {Iacovides, Ioanna and Cutting, Joe and Beeston, Jen and Cecchinato, Marta E. and Mekler, Elisa D. and Cairns, Paul},
title = {Close but Not Too Close: Distance and Relevance in Designing Games for Reflection},
year = {2022},
issue_date = {October 2022},
publisher = {Association for Computing Machinery},
address = {New York, NY, USA},
volume = {6},
number = {CHI PLAY},
url = {https://doi.org/10.1145/3549487},
doi = {10.1145/3549487},
journal = {Proc. ACM Hum.-Comput. Interact.},
month = oct,
articleno = {224},
numpages = {24}
}

@article{buffalo, title={A psychologically “embedded” approach to designing games for prosocial causes}, volume={9}, url={https://cyberpsychology.eu/article/view/4343}, DOI={10.5817/CP2015-3-5}, abstractNote={&lt;p&gt;Prosocial games often utilize a direct, explicit approach to engage players with serious real-life scenarios and present information about key societal issues. This approach, however, may limit a game’s persuasive impact and ability to produce beneficial outcomes, particularly when the apparent aims of the game trigger players’ psychological defenses or reduce players’ potential engagement with – and enjoyment of – the game experience. In contrast, the “Embedded Design” approach that we introduce here offers effective, evidence-based strategies for more stealthily or covertly delivering persuasive content in a game in a fashion that circumvents players’ psychological defenses and triggers a more receptive mindset. This paper provides an in-depth exploration of two key Embedded Design strategies: (1) intermixing: combining “on-topic” and “off-topic” game content in order to make the focal message or theme less obvious and more accessible and (2) obfuscating: using game genres or framing devices that direct players’ attention or expectations away from the game’s true aims. To illustrate the implementation and effectiveness of these strategies, we detail the design of two games that utilize a number of these techniques to reduce stereotypes and biases and present the methods and results of a set of empirical studies testing the prosocial impact of these games. In addition, we introduce a number of other Embedded Design strategies that have emerged in our work and discuss the most viable contexts for the use of this design approach.&lt;/p&#38;gt;}, number={3}, journal={Cyberpsychology: Journal of Psychosocial Research on Cyberspace}, author={Kaufman, Geoff and Flanagan, Mary}, year={2015}, month={Oct.}, pages={Article 5} }

@inproceedings{chimeria,
author = {Ortiz, Pablo and Harrell, D. Fox},
title = {Enabling Critical Self-Reflection through Roleplay with Chimeria: Grayscale},
year = {2018},
isbn = {9781450356244},
publisher = {Association for Computing Machinery},
address = {New York, NY, USA},
url = {https://doi.org/10.1145/3242671.3242687},
doi = {10.1145/3242671.3242687},
booktitle = {Proceedings of the 2018 Annual Symposium on Computer-Human Interaction in Play},
pages = {353–364},
numpages = {12},
location = {Melbourne, VIC, Australia},
series = {CHI PLAY '18}
}

@inproceedings{climate,
  title={Climate Change at Your Doorstep: An Experiment Using a Digital Game and Distance Framing},
  author={Fern{\'a}ndez Galeote, Daniel and Legaki, Nikoletta-Zampeta and Hamari, Juho},
  booktitle={Companion Proceedings of the 2024 Annual Symposium on Computer-Human Interaction in Play},
  pages={71--77},
  year={2024}
}

@article{moral,
  title={It’s my choice: The effects of moral decision-making on narrative game engagement},
  author={Ferchaud, Arienne and Beth Oliver, Mary},
  journal={Journal of Gaming \& Virtual Worlds},
  volume={11},
  number={2},
  pages={101--118},
  year={2019},
  publisher={Intellect}
}

@article{Bitter,
  author       = {Alex Bitter},
  title        = {Can you make it as an Uber driver? A new game simulates work in the gig economy},
  journal      = {Business Insider},
  year         = {2025},
  month        = {August 30},
  url          = {https://www.msn.com/en-us/money/companies/can-you-make-it-as-an-uber-driver-a-new-game-simulates-work-in-the-gig-economy/ar-AA1Lx9R9},
}

@misc{UberGame,
  author       = {{Financial Times}},
  title        = {The Uber Game},
  howpublished = {\url{https://ig.ft.com/uber-game/}},
  year         = {2017},
  note         = {Interactive journalism / news game.},
}

@inproceedings{ma,
  title={Toward Affective Empathy via Personalized Analogy Generation: A Case Study on Microaggression},
  author={Ju, Hyojin and Lee, Jungeun and Yang, Seungwon and Ok, Jungseul and Hwang, Inseok},
  booktitle={Proceedings of the 2025 CHI Conference on Human Factors in Computing Systems},
  pages={1--31},
  year={2025}
}

@article{qcae,
  title={The QCAE: A questionnaire of cognitive and affective empathy},
  author={Reniers, Renate LEP and Corcoran, Rhiannon and Drake, Richard and Shryane, Nick M and V{\"o}llm, Birgit A},
  journal={Journal of personality assessment},
  volume={93},
  number={1},
  pages={84--95},
  year={2011},
  publisher={Taylor \& Francis}
}

@article{gameful,
  title={Creating gameful design in mHealth: a participatory co-design approach},
  author={Jessen, Stian and Mirkovic, Jelena and Ruland, Cornelia M and others},
  journal={JMIR mHealth and uHealth},
  volume={6},
  number={12},
  pages={e11579},
  year={2018},
  publisher={JMIR Publications Inc., Toronto, Canada}
}

@article{corporate,
  title={Corporate social responsibility and crowdwashing in the gig economy},
  author={Cherry, Miriam A},
  journal={. Louis ULJ},
  volume={63},
  pages={1},
  year={2018},
  publisher={HeinOnline}
}

@article{schor,
  title={Consent and contestation: How platform workers reckon with the risks of gig labor},
  author={Schor, Juliet B and Tirrell, Christopher and Vallas, Steven Peter},
  journal={Work, employment and society},
  volume={38},
  number={5},
  pages={1423--1444},
  year={2024},
  publisher={SAGE Publications Sage UK: London, England}
}

@incollection{instability,
  title={Job instability, precarity, informality, and inequality: Labour in the gig economy},
  author={Rani, Uma and Gobel, Nora},
  booktitle={The Routledge handbook of the gig economy},
  pages={15--32},
  year={2022},
  publisher={Routledge}
}

@inproceedings{back,
author = {Tang, Joice and Andrus, McKane and So, Samuel and Tandon, Udayan and Monroy-Hern\'{a}ndez, Andr\'{e}s and Khovanskaya, Vera and Munson, Sean A. and Zachry, Mark and Ghoshal, Sucheta},
title = {Back to “ Back to Labor”: Revisiting Political Economies of Computer-Supported Cooperative Work},
year = {2023},
isbn = {9798400701290},
publisher = {Association for Computing Machinery},
address = {New York, NY, USA},
url = {https://doi.org/10.1145/3584931.3611285},
doi = {10.1145/3584931.3611285},
booktitle = {Companion Publication of the 2023 Conference on Computer Supported Cooperative Work and Social Computing},
pages = {522–526},
numpages = {5},
location = {Minneapolis, MN, USA},
series = {CSCW '23 Companion}
}

@inproceedings{reimagined,
author = {Zhang, Angie and Boltz, Alexander and Wang, Chun Wei and Lee, Min Kyung},
title = {Algorithmic Management Reimagined For Workers and By Workers: Centering Worker Well-Being in Gig Work},
year = {2022},
isbn = {9781450391573},
publisher = {Association for Computing Machinery},
address = {New York, NY, USA},
url = {https://doi.org/10.1145/3491102.3501866},
doi = {10.1145/3491102.3501866},
booktitle = {Proceedings of the 2022 CHI Conference on Human Factors in Computing Systems},
articleno = {14},
numpages = {20},
location = {New Orleans, LA, USA},
series = {CHI '22}
}

@techreport{wager,
  author       = {Niels Van Doorn},
  title        = {From a Wage to a Wager: Dynamic Pricing in the Gig Economy},
  institution  = {Autonomy},
  year         = {2020},
  type         = {Policy Brief},
  url          = {https://autonomy.work/wp-content/uploads/2020/09/VanDoorn.pdf?utm_source=chatgpt.com}
}

@inproceedings{turkopticon,
author = {Irani, Lilly C. and Silberman, M. Six},
title = {Turkopticon: interrupting worker invisibility in amazon mechanical turk},
year = {2013},
isbn = {9781450318990},
publisher = {Association for Computing Machinery},
address = {New York, NY, USA},
url = {https://doi.org/10.1145/2470654.2470742},
doi = {10.1145/2470654.2470742},
booktitle = {Proceedings of the SIGCHI Conference on Human Factors in Computing Systems},
pages = {611–620},
numpages = {10},
location = {Paris, France},
series = {CHI '13}
}

@inproceedings{self_tracking,
author = {Hernandez, Rie Helene (Lindy) and Song, Qiurong and Kou, Yubo and Gui, Xinning},
title = {"At the end of the day, I am accountable": Gig Workers' Self-Tracking for Multi-Dimensional Accountability Management},
year = {2024},
isbn = {9798400703300},
publisher = {Association for Computing Machinery},
address = {New York, NY, USA},
url = {https://doi.org/10.1145/3613904.3642151},
doi = {10.1145/3613904.3642151},
booktitle = {Proceedings of the 2024 CHI Conference on Human Factors in Computing Systems},
articleno = {382},
numpages = {20},
location = {Honolulu, HI, USA},
series = {CHI '24}
}

@book{psychosocial,
  title={Exposure to psychosocial risk factors in the gig economy: a systematic review},
  author={B{\'e}rast{\'e}gui, Pierre},
  number={2021.01},
  year={2021},
  publisher={Report}
}

@article{thelen, title={Regulating Uber: The Politics of the Platform Economy in Europe and the United States}, volume={16}, DOI={10.1017/S1537592718001081}, number={4}, journal={Perspectives on Politics}, author={Thelen, Kathleen}, year={2018}, pages={938–953}}

@incollection{handbook,
  author       = {Healy, Joshua and Pekarek, Andreas},
  title        = {Consumers in the Gig Economy: Resisting or Reinforcing Precarious Work?},
  booktitle    = {The Routledge Handbook of the Gig Economy},
  editor       = {Ness, Immanuel and Ovetz, Robert and Roque, Isabel and Swidler, Eva-Marie and Zwick, Austin},
  publisher    = {Routledge},
  year         = {2023},
  doi          = {10.4324/9781003161875-20}
}

@incollection{buttons,
  title={How Uber uses psychological tricks to push its drivers' buttons},
  author={Scheiber, Noam},
  booktitle={Ethics of data and analytics},
  pages={362--371},
  year={2022},
  publisher={Auerbach Publications}
}

@article{surveillance,
  title={The politics of platform power in surveillance capitalism: A comparative case study of ride-hailing platforms in China and the United States},
  author={Chan, Ngai Keung and Kwok, Chi},
  journal={Global Media and China},
  volume={7},
  number={2},
  pages={131--150},
  year={2022},
  publisher={SAGE Publications Sage UK: London, England}
}

@article{atom,
author = {Yao, Zheng and Weden, Silas and Emerlyn, Lea and Zhu, Haiyi and Kraut, Robert E.},
title = {Together But Alone: Atomization and Peer Support among Gig Workers},
year = {2021},
issue_date = {October 2021},
publisher = {Association for Computing Machinery},
address = {New York, NY, USA},
volume = {5},
number = {CSCW2},
url = {https://doi.org/10.1145/3479535},
doi = {10.1145/3479535},
journal = {Proc. ACM Hum.-Comput. Interact.},
month = oct,
articleno = {391},
numpages = {29}
}

@article{fareshare,
  title={FareShare: A Tool for Labor Organizers to Estimate Lost Wages and Contest Arbitrary AI and Algorithmic Deactivations},
  author={Rao, Varun Nagaraj and Dalal, Samantha and Schwartz, Andrew and Liaqat, Amna and Calacci, Dana and Monroy-Hern{\'a}ndez, Andr{\'e}s},
  journal={arXiv preprint arXiv:2505.08904},
  year={2025}
}

@article{happy,
  title={Uber happy? Work and well-being in the ‘gig economy’},
  author={Berger, Thor and Frey, Carl Benedikt and Levin, Guy and Danda, Santosh Rao},
  journal={Economic Policy},
  volume={34},
  number={99},
  pages={429--477},
  year={2019},
  publisher={Oxford University Press}
}

@article{rating_service,
  title={Platform-mediated reputation systems in the sharing economy and incentives to provide service quality: The case of ridesharing services},
  author={Basili, Marcello and Rossi, Maria Alessandra},
  journal={Electronic Commerce Research and Applications},
  volume={39},
  pages={100835},
  year={2020},
  publisher={Elsevier}
}

@article{manufactures,
author = {Lindsey D. Cameron},
title ={The Making of the “Good Bad” Job: How Algorithmic Management Manufactures Consent Through Constant and Confined Choices},
journal = {Administrative Science Quarterly},
volume = {69},
number = {2},
pages = {458-514},
year = {2024},
doi = {10.1177/00018392241236163},
URL = { https://doi.org/10.1177/00018392241236163
},
eprint = { 
        https://doi.org/10.1177/00018392241236163
}
}

@article{decent,
  title={Uber and employment in the Global South--not-so-decent work},
  author={Giddy, Julia K},
  journal={Tourism Geographies},
  volume={24},
  number={6-7},
  pages={1022--1039},
  year={2022},
  publisher={Taylor \& Francis}
}

@article{workaholism,
  title={Workaholism among young people in the ride-hailing travel economy},
  author={Adongo, Charles Atanga and Dayour, Frederick and Bukari, Shaibu and Akotoye, Evelyn Addison and Amissah, Eunice Fay},
  journal={Annals of Tourism Research Empirical Insights},
  volume={5},
  number={1},
  pages={100117},
  year={2024},
  publisher={Elsevier}
}

@misc{investoruber,
  title={Uber Announces Results for Fourth Quarter and Full Year 2023},
  author={Investor, Uber},
  year={2024}
}

@article{laundering,
author = {Michael David Maffie},
title ={The Perils of Laundering Control through Customers: A Study of Control and Resistance in the Ride-hail Industry},
journal = {ILR Review},
volume = {75},
number = {2},
pages = {348-372},
year = {2022},
doi = {10.1177/0019793920972679},
URL = { 
        https://doi.org/10.1177/0019793920972679
},
eprint = { 
        https://doi.org/10.1177/0019793920972679
}
}

@techreport{evolution,
  title={The Evolution of Platform Gig Work, 2012-2023},
  author={Garin, Andrew and Jackson, Emilie and Koustas, Dmitri and Miller, Alicia},
  year={2024},
  institution={SOI Working Paper}
}

@article{psychological,
  title={Dillema of Multi-Platform Ride-Hailing Drivers in Yogyakarta: Between Economic Demands and Psychological Impact in Achieving Well-Being},
  author={Nurrahmad, Daffa and Rahmasari, Salma and Pradana, Doni and Utomo, Syahriza and Putri, Mutiara and Afandi, Ardian},
  year={2023},
  publisher={Preprints}
}

@inproceedings{drives,
  title={Gamification in Ride-Hailing: What Drives a Driver to Drive},
  author={Hidajat, Taofik and Kusuma, Agung Hendra and Sulchan, Achmad},
  booktitle={The 3rd International Conference on Banking, Accounting, Management and Economics (ICOBAME 2020)},
  pages={241--244},
  year={2021},
  organization={Atlantis Press}
}

@article{org_justice, 
title={The real-time and carry-over effects of injustice on performance and service quality in a ridesharing driver scenario}, volume={42}, ISSN={1046-1310}, DOI={10.1007/s12144-022-04215-3}, abstractNote={The nature of gig work and its growth have important implications for organizational justice theory. Aspects of gig work, including the transactional compensation arrangement, strict algorithmic rating system, and power asymmetry between drivers and customers, have implications for understanding how dimensions of distributive, informational, and interpersonal injustice manifest and impact job performance in the gig context. An understanding of this topic can inform justice theory more broadly and help explain inconsistent findings in the literature. Here, we report the results of two studies examining the unique effects of these respective dimensions of injustice on emotions and, ultimately, the driving performance and service quality in a ridesharing service context. In Study 1, we modeled the passenger-driver interaction of the ridesharing context using a driving simulator in a laboratory setting to differentiate the real-time and carry-over effects of specific dimensions of injustice. The results from 99 participants showed that perceptions of interpersonal injustice increased anger and unhappiness during the ride, in turn impairing driving and service performance. Antecedent-focused emotion regulation strategies (ERS) reduced felt unhappiness. Moreover, unexpectedly, perceived distributive injustice as caused by the customer rating had opposite (direct versus indirect) effects on service performance in the subsequent ride. Study 2 was an online simulation vignette scenario with 294 participants. The results replicated the findings of Study 1 and revealed two moderators of the unexpected distributive justice-performance relationship. The online version contains supplementary material available at 10.1007/s12144-022-04215-3.}, number={36}, journal={Current Psychology (New Brunswick, N.j.)}, author={Lei, Xue and Kaplan, Seth A.}, year={2023}, pages={1–22} 
}

@article{rosenblat_discriminating,
  title={Discriminating tastes: Uber's customer ratings as vehicles for workplace discrimination},
  author={Rosenblat, Alex and Levy, Karen EC and Barocas, Solon and Hwang, Tim},
  journal={Policy \& Internet},
  volume={9},
  number={3},
  pages={256--279},
  year={2017},
  publisher={Wiley Online Library}
}

@incollection{dignity,
  author    = {Hodson, Randy},
  title     = {Dignity, Agency, and the Future of Work},
  booktitle = {Dignity at Work},
  publisher = {Cambridge University Press},
  year      = {2001},
  pages     = {259--273},
  doi       = {10.1017/CBO9780511499333.011}
}

@article{sabotage, title={How Gig Worker Responds to Negative Customer Treatment: The Effects of Work Meaningfulness and Traits of Psychological Resilience}, volume={12}, ISSN={1664-1078}, DOI={10.3389/fpsyg.2021.783372}, abstractNote={The negative interpersonal interaction between customers and platform gig workers has become a problem for platform owners and government. This study investigates the role of negative customer treatment in the context of gig work and its impact on gig workers’ sabotage behavior. A questionnaire survey approach was used in the study, collected three-wave survey data from 258 Chinese gig workers including food-deliver platform workers and app-based ride-hailing drivers. Both effects of the mediation and moderation were tested, all of which find support, using hierarchical multiple regression by SPSS22.0. Results indicate that negative customer treatment can also predict gig workers’ service sabotage through work meaningfulness. Furthermore, positive customer treatment acted as an effective safeguard against the effects of negative customer treatment on employee service sabotage. Trait psychological resilience can also mitigate the effects of a low level of work meaningfulness. The manuscript’s focus provides an interesting angle to the previous research, especially the inclusion of work meaningfulness and trait resilience, on negative customer treatment in the context of gig work. This study contributes to further broaden the perspective of conservation of resource (COR) theory for individual intrinsic motivation analysis. Practical implications for platform management and government governance have also been discussed in this manuscript.}, journal={Frontiers in Psychology}, author={Xiongtao, He and Wenzhu, Lu and Haibin, Luo and Shanshi, Liu}, year={2021}, pages={783372} }

@article{incivility, title={When passengers misbehave: exploring the pathways from customer incivility to service sabotage among ride-hailing drivers}, volume={39}, ISSN={0887-6045}, DOI={10.1108/jsm-10-2024-0551}, abstractNote={Grounded in conservation of resources theory, this study aims to examine the relationship between customer incivility and ride-hailing drivers’ (RHDs) service sabotage, testing emotional dissonance and ego depletion as serial mediators and emotional intelligence as a moderator. A three-wave, time-lagged data set collected from Pakistani RHDs through snowball sampling was analyzed using PROCESS macro in SPSS to test the moderated serial mediation model. The findings partially support the model, demonstrating that customer incivility significantly affects RHDs, resulting in service sabotage both directly and indirectly through the sequential mediation of emotional dissonance and ego depletion. Although emotional intelligence did not buffer the indirect effects of customer incivility, low emotional intelligence was found to intensify the customer incivility–service sabotage relationship, while high emotional intelligence neutralized this association. To the best of the authors’ knowledge, this study is among the first to offer a comprehensive examination of the negative impacts of customer incivility on RHDs in the gig economy. By introducing a moderated serial mediation framework, it provides novel insights into the psycho-emotional-behavioral consequences of customer mistreatment, addressing critical gaps in gig-economy service research and highlighting the role of emotional intelligence in mitigating adverse outcomes.}, number={5}, journal={Journal of Services Marketing}, author={Laeeque, Syed Harris and Ali, Madiha}, year={2025}, pages={426–441} }

@article{quality, title={Service Quality on Online Platforms: Empirical Evidence About Driving Quality at Uber}, DOI={10.2139/ssrn.5001370}, abstractNote={Online marketplaces have adopted new quality control mechanisms that can accommodate a flexible pool of providers. In the context of ride-hailing, we measure the effectiveness of these mechanisms, which include ratings, incentives, and behavioral nudges. Using telemetry data as an objective measure of quality, we find that drivers not only respond to user preferences but also improve their behavior after receiving warnings about their low ratings. Furthermore, we use data from a randomized experiment to show that informing drivers about their past behavior improves quality, especially for low-performing drivers. Lastly, we find that UberX drivers exhibit behavior comparable to that of UberTaxi drivers, suggesting that Uber’s new quality control mechanisms successfully maintain a high level of service quality.Institutional subscribers to the NBER working paper series, and residents of developing countries may download this paper without additional charge at www.nber.org.}, journal={SSRN Electronic Journal}, author={Athey, Susan and Castillo, Juan Camilo and Chandar, Bharat}, year={2024} }

@article{primed, title={Are We All Amazon Primed? Consumers and the Politics of Platform Power}, volume={53}, ISSN={0010-4140}, DOI={10.1177/0010414019852687}, abstractNote={This article articulates a distinctive source of political influence of some technology firms, which we call platform power. Platform power inheres in companies of economic scale that provide the terms of access through which large numbers of consumers access goods, services, and information. Firms with platform power benefit from a deference from policymakers, but this deference is not primarily a function of direct influence through lobbying or campaign contributions, nor does it come from the threat of disinvestment. Companies with platform power instead benefit from the tacit allegiance of consumers, who can prove a formidable source of opposition to regulations that threaten these platforms. Focusing on the critical role played by consumers in explaining the powers platform firms wield in the rich democracies lends insight as well into their distinctive vulnerabilities, which flow from events that split the consumer–platform alliance or that cue citizen, as opposed to consumer, political identities.}, number={2}, journal={Comparative Political Studies}, author={Culpepper, Pepper D. and Thelen, Kathleen}, year={2020}, pages={288–318} }

@article{disrupt, title={Disrupting Regulation, Regulating Disruption: The Politics of Uber in the United States}, volume={16}, ISSN={1537-5927}, DOI={10.1017/s1537592718001093}, abstractNote={Platform companies disrupt not only the economic sectors they enter, but also the regulatory regimes that govern those sectors. We examine Uber in the United States as a case of regulating this disruption in different arenas: cities, state legislatures, and judicial venues. We find that the politics of Uber regulation does not conform to existing models of regulation. We describe instead a pattern of “disruptive regulation”, characterized by a challenger-incumbent cleavage, in two steps. First, an existing regulatory regime is not deregulated but successfully disregarded by a new entrant. Second, the politics of subsequently regulating the challenger leads to a dual regulatory regime. In the case of Uber, disruptive regulation takes the form of challenger capture, an elite-driven pattern, in which the challenger has largely prevailed. It is further characterized by the surrogate representation of dispersed actors—customers and drivers—who do not have autonomous power and who rely instead on shifting alignments with the challenger and incumbent. In its surrogate capacity in city and state regulation, Uber has frequently mobilized large numbers of customers and drivers to lobby for policy outcomes that allow it to continue to provide service on terms it finds acceptable. Because drivers have reaped less advantage from these alignments, labor issues have been taken up in judicial venues, again primarily by surrogates (usually plaintiffs’ attorneys) but to date have not been successful.}, number={4}, journal={Perspectives on Politics}, author={Collier, Ruth Berins and Dubal, V.B. and Carter, Christopher L.}, year={2018}, pages={919–937} }

@article{power_resources, title={Power resources for disempowered workers? Re-conceptualizing the power and potential of consumers in app-based food delivery}, volume={63}, ISSN={0019-8676}, DOI={10.1111/irel.12340}, abstractNote={Consumers play an integral role in the labor process of app-based food delivery services through their consumption behaviors and performance ratings of workers. Some therefore see them as a potential ally of workers, whereas others view them as beholden by capital. This quantitative study uses power resource theory and a Rasch model to appraise consumers’ understandings and attitudes toward working conditions in this segment of the “gig” economy. Drawing on two surveys of 1820 Australian consumers, we find that consumers are a potential yet heterogenous coalitional power resource who may align with workers on certain entitlements like minimum wages.}, number={2}, journal={Industrial Relations: A Journal of Economy and Society}, author={Goods, Caleb and Veen, Alex and Barratt, Tom and Smith, Brett}, year={2024}, pages={107–131} }

@article{spektor, title={Charting the Automation of Hospitality: An Interdisciplinary Literature Review Examining the Evolution of Frontline Service work in the Face of Algorithmic Management}, volume={7}, DOI={10.1145/3579466}, abstractNote={Recent investments in automation and AI are reshaping the hospitality sector. Driven by social and economic forces affecting service delivery, these new technologies have transformed the labor that acts as the backbone to the industry-namely frontline service work performed by housekeepers, front desk staff, line cooks and others. We describe the context for recent technological adoption, with particular emphasis on algorithmic management applications. Through this work, we identify gaps in existing literature and highlight areas in need of further research in the domains of worker-centered technology development. Our analysis highlights how technologies such as algorithmic management shape roles and tasks in the high-touch service sector. We outline how harms produced through automation are often due to a lack of attention to non-management stakeholders. We then describe an opportunity space for researchers and practitioners to elicit worker participation at all stages of technology adoption, and offer methods for centering workers, increasing transparency, and accounting for the context of use through holistic implementation and training strategies.}, number={CSCW1}, journal={Proceedings of the ACM on Human-Computer Interaction}, author={Spektor, Franchesca and Fox, Sarah E. and Awumey, Ezra and Begleiter, Ben and Kulkarni, Chinmay and Stringam, Betsy and Riordan, Christine A. and Rho, Hye Jin and Akridge, Hunter and Forlizzi, Jodi}, year={2023}, pages={1–20} }

@article{located,
  title={Located accountabilities in technology production},
  author={Suchman, Lucy},
  journal={Scandinavian journal of information systems},
  volume={14},
  number={2},
  pages={7},
  year={2002}
}

@inproceedings{f2f,
author = {Spektor, Franchesca and Fox, Sarah E and Min, Somang and Sarfo, Grace and Stringam, Betsy and Riordan, Christine A. and Rho, Hye Jin and Begleiter, Ben and Forlizzi, Jodi},
title = {Working Together: Algorithmic Management and Peer Relationships in the Hospitality Industry},
year = {2025},
isbn = {9798400714856},
publisher = {Association for Computing Machinery},
address = {New York, NY, USA},
url = {https://doi.org/10.1145/3715336.3735704},
doi = {10.1145/3715336.3735704},
booktitle = {Proceedings of the 2025 ACM Designing Interactive Systems Conference},
pages = {3221–3234},
numpages = {14},
location = {
},
series = {DIS '25}
}

@article{guardian,
  title={Uber stripped of London licence due to lack of corporate responsibility},
  author={Butler, Sarah and Topham, Gwyn},
  journal={The Guardian},
  volume={23},
  pages={09--17},
  year={2017}
}

@article{tested,
  title={More than 500,000 people sign Uber petition to overturn London ban},
  author={Dave, P and Schomberg, W},
  journal={The Sydney Morning Herald},
  volume={24},
  year={2017}
}

@article{million,
  title={More than 500,000 sign petition to save Uber as firm fights London Ban},
  author={Farrer, M and Khomami, N},
  journal={The Guardian, Sept},
  volume={23},
  year={2017}
}

@article{awareness,
  title={From Awareness to Action: Compassionate Consumers as Agents of Change in the Gig Economy An Experimental Design Study on the Influence of Compassion on Prosocial Behavior in the Gig Economy.},
  author={Larsson, Kerstin},
  year={2023}
}

@article{soft_hard, title={The “soft” and “hard” sides of the sharing economy: a discussion of marketing, financial and socio-cultural aspects}, volume={24}, ISSN={1753-3627}, DOI={10.1504/ijbg.2020.106478}, abstractNote={This paper discusses three aspects of the sharing economy: the marketing aspect, the financial aspect, and the socio-cultural aspect. The discussion of the marketing aspect suggests that mindful consumption and advances in technology have made sharing economy possible and increasingly popular. This has modified consumption patterns, particularly in the tourism industry, and has changed traditional marketing channels. It is further suggested, that the increasing prevalence of sharing economy practices is directly linked to increases in revenue and improvements in customer welfare and job creation. Moreover, socio-cultural characteristics of societies have also been linked to the provision and use of sharing economy. More specifically, attitudes towards innovation, sharing, trust and environmental sensitivity have been identified as socio-cultural factors that are expected to influence the provision of sharing economy services, and individualism-collectivism, uncertainty avoidance, power distance and the internet appear to be key behaviour determinant variables.}, number={3}, journal={International Journal of Business and Globalisation}, author={Melanthiou, Yioula and Evripidou, Loukia and Epaminonda, Epaminondas and Komodromos, Marcos}, year={2020}, pages={330–346} }

@article{tipping,
  title={Explaining the decline of tipping norms in the gig economy},
  author={Duhaime, Erik P and Woessner, Zachary W},
  journal={Journal of Managerial Psychology},
  volume={34},
  number={4},
  pages={233--245},
  year={2019},
  publisher={Emerald Publishing Limited}
}

@article{csr,
  title={Does corporate social responsibility affect consumer boycotts? A cost--benefit approach},
  author={Zeng, Tian and Audrain-Pontevia, Anne-Fran{\c{c}}oise and Durif, Fabien},
  journal={Corporate Social Responsibility and Environmental Management},
  volume={28},
  number={2},
  pages={796--807},
  year={2021},
  publisher={Wiley Online Library}
}

@article{enhancing,
  title={Enhancing user engagement: The role of gamification in mobile apps},
  author={Bitri{\'a}n, Paula and Buil, Isabel and Catal{\'a}n, Sara},
  journal={Journal of Business Research},
  volume={132},
  pages={170--185},
  year={2021},
  publisher={Elsevier}
}

@article{self_brand,
  title={Gamified interactions: whether, when, and how games facilitate self--brand connections},
  author={Berger, Axel and Schlager, Tobias and Sprott, David E and Herrmann, Andreas},
  journal={Journal of the Academy of Marketing Science},
  volume={46},
  number={4},
  pages={652--673},
  year={2018},
  publisher={Springer}
}

@article{equity,
  title={Does gamification affect brand engagement and equity? A study in online brand communities},
  author={Xi, Nannan and Hamari, Juho},
  journal={Journal of Business Research},
  volume={109},
  pages={449--460},
  year={2020},
  publisher={Elsevier}
}

@article{mandatory, title={Mandatory Fun: Consent, Gamification and the Impact of Games at Work}, DOI={10.2139/ssrn.2277103}, abstractNote={In an effort to create a positive experience at work, managers have deployed a wide range of initiatives and practices designed to improve the affective experience for workers. One such practice is gamification, introducing elements from games into the work environment with the purpose of improving employees’ affective experiences. Games have long been played at work, but they have emerged spontaneously from the employees themselves. Here, we examine whether managerially-imposed games provide the desired benefits for affect and performance predicted by prior studies on games at work or whether they are a form of “mandatory fun.” We highlight the role of consent (Burawoy, 1979) as a psychological response to mandatory fun, which moderates these relationships and, in a field experiment, find that games, when consented to, increase positive affect at work, but, when consent is lacking, decrease positive affect. In a follow up laboratory experiment, we also find that legitimation and a sense of individual agency are important sources of consent.}, journal={SSRN Electronic Journal}, author={Mollick, Ethan R. and Rothbard, Nancy}, year={2014} }

@article{terrain, title={Algorithms at Work: The New Contested Terrain of Control}, volume={14}, ISSN={1941-6520}, DOI={10.5465/annals.2018.0174}, number={1}, journal={Academy of Management Annals}, author={Kellogg, Katherine C and Valentine, Melissa A and Christin, Angéle}, year={2020}, pages={366–410} }

@article{humanistic,
  title={Gamification in management: Between choice architecture and humanistic design},
  author={Deterding, Sebastian},
  journal={Journal of Management Inquiry},
  volume={28},
  number={2},
  pages={131--136},
  year={2019},
  publisher={SAGE Publications Sage CA: Los Angeles, CA}
}

@article{paternalism,
  title={Regulation for conservatives: Behavioral economics and the case for" asymmetric paternalism"},
  author={Camerer, Colin and Issacharoff, Samuel and Loewenstein, George and O'donoghue, Ted and Rabin, Matthew},
  journal={University of Pennsylvania law review},
  volume={151},
  number={3},
  pages={1211--1254},
  year={2003},
  publisher={JSTOR}
}

@article{positionality,
  title={Positionality, epistemology, and social justice in the classroom},
  author={Takacs, David},
  journal={Social justice},
  volume={29},
  number={4 (90},
  pages={168--181},
  year={2002},
  publisher={JSTOR}
}

@techreport{taxi_hire,
  title        = {Taxis and private hire vehicles in the UK transport system: how and why are they changing?},
  author       = {Marcus Enoch},
  institution  = {Government Office for Science, UK},
  year         = {2018},
  month        = dec,
  type         = {Foresight Future of Mobility: Evidence Review},
  url          = {https://assets.publishing.service.gov.uk/media/5c1a6060ed915d0c3d63f6d4/Taxisprivatehire.pdf},
  note         = {Information cut-off April 2018; © Crown copyright 2018}
}

@inproceedings{morality,
author = {Ma, Shuhao and Zimmerman, John and Fox, Sarah E and Nisi, Valentina and Nunes, Nuno Jardim},
title = {"My Sense of Morality Leads to My Suffering, Battling, and Arguing": The Role of Platform Designers in (Un)Deciding Gig Worker Issues},
year = {2024},
isbn = {9798400705830},
publisher = {Association for Computing Machinery},
address = {New York, NY, USA},
url = {https://doi.org/10.1145/3643834.3660713},
doi = {10.1145/3643834.3660713},
booktitle = {Proceedings of the 2024 ACM Designing Interactive Systems Conference},
pages = {3501–3514},
numpages = {14},
location = {Copenhagen, Denmark},
series = {DIS '24}
}

@article{resistance,
  title={Resistance towards autonomous vehicles (AVs)},
  author={Nordhoff, S},
  journal={Transportation Research Interdisciplinary Perspectives},
  volume={26},
  pages={101117},
  year={2024},
  publisher={Elsevier}
}

@article{karahalios,
  title={Auditing Algorithms From the Outside: Methods and Implications},
  author={Karahalios, Karrie}
}

@misc{join,
  author       = {Octopus Interactive},
  title        = {Apply to get a FREE Octopus game tablet — Join Play Octopus},
  howpublished = {\url{https://account.playoctopus.com/join}},
  note         = {Accessed: 2025-12-02}
}

@article{prosumer,
  title={Platform capitalism’s hidden abode: producing data assets in the gig economy},
  author={Van Doorn, Niels and Badger, Adam},
  journal={Antipode},
  volume={52},
  number={5},
  pages={1475--1495},
  year={2020},
  publisher={Wiley Online Library}
}

@article{cpr,
  title={The sharing economy and consumer protection regulation: The case for policy change},
  author={Koopman, Christopher and Mitchell, Matthew and Thierer, Adam},
  journal={J. Bus. Entrepreneurship \& l.},
  volume={8},
  pages={529},
  year={2014},
  publisher={HeinOnline}
}

@inproceedings{policycraft,
  title={PolicyCraft: Supporting Collaborative and Participatory Policy Design through Case-Grounded Deliberation},
  author={Kuo, Tzu-Sheng and Chen, Quan Ze and Zhang, Amy X and Hsieh, Jane and Zhu, Haiyi and Holstein, Kenneth},
  booktitle={Proceedings of the 2025 CHI Conference on Human Factors in Computing Systems},
  pages={1--24},
  year={2025}
}

@inproceedings{flash,
author = {Valentine, Melissa A. and Retelny, Daniela and To, Alexandra and Rahmati, Negar and Doshi, Tulsee and Bernstein, Michael S.},
title = {Flash Organizations: Crowdsourcing Complex Work by Structuring Crowds As Organizations},
year = {2017},
isbn = {9781450346559},
publisher = {Association for Computing Machinery},
address = {New York, NY, USA},
url = {https://doi.org/10.1145/3025453.3025811},
doi = {10.1145/3025453.3025811},
booktitle = {Proceedings of the 2017 CHI Conference on Human Factors in Computing Systems},
pages = {3523–3537},
numpages = {15},
location = {Denver, Colorado, USA},
series = {CHI '17}
}

@article{locus,
  title={Expanding the locus of resistance: Understanding the co-constitution of control and resistance in the gig economy},
  author={Cameron, Lindsey D and Rahman, Hatim},
  journal={Organization Science},
  volume={33},
  number={1},
  pages={38--58},
  year={2022},
  publisher={INFORMS}
}

@article{local,
  title={Off-grid? Resistant media operations by delivery gig workers in response to app-based tracking},
  author={Randerath, Sebastian and Friedrich, Kathrin},
  journal={Mobile Media \& Communication},
  volume={13},
  number={1},
  pages={214--233},
  year={2025},
  publisher={SAGE Publications Sage UK: London, England}
}

@inproceedings{bu_eu,
  title={A bottom-up end-user intelligent assistant approach to empower gig workers against AI inequality},
  author={Li, Toby Jia-Jun and Lu, Yuwen and Clark, Jaylexia and Chen, Meng and Cox, Victor and Jiang, Meng and Yang, Yang and Kay, Tamara and Wood, Danielle and Brockman, Jay},
  booktitle={Proceedings of the 1st Annual Meeting of the Symposium on Human-Computer Interaction for Work},
  pages={1--10},
  year={2022}
}

@inproceedings{replay,
author = {Holstein, Kenneth and Harpstead, Erik and Gulotta, Rebecca and Forlizzi, Jodi},
title = {Replay Enactments: Exploring Possible Futures through Historical Data},
year = {2020},
isbn = {9781450369749},
publisher = {Association for Computing Machinery},
address = {New York, NY, USA},
url = {https://doi.org/10.1145/3357236.3395427},
doi = {10.1145/3357236.3395427},
booktitle = {Proceedings of the 2020 ACM Designing Interactive Systems Conference},
pages = {1607–1618},
numpages = {12},
location = {Eindhoven, Netherlands},
series = {DIS '20}
}

@article{video_capture,
  title={Delivery workers’ visibility struggles: weapons of the gig,(extra) ordinary social media, and strikes},
  author={Bulut, Ergin and Ye{\c{s}}ilyurt, Adem},
  journal={Convergence},
  volume={30},
  number={1},
  pages={450--470},
  year={2024},
  publisher={SAGE Publications Sage UK: London, England}
}

@inproceedings{simulated_CA,
  title={Workers vs. The Algorithm: Simulating Collective Action in Gig-Economy Platforms},
  author={Lewandowska, Kristina},
  booktitle={NeurIPS 2025 Workshop on Algorithmic Collective Action}
}

@article{fairness,
  title={Fairness-enhancing vehicle rebalancing in the ride-hailing system},
  author={Guo, Xiaotong and Xu, Hanyong and Zhuang, Dingyi and Zheng, Yunhan and Zhao, Jinhua},
  journal={arXiv preprint arXiv:2401.00093},
  year={2023}
}

@inproceedings{quallm,
    title = "{Q}ua{LLM}: An {LLM}-based Framework to Extract Quantitative Insights from Online Forums",
    author = "Nagaraj Rao, Varun  and
      Agarwal, Eesha  and
      Dalal, Samantha  and
      Calacci, Dana  and
      Monroy-Hern{\'a}ndez, Andr{\'e}s",
    editor = "Chiruzzo, Luis  and
      Ritter, Alan  and
      Wang, Lu",
    booktitle = "Findings of the Association for Computational Linguistics: NAACL 2025",
    month = apr,
    year = "2025",
    address = "Albuquerque, New Mexico",
    publisher = "Association for Computational Linguistics",
    url = "https://aclanthology.org/2025.findings-naacl.74/",
    doi = "10.18653/v1/2025.findings-naacl.74",
    pages = "1355--1369",
    ISBN = "979-8-89176-195-7"
}

@article{planning,
  title={The effect of planning and monitoring as metacognitive strategies on Iranian EFL learners’ argumentative writing accuracy},
  author={Panahandeh, Esmaeil and Asl, Shahram Esfandiari},
  journal={Procedia-Social and Behavioral Sciences},
  volume={98},
  pages={1409--1416},
  year={2014},
  publisher={Elsevier}
}

@article{monitoring,
  title={A conceptual analysis of five measures of metacognitive monitoring},
  author={Schraw, Gregory},
  journal={Metacognition and learning},
  volume={4},
  number={1},
  pages={33--45},
  year={2009},
  publisher={Springer}
}

@inproceedings{metacognitions,
author = {Prather, James and Becker, Brett A. and Craig, Michelle and Denny, Paul and Loksa, Dastyni and Margulieux, Lauren},
title = {What Do We Think We Think We Are Doing? Metacognition and Self-Regulation in Programming},
year = {2020},
isbn = {9781450370929},
publisher = {Association for Computing Machinery},
address = {New York, NY, USA},
url = {https://doi.org/10.1145/3372782.3406263},
doi = {10.1145/3372782.3406263},
booktitle = {Proceedings of the 2020 ACM Conference on International Computing Education Research},
pages = {2–13},
numpages = {12},
location = {Virtual Event, New Zealand},
series = {ICER '20}
}

@article{south,
  title={The gig economy in Chile: Examining labor conditions and the nature of gig work in a Global South country},
  author={Arriagada, Arturo and Bonhomme, Macarena and Ib{\'a}{\~n}ez, Francisco and Leyton, Jorge},
  journal={Digital Geography and Society},
  volume={5},
  pages={100063},
  year={2023},
  publisher={Elsevier}
}
\newpage
\section{Appendix}
\begin{figure}[h!]
    \centering
\includegraphics[width=.9\linewidth]{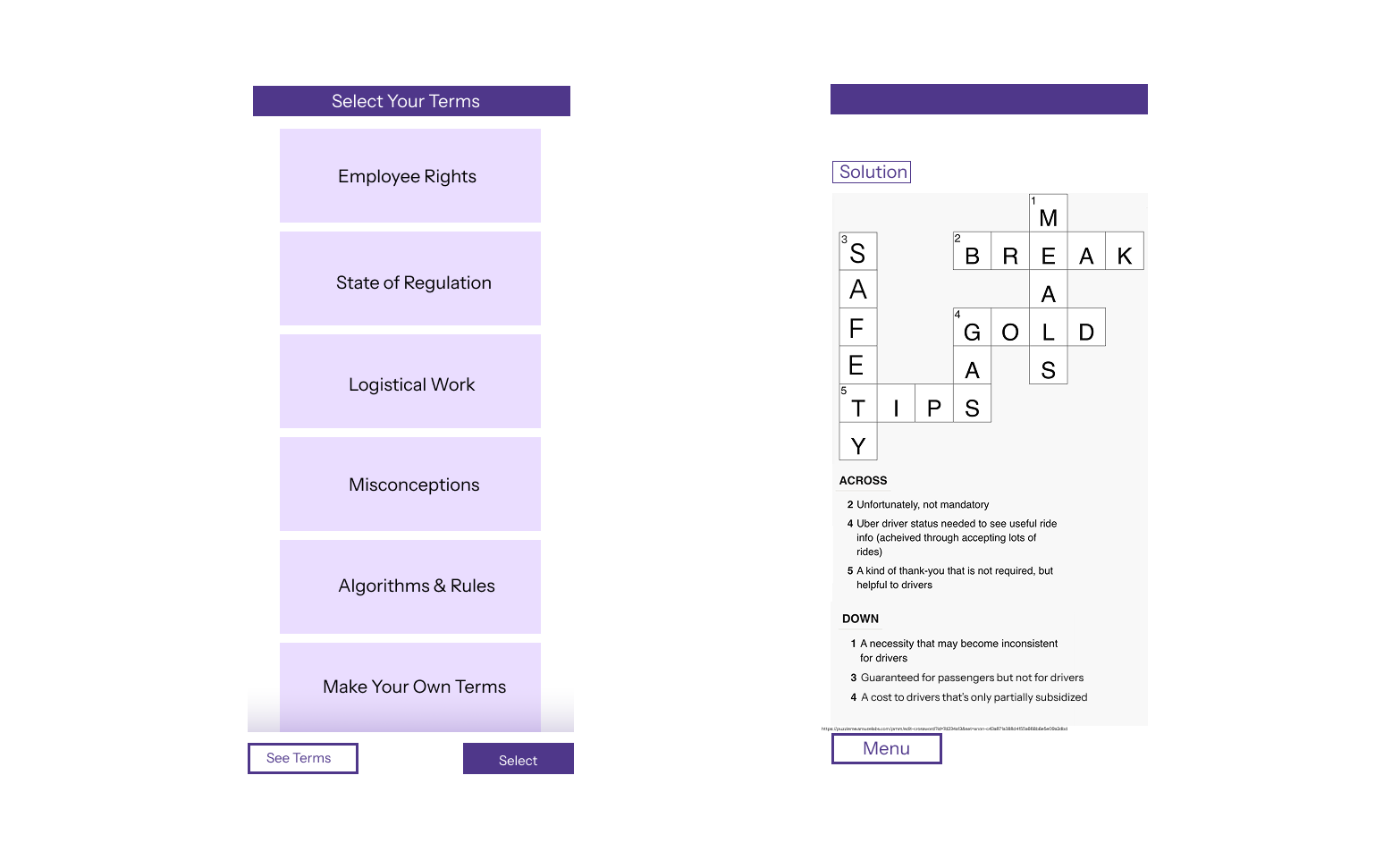}
    \caption{\elim{CrossRoads} contains driver-selected ridehail concepts}
    \label{crossroad}
\end{figure}
\FloatBarrier
\subsection{\elim{CrossRoads}} 
Resembling standard crossword puzzles, \elim{CrossRoads} embeds ridehail-related terms and clues to expose such knowledge to  passengers. In addition to incorporating ground-truth ridehail concepts, \elim{CrossRoads} contain mechanisms allowing drivers to pick and define their own terms and clues, affording them agency to select ridehail topics most relevant to their own experiences while improve replayability of the game for passengers across rides. The puzzle nature of the crossword orients players to focus on character order as opposed to the ridehail-related terms, but the small screensize of mobile and tablet devices limits content and therefore potential to introduce intermixing.
In this prototype, we embedded concrete definitions of concepts (fulfilling requirement for ground truth answers) related to driver rights, regulations, algorithmic management strategies, and logistical burdens.

While driver D2.1 saw potential in \elim{CrossRoads} as a ``\textit{good distraction}'' from work for riders, others found it ``\textit{boring}'', cognitively demanding, and criticized its lack of a ``\textit{social loop}'' to interact with the driver, as well as in failing to contribute to ``\textit{energy I'd want in a fun way}'' - D2.2. Combining these concerns with the difficulties of typing on a tablet keyboard, we decided to eliminate this prototype after the second driver workshop.

\begin{figure}[h!]
    \centering
\includegraphics[width=.5\linewidth]{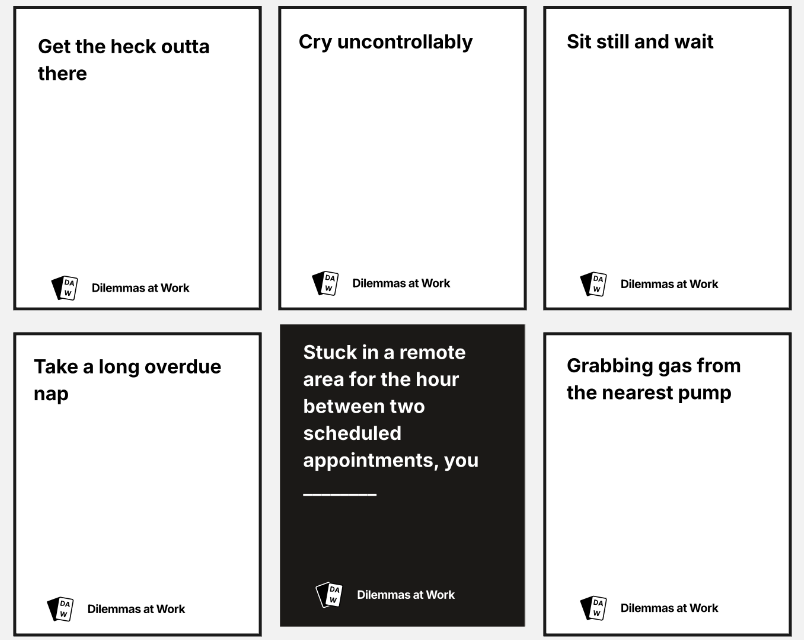}
    \caption{\elim{Dilemmas @ Work} contain black cards representing (ridehail) work dilemmas \& white cards with actions to take in response}
    \label{crossroad}
\end{figure}
\FloatBarrier
\subsection{\elim{Dilemmas @ Work}} 
Based off of the popular social party game \textit{Cards Against Humanity} and inspired by related applications of the card game towards discussion of contexts such as AI ethics \cite{license}, as well as driver-led advocacy \cite{cards}, we prototyped \elim{Dilemmas @ Work}, which adapted the black card deck to represent work dilemmas that drivers and traditional workers might face in their everyday labor -- leveraging intermixing (\S\ref{intermix}). Correspondingly, white cards depicted potential strategies for handling the various dilemmas presented in black. Designed for a physical social context, the random dealing of cards each round creates a replayable (\S\ref{replayable}) experience even among the same group of players. Meanwhile, the objective of humoring the dealing player of each round serves to obfuscate the concepts ridehail vulnerabilities.

Participants of the first driver workshop ranked \elim{Dilemmas @ Work} as the lowest among presented prototypes, explaining how its design to operate outside of a ride discourages engagement with the topic: ``\textit{I don't know that I would see many people actually doing it, if the purpose of this is to educate riders on the driver experience}'', especially since they believed ``\textit{the impetus for any of these games would be [with how they are played] during a ride}'' - DW1.1.

\subsection{Passenger Protocol}
\begin{enumerate}
    \item \textbf{[Character Card Activity]} Let’s start off by sharing how long you’ve been a rider, how frequently you ride, what platform(s) you use, and where?
    \item \ul{What do you know about rideshare?}
    \begin{enumerate}
        \item What aspects of rideshare driving do you know of that are maybe surprising or concerning?
        \item Other than driving, what other responsibilities do you think drivers have when working for their platform?
        \item Reflecting on your experiences as a passenger, what rides have you taken that are most memorable? Did you learn anything about rideshare driving that was interesting or surprising?
        \item What makes you tip a driver more, or give higher, 5-star ratings?
        \item On the other hand, what types of rides make you most stressed out or annoyed?
    \end{enumerate}
    \item \ul{Introducing Game Prototypes }
    \begin{enumerate}
        \item \textit{Ranking Game Heuristics}
        \begin{enumerate}
            \item Rank the games based on which you thought were the most fun. How you define fun is up to you, but please explain your rationale.
            \item Rank the games based on whether you could see yourself(selves) playing more than once?
            \item Please rank the games based on how well-integrated you thought the concepts of ridesharing were (most sneaky)? Would you have noticed the concepts if we didn’t point them out? How did it impact the gameplay?
            \item Rank based on how easy the games would be to play in an uber ride – potentially on an Octopus-like tablet
            \item Rank based on the amount of collaboration/interaction you could have/would want to have with the driver in each game.
            \item Rank based on which games are the most casual/easiest to pick up/stop at any point – easiest on the mind.
            \item Rank each one based on how likely you would be to recommend the games to a friend?
            \item Rank which ones make you more curious about rideshare driving conditions. Which ones serve as a launch point and make you want to learn more about them?
            \end{enumerate}
        \item \textit{Embedded Concepts}
        \begin{enumerate}
            \item Does this game capture the rideshare 
            \item What other aspects of ridesharing are you curious about? Could you imagine it as a part of the games? 
            \item Last row are for important concepts that don’t yet fit with the existing game prototypes
        \end{enumerate}
        concept(s) accurately? 
    \end{enumerate}
    \item \ul{Reflections}
    \begin{enumerate}
        \item Other than gamified experiences, what are other interactive modes of communicating driving conditions that you'd be receptive to?
        \item Were there any particular concepts that would be better implemented in this format rather than in a game?
        \item Take this time to revisit your character card rankings from the beginning, how did these change?
    \end{enumerate}
    \item Which driving conditions issues are you most curious or concerned about?
\end{enumerate}

\subsection{Driver Protocol}
\begin{enumerate}
    \item \ul{Why rideshare?}
    \begin{enumerate}
        \item Why did each of you start rideshare, and how long have you been driving?
        \item What are you hoping to learn or gain from the workshop today (can be from us or the other drivers)?
        \item  Now (as a group) choose the question that you would all like to share with passengers the most. Why this question?
        \item How often do riders talk to you about these with you?
        \item By contrast, which issues are not necessarily good ideas to be discussed with passengers or in a game?
    \end{enumerate}
    \item \ul{Introducing Game Prototypes}
    \begin{enumerate}
        \item Does this game capture the rideshare concept(s) accurately? 
        \item What other aspects of ridesharing can you imagine this game to include? 
        \item How does your experience shape the way you ordered the pieces of information?
        \item Would you want to subject passengers to similar decisions or experiences (even if simulated) as you?
        \item Are you interested in knowing more about drivers who identify differently than you? Why [what insights would this tell you if you knew this]?
    \end{enumerate}
    \item \ul{Reflections}
    \begin{enumerate}
        \item Jot down your thoughts, concerns and additional ideas about these proposed gaming experiences
        \item Choose the concern that you think would be the biggest blocker to riders adopting this game. How come?
    \end{enumerate}
\end{enumerate}
\end{document}